\documentclass[a4paper, 11pt, abstracton, titlepage, oneside]{scrartcl}
\makeatletter

\usepackage[T1]{fontenc}
\usepackage[utf8]{inputenc}
\usepackage[onehalfspacing]{setspace}
\usepackage[headsepline]{scrlayer-scrpage}
	\clearscrheadfoot 
\usepackage{layout}
\usepackage[english]{babel}
\usepackage[left=2cm,right=2cm,top=2.5cm,bottom=2.5cm]{geometry}

\usepackage{algorithm}
\usepackage[noend]{algpseudocode}
\usepackage{adjustbox}
\usepackage{eurosym}
\usepackage{lscape}
\usepackage{nicefrac}
\usepackage{array}
\usepackage{paralist}
\usepackage{authblk}
\usepackage{graphicx}
\usepackage{booktabs}
\usepackage{placeins}
\usepackage{amsmath}
\usepackage{mathtools}
\usepackage{bm}
\usepackage{amssymb}
\usepackage{color}
\usepackage[acronym]{glossaries}
\usepackage{pgfplots}
\usepackage[normal]{threeparttable}
\usepackage{ulem}
\usepackage{natbib}
\usepackage[titletoc,title]{appendix}
\usepackage[hyphens]{url} 
\usepackage{ellipsis}
\usepackage{breqn} 
\let\counterwithin\relax

\usepackage{chngcntr}  
\usepackage{enumerate}
\usepackage{theorem}
\usepackage{etoolbox}
\usepackage{multirow}
\usepackage{colortbl}
\usepackage{subcaption}
\newcommand\fnote[1]{\captionsetup{font=small}\caption*{#1}}
\usepackage{tikz}
\usetikzlibrary{arrows.meta}
\usepgfplotslibrary{groupplots}
\usepgfplotslibrary{dateplot}
\usepackage[utf8]{inputenc}
\ifdefined\DeclareUnicodeCharacter
\DeclareUnicodeCharacter{2212}{-}
\fi

\usepackage{amsthm}
\newtheorem{prop}{Proposition}

\ifdefined\mathleft
\else
\newcommand{\mathleft}{}
\fi	
\ifdefined\mathright
\else
\newcommand{\mathright}{}
\fi
\ifdefined\mathcenter
\else
\newcommand{\mathcenter}{}
\fi

\DeclareMathOperator*{\argmin}{arg\,min}

\makeatletter
\providecommand\phantomcaption{\caption@refstepcounter\@captype}
\makeatother

\newglossary{symbolslist}{1i}{1o} {Table of notation}
\newglossary{heuristlslist}{2i}{2o}{Heuristic variants}
\newglossary{problemlist}{3i}{3o}{Sequential Decision Processes}
\newglossary{appendixlist}{4i}{4o}{Appendix Notation}
\newglossary{soclist}{5i}{5o}{SoC model extension}

\makeglossaries

\newglossaryentry{low_avail}{name={\textit{low-15\%}},
	description={Low-density scenario},
	type=heuristlslist}
\newglossaryentry{avg_avail}{name={\textit{avg-60\%}},
	description={Medium-density scenario},
	type=heuristlslist}
\newglossaryentry{high_avail}{name={\textit{high-90\%}},
	description={High-density scenario},
	type=heuristlslist}
\newglossaryentry{ber-1}{name={\textit{BER-1}},
	description={Scenario corresponding to the city of Berlin with start 1},
	type=heuristlslist}
\newglossaryentry{sf-1}{name={\textit{SF-1}},
	description={Scenario corresponding to the city of San Francisco with start 1},
	type=heuristlslist}
\newglossaryentry{sf-2}{name={\textit{SF-2}},
	description={Scenario corresponding to the city of San Francisco with start 2},
	type=heuristlslist}
\newglossaryentry{direct}{name={\textit{direct}},
	description={$\mathcal{U}((C, 0) = \{ (c, w) : (c_k, c) \in \widetilde{\mathcal{A}}(C), w \in \mathcal{W} \}$},
	type=heuristlslist}
\newglossaryentry{direct_res}{name={\textit{direct/restricted}},
	description={$\mathcal{U}((C, 0) = \{ (c, w) : (c_k, c) \in \widetilde{\mathcal{A}}(C) ,\delta=0, w \in \mathcal{W }\}$},
	type=heuristlslist}
\newglossaryentry{res_5}{name={\textit{$T^r$-restricted}},
	description={$\mathcal{U}((C, 0)=\{(c, w) : w \in \mathcal{W}, c \in \widetilde{\mathcal{C}}, (\mathcal{W} = \mathcal{W}^0 \land \delta=0)  \lor (\mathcal{W} = \mathcal{W}^1 \land   ( \delta = 0 \lor (\delta = 1 \land w=1))), t_{c_k,c} \leq T^r \}$},
	type=heuristlslist}
\newglossaryentry{all}{name={\textit{complete}},
	description={$\mathcal{U}((C, 0)=\{ (c, w) : w \in \mathcal{W}, c \in \widetilde{\mathcal{C}},  (\mathcal{W} = \mathcal{W}^0 \land \delta=0)  \lor (\mathcal{W} = \mathcal{W}^1 \land   ( \delta = 0 \lor (\delta = 1 \land w=1))) \}$},
	type=heuristlslist}
\newglossaryentry{A}{name=\ensuremath{{\neg W/\neg C}},
	description={Network graph},
	type=problemlist}
\newglossaryentry{B}{name=\ensuremath{{W/\neg C}},
	description={Network graph},
	type=problemlist}
\newglossaryentry{C}{name=\ensuremath{{\neg W/C}},
	description={Network graph},
	type=problemlist}
\newglossaryentry{D}{name=\ensuremath{{W/C}},
	description={Network graph},
	type=problemlist}
\newglossaryentry{charge_action}{name=\ensuremath{l_c},
	description={Decision to charge at a charging station $c_c$},
	type=appendixlist}
\newglossaryentry{graph}{name=\ensuremath{{\mathcal{ G= (C, A)}}},
	description={Network graph},
	type=symbolslist}
\newglossaryentry{proba}{name=\ensuremath{p_c},
	description={Initial probability that charging station $c$ is available before any visit},
	type=symbolslist}
\newglossaryentry{consum}{name=\ensuremath{k_{\beta}(t, l)},
	description={Energy consumption model for $\beta$-type car},
	type=soclist}
\newglossaryentry{path}{name=\ensuremath{\mathcal{P}(0,n)},
	description={Searching path, starting from origin vertex $c_0$ and going through a sequence of charging station $(c_0, ..., c_n)$},
	type=symbolslist}
\newglossaryentry{cstation}{name=\ensuremath{c_c},
	description={Charge station located at vertex i},
	type=symbolslist}
\newglossaryentry{t}{name=\ensuremath{t_{c,c'}},
	description={Driving time on arc (c,c')},
	type=symbolslist}
\newglossaryentry{surface}{name=\ensuremath{\overline{S}},
	description={Maximal distance allowed between any vertex and the origin vertex $c_0$},
	type=symbolslist}
\newglossaryentry{dist_start}{name=\ensuremath{z_{0}(i)},
	description={Distance from the charging station $c$ to the starting vertex $0$.},
	type=symbolslist}
\newglossaryentry{soc}{name=\ensuremath{b_c},
	description={SoC of the vehicle currently located at vertex i},
	type=soclist}
\newglossaryentry{termin_cost}{name=\ensuremath{\beta_c},
	description={Termination penalty cost at an occupied station $c$},
	type=symbolslist}
\newglossaryentry{cs_cost}{name=\ensuremath{\gamma_c},
	description={Termination cost at an available station $c$},
	type=symbolslist}
\newglossaryentry{imm_cost}{name=\ensuremath{d(u,u(x))},
	description={Immediate cost induced by taking action $u$ in state $x \in \mathcal{S}$},
	type=symbolslist}
\newglossaryentry{ps}{name=\ensuremath{\rho(\pi, k)},
	description=Probability that at least one of the first $k$ first stations of policy $\pi$ is available,
	type=symbolslist}
\newglossaryentry{tn}{name=\ensuremath{t(\pi)},
	description={Driving time follwing policy $\pi$},
	type=symbolslist}
\newglossaryentry{Sn}{name=\ensuremath{S(\pi)},
	description={Set of feasible remaining vertices to visit after having followed $\pi$, not visited yet},
	type=symbolslist}
\newglossaryentry{Rn}{name=\ensuremath{R(\pi)},
	description={Set of feasible remaining vertices to visit after having driven through path $\mathcal{P}$, already visited},
	type=symbolslist}
\newglossaryentry{sequence}{name=\ensuremath{C},
	description={Ordered sequence of charging station visits},
	type=symbolslist}
\newglossaryentry{Tmax}{name=\ensuremath{\overline{T}},
	description={Maximal overall driving time},
	type=symbolslist}   
\newglossaryentry{Smin}{name=\ensuremath{Q_{MIN}},
	description={Minimial allowed SoC},
	type=soclist}
\newglossaryentry{Aspace}{name=\ensuremath{\mathcal{U} },
	description={Action space},
	type=symbolslist}   
\newglossaryentry{Sspace}{name=\ensuremath{\mathcal{S}},
	description={State space},
	type=symbolslist}
\newglossaryentry{avail}{name=\ensuremath{a_c},
	description={Binary random variable modelling $c$' availability},
	type=symbolslist}
\newglossaryentry{policy}{name=\ensuremath{\pi},
	description={Policy},
	type=symbolslist}
\newglossaryentry{policy_heur}{name=\ensuremath{\pi_{\mathcal{H}}},
	description={Policy found based on Heuristic \ensuremath{\mathcal{H}}},
	type=symbolslist}
\newglossaryentry{value}{name=\ensuremath{V^*},
	description={Optimal Value function},
	type=symbolslist}
\newglossaryentry{value_pol}{name=\ensuremath{V^{\pi}},
	description={Value function, based on policy $\pi$},
	type=symbolslist}
\newglossaryentry{state}{name=\ensuremath{x},
	description={State \ensuremath{\in \mathcal{S}}},
	type=symbolslist}
\newglossaryentry{action}{name=\ensuremath{u(x)},
	description={Action choosen in state $x \in \mathcal{S}$},
	type=symbolslist}
\newglossaryentry{waiting}{name=\ensuremath{W_c},
	description={Expected waiting time at an occupied charging station $c$},
	type=symbolslist}
\newglossaryentry{charging}{name=\ensuremath{L_c},
	description={Expected charging time at a charging station $c$},
	type=symbolslist}
\newglossaryentry{wait_action}{name=\ensuremath{w},
	description={Decision to wait (and charge) at the last visited station $C$ if occupied},
	type=symbolslist}
\newglossaryentry{search_cost}{name=\ensuremath{\alpha(\pi)},
	description={Total search cost for following policy $\pi$},
	type=symbolslist}
\newglossaryentry{label}{name=LH,
	description={Labeling heuristic algorithm},
	type=symbolslist}
\newglossaryentry{rollout}{name=RO,
	description={Rollout algorithm},
	type=symbolslist}
\newglossaryentry{exact}{name=LE,
	description={Labeling algorithm},
	type=symbolslist}
\newglossaryentry{naive}{name=N,
	description={Naive search method},
	type=symbolslist}
\newglossaryentry{greedy}{name=G,
	description={Greedy search method},
	type=symbolslist}
\newglossaryentry{rollout_des}{name=RO,
	description={Rollout algorithm},
	type=symbolslist}
\newglossaryentry{label_des}{name=LH,
	description={Rollout algorithm},
	type=symbolslist}
\newglossaryentry{exact_des}{name=LE,
	description={Rollout algorithm},
	type=symbolslist}
\newacronym{abk:tsp}{TSP}{traveling salesman problem}

\definecolor{lightblue}{rgb}{0.60784,0.76078,0.90196}
\definecolor{darkblue}{rgb}{0.26667,0.44706,0.76863}
\definecolor{lightgreen}{rgb}{0.66275,0.81569,0.55686}
\definecolor{darkgreen}{rgb}{0.43922,0.67843,0.27843}
\definecolor{orange}{rgb}{0.92941,0.49020,0.19216}
\definecolor{yellow}{rgb}{1.00000,0.75294,0.00000}
\definecolor{grey}{rgb}{0.64706,0.64706,0.64706}
\definecolor{purple}{rgb}{0.51373,0.23529,0.04706}

\DeclareOldFontCommand{\rm}{\normalfont\rmfamily}{\mathrm}
\DeclareOldFontCommand{\sf}{\normalfont\sffamily}{\mathsf}
\DeclareOldFontCommand{\tt}{\normalfont\ttfamily}{\mathtt}
\DeclareOldFontCommand{\bf}{\normalfont\bfseries}{\mathbf}
\DeclareOldFontCommand{\it}{\normalfont\itshape}{\mathit}
\DeclareOldFontCommand{\sl}{\normalfont\slshape}{\@nomath\sl}
\DeclareOldFontCommand{\sc}{\normalfont\scshape}{\@nomath\sc}
\makeatother
\counterwithin*{equation}{section}

\bibpunct[, ]{(}{)}{,}{a}{}{,}%
\newcolumntype{L}[1]{>{\raggedright\let\newline\\\arraybackslash\hspace{0pt}}p{#1}}
\newcolumntype{C}[1]{>{\centering\let\newline\\\arraybackslash\hspace{0pt}}p{#1}}
\newcolumntype{R}[1]{>{\raggedleft\let\newline\\\arraybackslash\hspace{0pt}}p{#1}}
\def\VRHDW#1#2#3{\vrule height #1 depth #2 width #3}
\newcommand{\up}{\VRHDW{1.5em}{0em}{0em}} 
\newcommand{\down}{\VRHDW{0em}{0.625em}{0em}} 

\addtokomafont{captionlabel}{\footnotesize\bfseries} 
\addtokomafont{caption}{\footnotesize\bfseries} 

\AtBeginDocument{%
	\abovedisplayskip=2.5pt plus 0pt minus 0pt
	\abovedisplayshortskip=2.5pt plus 0pt minus 0pt
	\belowdisplayskip=2.5pt plus 0pt minus 0pt
	\belowdisplayshortskip=2.5pt plus 0pt minus 0pt
}

\newsavebox{\abstractbox}
\renewenvironment{abstract}
{\begin{lrbox}{0}\begin{minipage}{\textwidth}
			\begin{center}\normalfont\sectfont\abstractname\end{center}\quotation}
		{\endquotation\end{minipage}\end{lrbox}%
	\global\setbox\abstractbox=\box0 }

\makeatletter
\expandafter\patchcmd\csname\string\maketitle\endcsname
{\vskip\z@\@plus3fill}
{\vskip\z@\@plus2fill\box\abstractbox\vskip\z@\@plus1fill}
{}{}
\makeatother

\makeatletter
\def\BState{\State\hskip-\ALG@thistlm}
\makeatother

\counterwithin*{equation}{section}

\newenvironment{myfont}{\fontfamily{ccr}\selectfont}{\par}
\DeclareTextFontCommand{\textmyfont}{\myfont}

\begin{document}

\definecolor{lightblue}{rgb}{0.60784,0.76078,0.90196}
\definecolor{darkblue}{rgb}{0.26667,0.44706,0.76863}
\definecolor{lightgreen}{rgb}{0.66275,0.81569,0.55686}
\definecolor{darkgreen}{rgb}{0.43922,0.67843,0.27843}
\definecolor{orange}{rgb}{0.92941,0.49020,0.19216}
\definecolor{yellow}{rgb}{1.00000,0.75294,0.00000}
\definecolor{grey}{rgb}{0.64706,0.64706,0.64706}
\definecolor{purple}{rgb}{0.51373,0.23529,0.04706}

\newcommand{\orange}[1]{\textcolor{orange}{#1}}

\newacronym{cs}{CS}{charging station}
\newacronym{dp}{DP}{dynamic programming}
\newacronym{abk:eu}{EU}{European Union}
\newacronym{abk:ecv}{ECV}{electric commercial vehicle}
\newacronym{ev}{EV}{electric vehicle}
\newacronym{abk:ghg}{GHG}{greenhouse gas}
\newacronym{abk:icev}{ICEV}{internal combustion engine vehicles}
\newacronym{mdp}{MDP}{Markov decision process}
\newacronym{abk:rcspp}{RCSPP}{resource constraint shortest path problem}
\newacronym{abk:ref}{REF}{resource extension function}
\newacronym{sdp}{SDP}{stochastic dynamic programming}
\newacronym{Soc}{SoC}{state of charge}
\newacronym{scps}{SCPS}{stochastic charge pole search}


\title{\large Electric Vehicle  Charging Station Search in Stochastic Environments}

\author[1,2,3]{\normalsize Marianne Guillet}
\author[1]{\normalsize Gerhard Hiermann}
\author[2]{\normalsize Alexander Kr{\"o}ller}
\author[1]{\normalsize Maximilian Schiffer}
\affil{\small 
	TUM School of Management, Technical University of Munich, 80333 Munich, Germany
	
	\textsuperscript{2} TomTom Location Technology Germany GmbH, 12435 Berlin, Germany

	\textsuperscript{3} Chair of Operations Management, RWTH Aachen University, 52072 Aachen, Germany
				
	\scriptsize
	marianne.guillet@tomtom.com,
	gerhard.hiermann@tum.de,
	alexander.kroeller@tomtom.com,
	schiffer@tum.de}

\date{}

\lehead{\pagemark}
\rohead{\pagemark}

\begin{abstract}
\begin{singlespace}
{\small\noindent  Electric vehicles are a central component of future mobility systems as they promise to reduce local noxious and fine dust emissions and CO\textsubscript{2} emissions, if fed by clean energy sources. However, the adoption of electric vehicles so far fell short of expectations despite significant governmental incentives. One reason for this slow adoption is the drivers’ perceived range anxiety, especially for individually owned vehicles. Here, bad user-experiences, e.g., conventional cars blocking charging stations or inconsistent real-time availability data, manifest the drivers' range anxiety. Against this background, we study stochastic search algorithms, that can be readily deployed in today’s navigation systems in order to minimize detours to reach an available charging station. We model such a search as a finite horizon Markov decision process and present a comprehensive framework that considers different problem variants, speed-up techniques, and three solution algorithms: an exact labeling algorithm, a heuristic labeling algorithm, and a rollout algorithm. Extensive numerical studies show that our algorithms significantly decrease the expected time to find a free charging station while increasing the solution quality robustness and the likelihood that a search is successful compared to myopic approaches.\\
\smallskip}
\end{singlespace}
\end{abstract}

\maketitle
\section{Introduction}
\label{sec:intro}
All around the world, governments and companies try to foster the adoption of \glspl{ev}, which are seen as a central component of future sustainable mobility systems that can play a major role in reducing noise, noxious, and carbon emissions, if fed by clean energy sources. While governments introduced national programs to support individual \gls{ev} purchase, e.g., by means of ecological bonuses (France), tax exemption (Germany), or subsidies (Netherlands), private companies invested in public charging infrastructure \citep{EDF2019, Froehlich2019}, and major logistic service providers progressively integrated \glspl{ev} to realize sustainable transportation \citep{DHL2019, Amazon2019}. 
Yet the market uptake for \glspl{ev} fell short in expectation in most countries. 
One of the main obstacles remaining, in particular for privately-owned \glspl{ev}, is the customer's perceived range-anxiety, triggered by missing charging infrastructure and incomplete or non-standardized information about charging services available.
Hence, facilitating a reliable \gls{ev} charging process is crucial to decrease range anxiety in order to invigorate the adoption of \glspl{ev} \citep{Bonges2016}. 

Different stakeholders focus on different measures to facilitate a reliable charging experience for users.
At strategic level, municipalities try to facilitate reliable charging options by improving the overall charging infrastructure, e.g., by building new charging stations or increasing the capacity of existing ones. Further, enforcing new pricing schemes that support a higher turnover rate of (re)charged vehicles may help to save additional capacities \citep{Bonges2016}. However, infrastructure investments require long planning lead  times and pricing schemes can hardly be realized without standardization and agreements between charging station operators. At operational level, some online map services exist and aim to help \gls{ev} drivers to locate available charging stations. Unfortunately, such services struggle with data inaccuracy, e.g., an incomplete coverage with real-time status data. Even worse, charging stations can be blocked by non-charging vehicles (so-called ``ICEing'') without the inaccessibility being reflected in status data. An empirical study in Berlin revealed a high correlation between such inaccessibility and parking availability, showing that in areas without available parking it is three times more likely for a charging station to be illegitimately blocked than legitimately used (see Table~\ref{tab:parking}).

\begin{table}[bp]
	\centering
	\begin{threeparttable}	
	\caption{Parking availability on data accuracy\label{tab:parking}}
	{\setlength{\tabcolsep}{0.15cm}\footnotesize
		\begin{tabular}{lrrr}\toprule
			Legal parking availability in vicinity & $\geq 2$ spots & 1       & 0\\\midrule
			Blocked by charging EV                 &         5\% & 14\% & 10\%\\
			Blocked without connection  (ICEing)        &         1\% &   9\% & 34\%\\
			Available                                        &        94\%& 77\% & 56\%\\
			\bottomrule
		\end{tabular}}
		\begin{tablenotes}
		\small
		\item Internal TomTom study in Berlin, 2019. The study shows how often charge points were legitimately used (i.e.,``blocked'' and actually used), actually available (correct ``free''), or illegitimately blocked by a vehicle. The data relates to the amount of free parking spots found in the immediate vicinity. For most of the stations, real-time availability data was not available.
		\end{tablenotes}
	\end{threeparttable}
\end{table}

 Drivers may find a selected and seemingly free station to be blocked when arriving there. Such negative user experiences may require additional detours with an already depleted battery in order to find a suitable charging station and thus may even increase a driver's range anxiety. To this end, a reliable stochastic search algorithm that helps to route drivers to an available charging station as convenient as possible constitutes a valuable algorithmic component for today's map and navigation services and may help to reduce the drivers' range anxiety.
      
The goal of this paper is to develop such an algorithmic solution that can be readily deployed as a built-in implementation in today's navigation systems and map services. In the remainder of this section, we first review the related literature, before we detail our contribution and elaborate the organization of this paper.

\subsection{Literature review}
We now survey literature on \gls{ev} routing with uncertain charging stations before we focus on related search problems with stochastic resources.  

\cite{Sweda2017} were the first to study probabilistic charging station availability as a multi-stop shortest path problem under uncertainty. They used a dynamic programming approach on a grid network and considered waiting times at unavailable stations. \cite{Kullman2017} extended this work by modeling an electric vehicle routing problem with uncertain charging at public charging stations, as a \gls{mdp}. Given the large action and state spaces, they proposed an approximate stochastic dynamic programming approach with a look-ahead procedure \citep{Goodson2017}. \cite{Jafari2017} studied a multicriteria stochastic shortest path problem for \glspl{ev}, in which stochasticity relates to travel time and energy consumption. 

Focusing on related stochastic search problems, \cite{Arndt2016} studied a probabilistic routing problem for on-street parking search with parking spots as stochastic resources and incorporated user preferences for stops closer to their destination. Besides showing the problem's NP-completeness, they proposed a brand-and-bound algorithm for small problem spaces.
More generally, \cite{Guo2018} studied a probabilistic spatio-temporal resource search problem, in which resources have a general usage cost but the resource seeker is not allowed to wait for an occupied resource to become available again. Contrary to \cite{Arndt2016} in which resources observations are persistent during the search, stations can become available again after a defined time threshold. \cite{Guo2018} proposed a value iteration solution procedure that remains tractable by making a fast recovery assumption at the instance level, which keeps the state space small.
\cite{Schmoll2018} studied a dynamic resource routing problem under reliable real-time information, which requires fast (re)computations as resources frequently change their occupancy state. Assuming a large problem space, they used real-time dynamic programming to maintain utility values for likely states to determine the best action on-the-fly. 

As can be seen, related research on stochastic charging station search is still scarce. Approaches that are specifically tailored to \gls{ev} charging station search are limited to finding best cost paths between an origin and a destination rather than on an open search with an a-priori unknown destination. Further these approaches lack the consideration of sufficient real-world problem variants and heterogeneous station characteristics. Literature on stochastic resource search focus on an open search concept but lack EV and charging station specific characteristics. Moreover, the algorithmic solutions that have been proposed in this domain remain limited to a pure academic interest on artificial and small instances. Accordingly, these approaches cannot be embedded into real-time environments.

\subsection{Contribution}
With this work, we close the research gaps outlined above by providing a profound algorithmic framework that covers a wide range of problem variants in the context of stochastic \gls{ev} charging station search, which can readily be deployed into real-time environments for online optimization purposes. Specifically, our contribution is fourfold. First, we formalize the problem of stochastic \gls{ev} charging station search as a new stochastic search problem. We model this problem as a discrete finite horizon \gls{mdp} and consider two objectives: minimizing i)~the time until charging, and ii)~the time until completion in case of heterogeneous charging times. Moreover, we present different problem variants in which a driver may wait or not at occupied stations and charging stations might be homogeneous or heterogeneous. Second, we prove the complexity of this problem class and show that it is NP-hard. Third, we develop an algorithmic tool chain that consists of three algorithms: an exact labeling algorithm for which we present a cost decomposition of the Markovian policy to derive effective dominance rules; a heuristic variant of this labeling algorithm which is amenable for real-time application; and a rollout algorithm. Additionally, we present multiple speed-up mechanisms, e.g., reduced action spaces and sharpened dominance relations. Fourth, we provide extensive numerical studies that base on real-world data for the cities of Berlin and San Francisco.  Our results show that compared to myopic approaches, our search algorithms decrease the average time spent finding an available station by up to 44\%. We further benchmark the performance of our heuristic algorithms against the exact algorithm and show that a combined, case-dependent usage of both allows for effective real-time application. 

\subsection{Organization}
The remainder of the paper is as follows. In Section~\ref{sec:model}, we introduce the \gls{scps}~problem. Section~\ref{sec:method} formalizes our problem as an \gls{mdp} and consecutively develops the corresponding algorithmic framework. In Section~\ref{sec:exper_design}, we describe our case study and the experimental design. Section~\ref{sec:experiments_v2} discusses our numerical results. Section~\ref{sec:conclusion} concludes this paper and provides an outlook on future research. To keep this paper concise, we shift all proofs to Appendix~\ref{sec:appendixProofs}.

\section{Problem definition and representation}
\label{sec:model}
In this section, we introduce the \gls{scps} problem for which we specify four problem variants, each corresponding to a distinct real-world scenario.
\subsection{Problem setting}\label{subsec:problemDefinition}
We focus on a routing problem with stochastic charging station availability, where a driver starts at a given location and seeks to find an unoccupied charging station to recharge her vehicle. Her objective is to minimize her total expected cost, which consists of the driving time during the search and additional factors, e.g., the time until the charging is complete, or the time spent walking from the charge point to the actual destination. We consider this search to be spatially and temporally bounded to account for the driver's limited time budget and penalize unsuccessful searches with a high termination cost to model the resulting discomfort.

Formally, we define this problem on a directed and complete graph \gls{graph} that consists of a set of vertices $\mathcal{C}$ and a set of arcs $(c, c') \in \mathcal{A}$. Each vertex $c\in \mathcal{C}$, except the vertex where the search starts, represents a charging station. The driver starts her search at a designated start vertex $c_0\in\mathcal{C}$ at time $0$, with a maximum search time of $\gls{Tmax}$. We assume that the driver is willing to charge at any vertex in $\mathcal{C}$ and consider a limited search radius by restricting $\mathcal{C}$ to the stations within a maximum distance \gls{surface} around the start vertex. Driving from $c$ to $c'$ takes $t_{c,c'}\geq 0$ units of time. We model the availability of a station $c\in\mathcal{C}$ at time $t\in [0,\gls{Tmax}]$ as a visit-dependent binary random variable $a_c\in\{0,1\}$, which the driver observes by visiting $c$. Without prior knowledge, i.e., when visiting a station for the first time, the probability that the station is free ($a_c=1$) is a constant $p_c$. 
Table~\ref{tab:symbols} summarizes this notation.

We introduce two time-based penalties: $\gamma_c$ is the time-equivalent usage cost for using pole $c$ if it is available upon arrival; $\beta_c$ denotes the cost for unsuccessfully terminating the search at $c$ in case $c$ is occupied, which may happen if all vertices in $\mathcal{C}$ have been unsuccessfully visited, or it is impossible to visit another candidate within the remaining time budget $\gls{Tmax}$.

\begin{table}[bp]
	\centering
	\caption{Notation used to define the \gls{scps} \label{tab:symbols}}
	{\setlength{\tabcolsep}{0.08cm}\footnotesize
	\begin{tabular}{rl}
			\toprule
			\gls{graph} & \glsdesc{graph} \\
			$\mathcal{W}$ & Permissible values for waiting decisions\\
			\gls{charging} & \glsdesc{charging} \\
			\gls{surface} & \glsdesc{surface} \\
			\gls{Tmax} & \glsdesc{Tmax} \\
			\gls{waiting} & \glsdesc{waiting} \\
			\gls{avail} & \glsdesc{avail} \\
			\gls{proba} & \glsdesc{proba} \\
			\gls{t} & \glsdesc{t} \\
			\gls{termin_cost} & \glsdesc{termin_cost} \\
			\gls{cs_cost} & \glsdesc{cs_cost} \\  
			\bottomrule
	\end{tabular}}
	{}
\end{table}

We define a solution to our problem as an ordered sequence of charging station visits~$C=(c_0,\ldots,c_n)$. In practice, a driver that uses this solution starts at $c_0$ an visits the charging stations in the given sequence up to the first available charging station. Accordingly, our search terminates either successfully at any $c \in C$ (with a total cost of driving time until that vertex, plus extra cost~$\gamma_c$), or unsuccessfully at $c_n$ (incurring driving time along $C$ plus cost~$\beta_{c_n}$). We denote by $\alpha$ the expected cost of solution~$C$.

\paragraph{Problem setting variants:}
In practice, different system characteristics and user preferences may require to solve different problem variants, e.g., some drivers may want to find a charging station as fast as possible without considering station heterogeneity, while others may want to finish charging as quickly as possible. Such differences can be incorporated in our problem through the generic penalties $\beta_{c}$ and $\gamma_{c}$, and an additional parameter $\mathcal{W}$, which states whether it is permitted to finish the search by waiting at a station $\mathcal{W} = \mathcal{W}^1$ or not $\mathcal{W} = \mathcal{W}^0$. 

We identify four different problem variants that reflect the most common use cases in practical applications. Table~\ref{tab:heur} summarizes these variants and states the respective realizations as a triple $(\beta_{c}, \gamma_{c}, \mathcal{W})$.
In variant \gls{A}, the driver is neither allowed to wait at stations nor do stations have (heterogeneous) usage costs. To this end, we consider a constant penalty cost $\overline{\beta}$ for an unsuccessful terminated search. While stations remain homogeneous, the driver is allowed to wait in variant \gls{B}. We assume that an expected waiting duration $W_c$ is known for every station $c$, and to be a constant independent of arrival time and earlier observations. In variants \gls{C} and \gls{D}, stations have heterogeneous charging durations and the driver seeks to minimize her total time to finish charging up to a preferred charge level. 
While the driver is not allowed to wait in \gls{C} (i.e., the constant penalty cost $\overline{\beta}$ is induced in case of failure), this constraint is relaxed in \gls{D}.

\begin{table}[tb]
	\centering
	\caption{Summary of problem setting variants and parameters $(\beta_{c}, \gamma_{c}, \mathcal{W})$\label{tab:heur}}
	{\setlength{\tabcolsep}{0.08cm}\footnotesize
	\begin{tabular}{ p{5cm}  p{5cm}  p{5cm}}
		\toprule
		& No waiting allowed ($\neg W$) &  Waiting allowed ($W$)\\
		\midrule
		Charging time insensitive ($\neg L$) &\textbf{ \gls{A} }= $(\overline{\beta}, 0, \mathcal{W}^0)$ &\textbf{\gls{B} } =  $(W_c, 0, \mathcal{W}^1)$ \\
		Charging time sensitive ($L$) & \textbf{\gls{C} }=  $(\overline{\beta}, L_c, \mathcal{W}^0)$ &\textbf{\gls{D}} =  $(W_c + L_c, L_c, \mathcal{W}^1)$   \\
		\bottomrule
	\end{tabular}}
	{}
\end{table}

\subsection{Discussion}\label{subsec:modelDiscussion}

While our model captures the essential characteristics of the underlying real-world problem, some comments are in order. 
First, we differentiate charging stations based on the charging duration in problem variants~\gls{C} and \gls{D}. The model is however not restrictive and allows charging stations to be heterogeneous with respect to any other criteria, including walking distance to a destination location, price for charging, etc. Yet one needs to convert non time-based usage costs (e.g., charging prices) to time-based costs to ensure search cost compatibility.
Second, we assume an occupied station to remain occupied during the search for most problem variants. This seems to be a plausible assumption based on the large difference between typical search times (minutes) and charging durations (hours) in an urban setting. However, we also present a problem variant with recovering probabilities in Section~\ref{subsec:timeDependent} and discuss its impact in Section~\ref{sec:experiments_v2}. Third, our basic model is agnostic of the energy spent during the search. Here, a similar argument holds: given that a search typically covers only a small distance radius compared to a vehicle's range, we assume this effect to be negligible. Nevertheless, we present a problem variant which considers the energy spent during the search. Our results for this problem variant show no significant impact of the respective energy consumption and thus confirm our assumption. 

\section{Methodology}\label{sec:method}

We now introduce an \gls{mdp} representation for the \gls{scps} problem (Section~\ref{subsec:problemRepresentation}) and a corresponding algorithmic framework. As the \gls{scps} problem is NP-hard (see Appendix~\ref{sec:hardness}) we develop one exact and two heuristic algorithms. We first present these algorithms for our basic problem variants and focus on an exact and a heuristic labeling algorithm in Section~\ref{subsec:method:A}, before we present a rollout algorithm in Section~\ref{subsec:method:B}. We then show how these algorithms must be modified to account for time dependent recovery probability functions (Section~\ref{subsec:timeDependent}) and search-related energy consumption (Section~\ref{subsec:chargingConsumption}).

\subsection{\glsentrylong{mdp}}\label{subsec:problemRepresentation}
To model the \gls{scps} problem as a finite-horizon \gls{mdp} with multiple decisions over a time-budgeted process, we use the additional notation as summarized in Table~\ref{tab:mdpsymbols}. 

\begin{table}[tpb]
	\centering
	\caption{Notation used to define the MDP \label{tab:mdpsymbols}}
	{\setlength{\tabcolsep}{0.08cm}\footnotesize
		\begin{tabular}{rl}	
			\toprule
			\gls{Sspace} & \glsdesc{Sspace} \\
			\gls{Aspace} & \glsdesc{Aspace} \\
			\gls{state} & \glsdesc{state} \\
			\gls{action} & \glsdesc{action} \\
			\gls{wait_action} & \glsdesc{wait_action} \\
			\gls{imm_cost} & \glsdesc{imm_cost} \\
			\gls{policy} & \glsdesc{policy} \\
			\gls{value_pol} & \glsdesc{value_pol} \\
			\gls{value} & \glsdesc{value} \\
			$a$ & Binary variable modeling the availability of the last-visited station\\
			$\delta$ & Binary variable indicating whether the selected station $c$ in $x=(C,0)$ is already contained in $C$\\
			\gls{tn} & \glsdesc{tn} \\
			\gls{ps} & \glsdesc{ps} \\
			\gls{search_cost} & \glsdesc{search_cost} \\
			\bottomrule
	\end{tabular}}
	{}
\end{table}
\subsubsection{Model variables and transition functions}\label{subsubsec:preliminaries}
We define a state $x$ as a tuple $(C,a)$, where $C$ is an ordered sequence of stations $C = (c_0, ..., c_k)$ that have already been visited, with the driver being located at $c_k$. Let $a$ be the binary realization of the availability of $c_k$, indicating whether $c_k$ is free (a=1) or occupied (a=0). Then, the state space results to
\[ \mathcal{S} = \{ (C , a) :\; C=(c_0,\ldots,c_k), c_j \in \mathcal{C}\; \forall j, a \in \{0,1\}\}. \]
We denote by $u(x)= (c, w)$ a possible action taken in state $x \in \mathcal{S}$ to transition to state $x'$, with $w\in \mathcal{W}$ being a binary that indicates whether to wait at the current station ($w = 1$) or not ($w = 0$), and $c$ stating the next station to visit. We note that, either $\mathcal{W} = \mathcal{W}^0 = \{0\}$ for problem variants without waiting or $\mathcal{W} = \mathcal{W}^1 = \{0,1\}$ for problem variants that allow to wait at a charging station.
Since we assume a station availability status to be persistent during the search, we further assume that for problem variants without waiting, each station should be visited at most once. For  problem variants with waiting, given the heterogeneous termination costs $\beta_c$, we assume that the last station can be visited twice. 
Actions are only taken at occupied stations: if $a = 1$, $x$ is a termination state and the search finishes as the driver charges at the found unoccupied station. Let $\widetilde{\mathcal{C}}(C)$ be the restricted set of charging station vertices, reachable in less than $\gls{Tmax} - \sum_{i=0}^{k-1}t_{c_i,c_{i+1}}$. 
Then, the action space for a state $x=(C,0)$ results to
\begin{equation}
\mathcal{U}(x) = \{ (c, w) : w \in \mathcal{W}, c \in \widetilde{\mathcal{C}},  (\mathcal{W} = \mathcal{W}^0 \land \delta=0)  \lor (\mathcal{W} = \mathcal{W}^1 \land   ( \delta = 0 \lor (\delta = 1 \land w=1))) \},
\end{equation}
with $\delta$ being the binary variable that indicates whether $c$ is already included in $C$ ($\delta=1$) or not ($\delta=0$).
If $\widetilde{\mathcal{C}}(C)$ is empty, no more station can be reached under the given time constraints and we refer to this state as a forced termination state.

We define the function $p_t( x' \;|\; x, u)$ to describe the probability that following action $u\in\mathcal{U}(x)$ from state $x\in\mathcal{S}$ would result in state $x'\in\mathcal{S}$, such that $\sum_{x' \in \mathcal{S}} p_t(x'| x, u) = 1$ and consider two cases:
\begin{enumerate}
	\item If the selected action is to wait, then the driver stays at the current station. The transition is deterministic and $p_t((C, 1)| (C, 0 ), (c_k,1)) = 1$ for a path $C$ ending in $c_k$.
	\item If the selected action is to continue searching, the selected next station $c$ is either available or occupied with respect to $\tilde{p}_{c}(C)$. Hence $p_t((C', 1)| (C, 0 ), (c,0)) = \tilde{p}_{c}(C)$ and  $p_t((C', 0)| (C, 0 ), (c,0)) = 1 - \tilde{p}_{c}(C)$, where $C'$ is the sequence $C$ extended by $c$.
\end{enumerate}

We denote by $d(x, u)$ the immediate induced cost for taking action $u = (c, w)$ in state $x=(C,a)$, which depends on the realized availability $a$ at the last station $c_k$ and the respective action
\begin{equation}
\begin{split}
d(x, u)  =  (1-a) w \beta_{c_k} + a \gamma_{c_k} +(1-w)(1-a)t_{c_k,c}.
\end{split}
\end{equation}
We note that if $a=1$, the driver charges at station~$c_k$, and $d(x, u)$ corresponds to the station usage cost~$\gamma_{c_k}$. If $a=0$ and the driver waits at the station, the search terminates and cost~$\beta_{c_k}$ results. If $a=0$ and the search continues at the next station~$c$, the cost results to the travel time~$t_{c_k,c}$.	

Finally, we define a policy~$\pi$ as a function mapping a state $x\in \mathcal{S}$ to an action $\pi(x) \in \mathcal{U}(x)$. Accordingly, $\pi$ implicitly describes a search path $C(\pi) = (c_0, ..., c_n)$ with $\pi(c_i,0) = (c_{i+1},0) ~ \forall i=0,\ldots,n-1$ and state $x_n = (C(\pi), a)$ that terminates the search at vertex $c_n$, either because the driver runs out of time or because she decides to wait with $\pi(x_n) = (c_n, 1)$. We refer to $C(\pi)$ as $C$ and to $\tilde{p}_{c}(C)$ as $\tilde{p}_{c}$ to keep the notation concise. 

\subsubsection{Cost function}\label{subsubsec:costFunction}
We now analyze the cost function  $V^{\pi} (C, a)$ that describes the expected cost for following a policy $\pi$, from a start state $(C, a)$. Then $V^{\pi} (x_0)$, with $x_0 =(c_0, 0) $ represents the expected search cost for the driver when following policy $\pi$ from starting at vertex $c_0$ and the objective is to find a policy $\pi$ that minimizes $V^{\pi} (x_0)$. Then, the cost function  $V^{\pi} (C, a)$ can be expressed as follows
\begin{equation}\label{eq:cost_func}
\begin{split}
V^{\pi} (C, a)
&= a \gamma_{c_k} +  (1-a)\left[  w \beta_{c_k}  + (1-w)( t_{c_k,c_{k+1}} + \tilde{p}_{c_{k+1}} V^{\pi} (C',  1)  + (1 - \tilde{p}_{c_{k+1}})  V^{\pi}(C',  0) )\right],
\end{split}
\end{equation}
 with $C=(c_0,\ldots,c_k)$, $u(C,a)=(c_{k+1},w)$, and $C'=C\cup\{c_{k+1}\}$. For both realizations of $a$, Equation~\ref{eq:cost_func} can be simplified:
\begin{eqnarray}
V^{\pi} (C, 1)    &=& \gamma_{c_k} \;,\label{eq:cost_avail}\\
V^{\pi} (C, 0)    &=& w \beta_{c_k} + (1-w) \left[  t_{c_k,c_{k+1}}  + (1 - \tilde{p}_{c_{k+1}})  V^{\pi}(C',  0)  + \tilde{p}_{c_{k+1}} \gamma_{c_{k+1}}\right] \;.\label{eq:cost_null}
\end{eqnarray}

\subsubsection{Cost structure variants}\label{subsubsec:costVariants}
For each problem variant as introduced in Section~\ref{subsec:problemDefinition}, the cost functions $V^{\pi} (C, 0)$, $V^{\pi} (C, 1)$, and $V^{\pi}(x_k)$ for a termination state $x_n=(C, a)$ can be expressed as follows, with $C'$ denoting the extension of $C$ by station $c_{k+1}$ for non-termination states.
\begin{description}
	\item[No waiting, without charging (\gls{A}):]
	\begin{equation}\begin{split}
	V^{\pi} (C, 0)  &= t_{c_k,c_{k+1}}  + (1 - \tilde{p}_{c_{k+1}}) V^{\pi}(C',  0) \;,\\
	V^\pi (C, 1)   &= 0 \;,\\
	V^\pi (x_n) &= (1- \tilde{p}_{c_n}) \overline{\beta} \;.
	\end{split}\end{equation}
	\item[Waiting permitted, without charging (\gls{B}):]
	\begin{equation}\begin{split}
	V^{\pi} (C, 0)    &= w W_{c_k} + (1-w) [  t_{c_k,c_{k+1}}  + (1 - \tilde{p}_{c_{k+1}})  V^{\pi}(C',  0) ] \;,\\
	V^\pi(C, 1) &= 0 \;,\\
	V^\pi (x_n) &= (1-\tilde{p}_{c_n}) W_{c_n} \;.
	\end{split}\end{equation}
	\item[No waiting, with charging (\gls{C}):] 
	\begin{equation}\begin{split}
	V^{\pi} (C, 0) &= t_{c_k,c_{k+1}}  + (1 - \tilde{p}_{c_{k+1}})  V^{\pi}(C',  0)  + \tilde{p}_{c_{k+1}}L_{c_{k+1}} \;,\\
	V^\pi(C, 1) &= L_{c_k} \;,\\
	V^\pi (x_n) &= (1-\tilde{p}_{c_n}) \overline{\beta} + \tilde{p}_{c_n} L_{c_n} \;.
	\end{split}\end{equation}
	\item[Waiting permitted, with charging (\gls{D}):]
	\begin{equation}\begin{split}
	V^{\pi} (C, 0)  &= w (L_{c_k} + W_{c_k}) + (1-w) [  t_{c_k,c_{k+1}}  + (1 - \tilde{p}_{c_{k+1}})  V^{\pi} (C',  0)  + \tilde{p}_{c_{k+1}} L_{c_{k+1}}] \;,\\
	V^\pi(C, 1) &= L_{c_k} \;,\\
	V^\pi(x_n) &= (1-\tilde{p}_{c_n}) W_{c_n} + L_{c_n} \;.\\
	\end{split}\end{equation}
\end{description}

\subsubsection{Cost function expansion}
We now expand $V^{\pi}$ to derive an explicit evaluation of the search cost.
To simplify notation, we use $C_{[i:j]}$ to denote a sub-sequence $(c_i,\ldots,c_j)$ of a sequence $C=(c_0, ..., c_i, ..., c_j, ..., c_n)$. 
\begin{prop}\label{prop:explicitCost}
	Let $C = (c_0,\ldots,c_n)$.
	Then, the cost for being in state  $x_k =(C_{[0:k]} , 0)$, with $k<n$, following policy $\pi$ until the termination state $x_n = (C, a)$ expands as follows
	\begin{equation}\label{eq:explicitCost}
	\begin{aligned}
	V^{ \pi}(x_k) &= \prod_{i=k}^{n} (1-\tilde{p}_{c_i}(C_{[k:i-1]})) \beta_{c_n}   + \sum_{i = k}^{n-1} \left[ t_{c_i,c_{i+1}}  \prod_{j=k}^{i} (1-\tilde{p}_{c_j}(C_{[k:j-1]}))\right] \\
	&+ \sum_{i = k}^{n} \left[ \gamma_{c_i} \tilde{p}_{c_i}(C_{[k:i-1]}) \prod_{j=k}^{i-1} (1-\tilde{p}_{c_j}(C_{[k:i-1]}))\right] \;.\\
	\end{aligned}
	\end{equation}
\end{prop}
Given that policies encode solutions, we can express the solution cost $\alpha$ for a policy, and set $\alpha(\pi) = V^{\pi}(x_0)$. Following Equation \ref{eq:explicitCost} and Proposition~\ref{prop:explicitCost}, this yields
\begin{equation}\label{eq:alpha_cost}
\begin{split}
\alpha(\pi) &= \prod_{i=0}^{n} (1-\tilde{p}_{c_i}(C_{[0:i-1]})) \beta_{c_n} + \sum_{i = 0}^{n-1} \left[ t_{c_i,c_{i+1}}  \prod_{j=0}^{i} (1-\tilde{p}_{c_j}(C_{[0:j-1]}))\right] \\
&+ \sum_{i = 0}^{n} \left[ \gamma_{c_i} \tilde{p_{c_i}}(C_{[0:i-1]})\prod_{j=0}^{i-1} (1-\tilde{p_{c_j}}(C_{[0:j-1]}))\right]
\;.
\end{split}
\end{equation}

For a more concise notation, let $\bar{\rho}(\pi, k)$ be the probability that a driver fails in finding at least one free station in $C_{[0:k]}$, while $\rho(\pi,k) = 1 - \bar{\rho}(\pi,k)$ is the probability that she succeeds in finding at least one free station in $C_{[0:k]}$
\begin{equation}\label{eq:rho}
\begin{split}
\bar{\rho}(\pi, k) = \prod_{i=0}^{k} (1-\tilde{p}_{c_i}(C_{[0:i-1]})) \;,
\end{split}
\end{equation}
with $\tilde{p}_{c_i}(C_{[0:i-1]})$ denoting the likelihood that $c_i$ is available after having visited all previous stations from $c_0$ to $c_{i-1}$.
Furthermore, let 
\begin{equation}\label{eq:def_A}
A(\pi) = \sum_{i = 0}^{n-1} [ t_{c_i,c_{i+1}} \bar{\rho}(\pi, i)] + \sum_{i = 0}^{n} \gamma_{c_i} \tilde{p}_{c_i}(C_{[0:i-1]}) \bar{\rho}(\pi, i-1) \;.
\end{equation}
Then, we rewrite Equation~\ref{eq:alpha_cost} as 
\begin{equation}\label{eq:alpha_cost_simple}
\begin{split}
\alpha(\pi) &= \bar{\rho}(\pi, n) \beta_{c_n}  + A(\pi) \;.
\end{split}
\end{equation}
We denote by $t(\pi) = \sum_{i = 0}^{n-1} t_{c_i,c_{i+1}}$ the accumulated driving time for all stations in $C$ and note that for a feasible solution, $t(\pi) \leq \gls{Tmax}$ holds.

\subsection{Dynamic programming based labeling algorithms}\label{subsec:method:A}
To find an optimal solution for the \gls{scps} problem, we develop a dynamic programming based labeling algorithm. Similar to solving multi-criteria constrained shortest path problems, we propagate partial policies  in order to find an in-expectation cost-optimal policy.
Herein, we use a dominance criterion to withdraw non-promising partial policies early to keep the explored search space as small as possible. Formally, we associate each partial policy $\pi_c$ with a label $L_c$ whose resources depend on the problem variant. 
For no-waiting variants \gls{A} and \gls{C}, a label $L_c = (t_c, A_c, \rho_c, \alpha_c, S_{c})$ consists of the accumulated driving time $t_c$, the partial cost $A_c$ (cf. Equation~\ref{eq:alpha_cost_simple}), the likelihood $\rho_c$ to successfully finish the search up to vertex $c$ (cf. Equation~\ref{eq:rho}),  the total cost $\alpha_c$, and the set of reachable and non-visited poles $S_{c}$. For waiting variants \gls{B} and \gls{D}, we add an additional resource $R_{c}$ that denotes the set of reachable but visited poles.

To describe our labeling algorithm, we denote by $\mathcal{L}^a$ the set of active labels and by $L_0$ the initial label that corresponds to our start location. Let $\mathcal{F}_{cc'}(L)$ be a set of~\glspl{abk:ref} that expand a label $L$ whose partial policy ends at vertex $c$ to a label $L'$ whose partial policy ends at vertex $c'$. Let $\alpha(L)$ be the cost associated with label $L$. We define $\delta^{+}(L)$ as a function that returns a set of tuples $(c,c')$ which denotes all feasible physical successor locations $c\in \mathcal{C}$ for a label $L$ whose partial policy ends at $c\in\mathcal{C}$.

 Using this notation, Figure~\ref{alg:dplab} shows a pseudo code of our dynamic programming algorithm. We initialize our list of active labels $\mathcal{L}^a$ and our so far best found solution $L^*$ with $L_0$~(l.1) and start propagating labels until our search terminates when $\mathcal{L}^a$ is empty~(l.2). We then process labels in $\mathcal{L}^a$ in cost increasing order~(l.3). Once a label got selected for propagation, we remove it from $\mathcal{L}^a$~(l.4) and propagate it considering all of its feasible successors~(l.5) using the~\glspl{abk:ref} (l.6). We check whether a newly created label $L'$ is dominated by an existing label in $\mathcal{L}^a$~(l.7). If this is not the case, we remove labels which are dominated by $L'$ from  $\mathcal{L}^a$~(l.8) and add $L'$ to $\mathcal{L}^a$ respectively~(l.9).
\begin{figure}[bp]
	\footnotesize
	\caption{Dynamic programming based labeling algorithm.}
	\label{alg:dplab}
	\begin{algorithmic}[1]   
		\State $\mathcal{L}^a\gets\{L_0\}$, $L^*\gets L_0$
		\While{$\mathcal{L}^a\neq \emptyset$}
		\State $L\gets \texttt{costMinimumLabel(}\mathcal{L}^a\texttt{)}$
		\State $\mathcal{L}^a\gets \mathcal{L}^a\setminus\{L\}$
		\For{$(c,c') \in \delta^+(L)$}
		\State $L' \gets \mathcal{F}_{ij}(L)$
		\If{\texttt{isNotDominated($L',\mathcal{ L}^a$)}}
		\State \texttt{dominanceCheck($\mathcal{L}^a,L'$)}
		\State $\mathcal{L}^a\gets\mathcal{L}^a\cup\{L'\}$
		\If{$((w' == 1) \lor (\delta^+(L') = \emptyset) )\land \alpha(L') < \alpha(L^*)$}
		\State $L^*\gets L'$
		\EndIf
		\EndIf
		\EndFor
		\EndWhile
		\State \Return $L^*$
	\end{algorithmic}
\vspace{-0.5cm}
\end{figure}
 Whenever the newly created label is a termination label, i.e., its corresponding state indicates that the driver should wait or there are no feasible successors left, we check if the found label improves the so far best found label and update $L^{*}$ accordingly (l.10\&11).

In the remainder of this section, we detail the \glspl{abk:ref} used to extend a label and the dominance criterion to discard a dominated label.

\paragraph{Resource extension functions:} To extend a label $L$ which corresponds to a partial policy ending at vertex $c\in\mathcal{C}$ to a new label $L'$ which corresponds to a partial policy ending at vertex $c'\in\mathcal{C}$, we write $L' \gets \mathcal{F}_{cc'}(L)$ and use the following set $\mathcal{F}_{cc'}$ of \glspl{abk:ref} :
\begin{align}
A_{c'} &= A_{c} +(1- \rho_c) ( t_{c,c'} + \tilde{p}_{c'}\gamma_{c'} ) \label{eq:increment_1} \\
1- \rho_{c'} &= (1- \rho_c) (1-\tilde{p}_{c'})  \label{eq:increment_2} \\
t_{c'} &= t_c + t_{c,c'} \label{eq:increment_3} \\
S_{c'} &= S_c \setminus{\{c'\}} - \bigcup_{ \tiny{d \in S_c, t_{c'} + t_{c',d} > \gls{Tmax}}} \{d\} \label{eq:increment_4}\\
R_{c'}  &= R_c \cup{\{c'\}} - \bigcup_{d \in R_c, t_{c'} + t_{c',d} > \gls{Tmax}} \{d\} \label{eq:increment_5}\\
\alpha_{c'}  &= A_{c'} + (1- \rho_{c'})\beta_{c'}  \label{eq:increment_6}
\end{align}

Equation~\ref{eq:increment_1} propagates partial cost $A_c$ along arc $(c,c')$ with respect to its definition (cf. Equation~\ref{eq:def_A}) considering the arc-dependent driving time $t_{c,c'}$, the availability probability $\tilde{p}_{c'}$ and usage cost $\gamma_{c'}$ of vertex $c'$. Equation~\ref{eq:increment_2} propagates the success rate $\rho_c$ (cf. Equation~\ref{eq:rho}) based on the availability probability of vertex $c'$.
The accumulated driving time  $t_{c'} $ is straightforwardly propagated along arc $(c,c')$ considering the arc-dependent driving time $t_{c,c'}$ (Equation~\ref{eq:increment_3}).
To obtain the set $S_{c'}$ from $S_{c}$ (Equation~\ref{eq:increment_4}), we remove from $S_{c}$ the last visited vertex $c'$ and all vertices $d \in S_{c}$ that cannot be reached from vertex $c'$ within the remaining search time $\gls{Tmax} - t_{c'}$, i.e., such that $t_{c',d} \geq \gls{Tmax} - t_{c'} $. To obtain $R_{c'}$ (Equation~\ref{eq:increment_5}), we insert the visited vertex $c'$ in $R_{c}$ and similarly substract all vertices $d \in R_{c}$ that cannot be reached from vertex $c'$ within the remaining search time. We propagate cost straightforwardly by Equation~\ref{eq:increment_6}.

\paragraph{Dominance criterion:}
Equation~\ref{eq:alpha_cost_simple} decomposes the non-monotonous cost $\alpha(\pi)$ into monotonous resources $A(\pi)$ and $\rho(\pi)$ that partly define a label. 
Given the monotonicity of $A(\pi)$, $\rho(\pi)$, and $t(\pi)$, we can then consider two partial policies $\pi_1$ and $\pi_2$ that end with the same vertex visit $c$ and their associated labels $L_1$ and $L_2$ and say that for problem variants \gls{A} and \gls{C}, $L_1$ dominates $L_2$ ($L_1 \succ L_2$), if the following conditions are true:
\mathleft
\begin{equation} \label{eq:dom_1}
1- \rho_c(\pi_1) \leq 1- \rho_c(\pi_2)
\end{equation}
\begin{equation} \label{eq:dom_2}
A_{c}(\pi_1) \leq A_{c}(\pi_2)
\end{equation}
\begin{equation} \label{eq:dom_3}
t_c(\pi_1) \leq t_c(\pi_2)
\end{equation}
\begin{equation} \label{eq:dom_4}
\forall c' \in S_c(\pi_2), c' \in S_c(\pi_1)
\end{equation}
\begin{equation} \label{eq:dom_5_new}
\forall c' \in S_c(\pi_1), \bar{T} - t_{c}(\pi_1) + p_{c'}(\gamma_{c'} - \bar{\beta})  \leq 0 
\end{equation}
\mathcenter
Here, conditions \eqref{eq:dom_1}--\eqref{eq:dom_3}\&\eqref{eq:dom_5_new} ensure that the cost and duration of $\pi_1$ is smaller than the cost of $\pi_2$. Conditions \eqref{eq:dom_3}\&\eqref{eq:dom_4} check whether all non-visited vertices reachable by $\pi_2$ can be reached from $\pi_1$ as well. Condition \eqref{eq:dom_5_new} checks whether all reachable stations from $c$ for $\pi_1$ contribute to decrease $\alpha_c(\pi_1)$. This check is necessary for settings where $\pi_1$ can be extended with $k$ more stations than $\pi_2$ to detect corner cases in which these $k$ additional stations may increase $\alpha(\pi_1)$ to a value larger than $\alpha(\pi_2)$.

For problem variants \gls{B} and \gls{D}, we recall that a search can terminate before the whole time budget is spent and that a station terminating the search might have been visited earlier. Accordingly we drop condition~\ref{eq:dom_5_new} and introduce condition~\ref{eq:dom_5} that checks whether all visited vertices  reachable by $\pi_2$ can be reached from $\pi_1$ as well. Recalling that the label reads $L_c = (t_c, A_c, \rho_c, \alpha_c, S_{c}, R_{c})$, we then say that for problem variants $\gls{B}$ and \gls{D}, $L_1 \succ L_2$ if conditions~\eqref{eq:dom_1}--\eqref{eq:dom_4} are true and 
\begin{equation} \label{eq:dom_5}
\forall c' \in R_c(\pi_2), c' \in R_c(\pi_1) \cup S_c(\pi_1)
\end{equation}
holds.
While Algorithm~\ref{alg:dplab} solves the \gls{scps} problem optimally with the dominance criterion above, one may consider to drop some of the dominance conditions to obtain a heuristic dominance criterion that withdraws more labels at the price of losing optimality. In the remainder of this paper, we study a heuristic labeling algorithm, where we preserve conditions \eqref{eq:dom_1}\&\eqref{eq:dom_2} to obtain cost dominance but neglect conditions \eqref{eq:dom_3},\eqref{eq:dom_4},\eqref{eq:dom_5_new}\ and \eqref{eq:dom_5}. In this context, the label definition simplifies to $L_c = (A_c, \rho_c, \alpha_c)$. We provide evidences to the selection of this heuristic dominance criterion in Section~\ref{subsec:sensitivities}.

\subsection{Rollout algorithm}\label{subsec:method:B}
In this section, we introduce a rollout algorithm, built as a forward dynamic programming procedure. We identify the best action at a given state as the one that yields minimal approximated cost. Here, the core idea of the cost approximation is to greedily expand the current policy from each candidate action until a defined horizon to obtain an associated cost value via backpropagation. We first detail the main procedure for \gls{A} and \gls{C} variants before we show how to account for additional waiting decisions in problem variants \gls{B} and \gls{D}.
We apply a one-step decision rollout strategy \citep[cf.][]{Goodson2017} whose complexity equals a post-state decision rule as the approximation reduces to $w = 0$ decisions.

Let $k$ be the index of the $k^{th}$ decision epoch and let $x_k$ be the non terminated epoch's state with $x_k = (C=(c_0, c_1, ..., c_k), 0)$ and $c_k$ being the last station visited in epoch $k$. 
Let $x_{k+1} = (C' , 0)$ be the state in the $(k+1)^{th}$ epoch that results from action $u  = (c,0)$ at epoch $k$, with $u \in \mathcal{U}(x_k)$. Using this additional notation, Figure~\ref{alg:rollout} details the pseudo-code of our algorithm.

\begin{figure}[bp]
	\footnotesize
	\caption{Forward programming based algorithm.}
	\label{alg:rollout}
	\begin{algorithmic}[1]  
		\State $c_k \gets c_0$, $C \gets (c_0)$, $x_k \gets (C,0)$, $t \gets 0$ 
		\While{$t  \leq \gls{Tmax}$}
		\State $c^* \gets 0$, $x^* \gets 0$, $C^* \gets 0$, $Q \gets \infty$
		\For{$(c,0) \in \mathcal{U}(x)$} 
		\State $x_{k+1}\gets (C', 0)$
		\State $V \gets \texttt{greedyCost}(x_{k+1}) $
		\State $Q(x_k,c, x_{k+1}) \gets t_{c_k,c} + (1 - \tilde{p}_{c}) V + \tilde{p}_{c}\gamma_{c}$
		\If{$Q(x_k,c, x_{k+1}) < Q $}
		\State  $Q \gets Q(x_k,c, x_{k+1}) $
		\State $c^* \gets c$, $x^* \gets x$, $C^* \gets C'$
		\EndIf
		\EndFor
		\State $C \gets C^* $, $x_k \gets x^*$, $t \gets t + t_{c_k,c^*}$, $c_k \gets c^*$
		\EndWhile
		\If{$w' == 1$}
		\State $C \gets \texttt{refinePolicy}(C)$
		\EndIf
		\State \Return $C$
	\end{algorithmic}
\end{figure}
We initialize the sequence of station visits $C$ with the start vertex~(l.2) and expand the sequence until the time horizon is reached~(l.3). From the current state $x_k$, we seek to determine the next best action $(c^*,0)$ and initialize the variables that encode it (l.4). 
For all possible succeeding states $x_{k+1}$~(l.5), we use a heuristic policy $\tilde{\pi}_{c}$ that bases on a greedy procedure to propagate state $x_{k+1}$ up to a forced terminated state $x_{k+K}$, with a look-ahead of $K$ epoch extensions. For all propagated states  $x_l$  with $ l \in [k+1, K-1]$, the greedy procedure chooses the action $(c_{l+1}, 0)$, with station $c_{l+1}$ being selected based on availability probabilities of vertices reachable from $c_{l}$ and the driving time to each of these vertices. 
We then use these anticipated states to evaluate the expected value of the policy-specific cost $V^{\tilde{\pi}_c}(x_{k+1})$ for the candidate state~$x_{k+1}$. We define $\texttt{Greedycost}(x_{k+1})$ as the function that carries out the greedy expansion from $x_{k+1} = (C',0)$ and returns $V^{\tilde{\pi}_c}(x_{k+1})$~(l.6\&l.7).
Repeating the greedy procedure and cost evaluation for all possible next actions~$u=(c,0) \in \mathcal{U}(x_k)$ allows us to find the action $u = (c, 0)$ that minimizes the cost to transition from state $x_k$ to state $x_{k+1}$~(l.8) that we define as $Q(x_k,c, x_{k+1})$ (cf. Equation~\ref{eq:transition_cost}). Eventually, the selected action $u = (c^*, 0)$ yields  minimal cost $Q(x_k,c^*, x_{k+1})$~(l.10-l.12). Only for variants \gls{B} and \gls{D}, $\texttt{refinePolicy(C)}$ bases on the resulting visits sequence $C$ to introduce the waiting decision at the best stage $1 \leq k \leq n$, with $n$ being the total amount of station visits in $C$~(l.13).

In this setting, the reduced action space for problem variants \gls{A} and \gls{C} reads
\begin{equation}
\tilde{\mathcal{U}}(x_k) = \{ (c, w) : c \in \mathcal{C} , w = 0 \}
\end{equation}
and we calculate the cost for being in state $x_{k+1}$, $V^{\tilde{\pi}_{c}}(x_{k+1})$, based on the greedy policy $\tilde{\pi}_{c}$ as
\begin{equation}\label{eq:greedy_cost}
V^{\tilde{\pi}_{c}}(x_{k+1}) = \prod_{l=k+1}^{K} (1-\tilde{p}_{c_l}) \beta_{c_l}  + \sum_{l = k+1}^{K-1} [ d(x_l, \tilde{\pi}_{c}(x_{l})) ]	 \prod_{m=k+1}^{l} (1-\tilde{p}_{c_m}),
\end{equation}
which allows to derive $Q(x_k,c, x_{k+1})$ as
\begin{equation}\label{eq:transition_cost}
Q(x_k,c, x_{k+1}) = t_{c_k,c} + (1 - \tilde{p}_{c}) V^{\tilde{\pi}_c}(x_{k+1}) + \tilde{p}_{c}\gamma_{c} .
\end{equation}

For  problem variants \gls{B}\&\gls{D}, we first calculate the no-waiting case and compute policy~$\pi$ with the associated ordered sequence of charging stations $C = (c_0, ..., c_n)$. We denote $\pi$ as an intermediate policy and introduce $\pi_S$ representing the final search policy. Then, $\texttt{refinePolicy(C)}$ calculates $\pi_S$ using the intermediate policy~$\pi$ while permitting wait-actions. 
For each intermediary charging station $c_k$ at the $k^{th}$ decision epoch, $c_k \in (c_0, ..., c_n), k \not= n$, $\pi$ provides a sub sequence of charging stations to visit until the end of the search $(c_{k+1}, ..., c_n)$ and thus the policy-specific cost value~$V^{\pi}(x_k)$, associated to policy~$\pi$ and state~$x_k = ((c_0, ..., c_k), 0)$. 
We aim to quantify for each station~$c_k$ whether the termination cost $\beta_{c_k}$ is actually lower that the expected cost for continuing the search $V^{\pi}(x_k)$. If this is the case, the optimal decision is to wait and we refine $\pi$ into $\pi_S$ with $\pi_S(x_k) = (c_k, w=1)$ and $V^{\pi_S}(x_l) \leq V^{\pi}(x_l) \forall l \in [0,n]$.

We define $\pi_S$ as
\begin{equation} 
\begin{aligned}
\pi_S(x_k)  &=  \argmin_{\pi(x_{k}), (c_k, w=1)}   w  \beta_{c_k} +( 1-w)[ t_{c_k,c_{k+1}} + (1 - \tilde{p}_{c_{k+1}}) \min(V^{\pi}(x_{k+1}) , V^{\pi_S}(x_{k+1}))+ \tilde{p}_{c_{k+1}} \gamma_{c_{k+1}}],
\end{aligned}
\end{equation}
where $ \pi(x_{k}) = (c_{k+1}, w=0)$.
If there exists an index $k$ such that $0 \leq k < n $, $x_k = ((c_0, ..., c_k),0)$ and $\pi_S(x_k) = (c_k, 1)$, then state $x_k$ terminates the search, as the driver will wait at $c_k$ if $c_k$ is not immediately available. In this case $\pi_S$ encodes the solution $(c_0, ..., c_k)$.

\subsection{Time-dependent probability recovery function}\label{subsec:timeDependent}
In the basic setting of the \gls{scps} problem, we assume that a station does not change its availability during the search's time horizon and restrict the amount of visits to each individual station. We now relax these assumptions and show which modifications are necessary to consider a recovery function $r_c$ that allows the availability of an occupied station $c$ to recover over time.
In this case, we define $\tilde{p_c}(C)$ as follows
\begin{equation}\label{eq:proba_visible}
\tilde{p_c}(C) = 
\begin{cases}
p_c , & \text{if } c \notin C\\
r_c(\Delta_c)  & \text{otherwise},
\end{cases}
\end{equation}
where $\Delta_c = \sum_{j=\ell}^{k-1}t_{c_j,c_{j+1}}+t_{c_k,c}$, with $l$ denoting the position of the last visit to $c$ in $C$.

We still consider a charging station~$c$ to be initially available at a probability of $p_c$. When it is blocked at the driver's arrival time~$t_c = \sum_{j=0}^{\ell-1}t_{c_j,c_{j+1}}$, it may become  available over time, i.e., the availability probability of $c$ recovers according to  $r_c(\Delta_c)$, which denotes a station's availability probability for an arbitrary point in time $t = \Delta_c + t_c$ that remains after the first visit and before the end of the time horizon. We specify $r_c(\Delta)$ for any $\Delta$ based on \cite{Schmoll2018} as 
 \begin{equation}\label{eq:time}
 \begin{split}
 r_c(\Delta)  &= \frac{\mu_c}{\lambda_c + \mu_c} (1 - e^{-(\mu_c + \lambda_c)(\Delta)}).
 \end{split}
 \end{equation}
Here, $\frac{1}{\lambda_c}$ and $\frac{1}{\mu_c}$ denote the average time station $c$ remains available, respectively occupied and remain constant over the search's time horizon. We can then express $p_c$ as a function of $\lambda_c$ and $\mu_c$, with $p_c =  \frac{\mu_c}{\lambda_c + \mu_c}$ \citep[cf.][]{Josse2015} and simplify Equation~\ref{eq:time} to
\begin{equation}
 r_c(\Delta_c)  = p_c(1 - e^{-(\frac{\mu_c}{p_c})(\Delta_c)})\;.\\ 
\end{equation}

We now assume that stations can be visited as many times as needed. Then, the action space for a state~$x=(C,0)$ is slightly modified as follows,
\[\mathcal{U}(x) = \{ (c, w) : c\in \widetilde{\mathcal{C}}(C), w \in \mathcal{W} \}. \]

In the remainder of this section, we outline the changes that are necessary to adapt the labeling algorithm to such a setting. Apart from the new availability probability definition $\tilde{p}_c(C)$, no modifications are necessary for the rollout algorithm.
\paragraph{Modifications for the Labeling Algorithm:}
The \gls{mdp} definition remains unchanged, because we can determine the arrival time at a station based on the arrival time at the preceding station extended by the driving time that remains deterministic. Thus $\Delta_c$ and $\tilde{p_c}(C)$ can be calculated from $C$ without any further information.
However, we need to modify the initial dominance criterion \eqref{eq:dom_1}--\eqref{eq:dom_5_new}(resp. \eqref{eq:dom_1}--\eqref{eq:dom_5}) as the optimal solution may now contain multiple visits to any charging station, independent on the problem variant.

We consider two partial policies~$\pi_1$ and $\pi_2$ that end with the same visit at vertex~$c$ and their associated labels~$L_1$ and $L_2$. We recall that $R_c(\pi_1)$ denotes the set of charging stations which have been already visited by $\pi_1$. 
We then say $L_1 \succ L_2$, if \eqref{eq:dom_1}--\eqref{eq:dom_5} hold and 
\mathleft
\begin{equation}\label{eq:dom_6}
\frac{1- \rho_c(\pi_1)}{1- \rho_c(\pi_1)} \leq \prod_{c' \in R_c(\pi_1)}  \frac{(1 - p_{c'}(1- e^{-(\mu + \lambda)(\Delta_{c'} (\pi_2)) }))}{(1 - p_{c'}(1- e^{-(\mu + \lambda)  (\Delta_{c'} (\pi_1) ) })) }
\end{equation}
\mathcenter
holds. Here, Equation~\ref{eq:dom_6} accounts for different probabilities of charging stations that have already been visited in both policies at different points in time by considering the biggest possible difference of probability values for $c'$ between both paths.
We note that we leave the heuristic variant of the dominance criterion unchanged in this context.

\subsection{Integrating search related energy consumption}\label{subsec:chargingConsumption}
In this setting, we assume that the vehicle starts its search with an initial \gls{Soc}~$b_{0}$, which reduces over the course of the search depending on the driven distances. Then, longer driving distances result in higher energy consumption and an additional trade-off results between visits to far-distanced stations with a high availability probability and visits to near-distanced stations with medium to low availability probabilities. Here, a higher availability probability may be counterbalanced by traveling longer distances, which increases the energy that must be recharged or limits future visits to potential stations accordingly.

To account for this setting, we introduce additional notation and denote the energy consumed when traversing arc~$(c,c')$ as $k_{c,c'}$ and a vehicle's \gls{Soc} after having visited all stations in $C = (c_0, ..., c_k)$ as $b$. In this setting, we can only transition from state~$(C,b,0)$ using arc~$(c_k,c_{k+1})$ if  $b~\geq k_{c_k,c_{k+1}} + \underline{b}$, with $\underline{b}$ denoting the minimum feasible state of charge a vehicle must keep. Analogously to common monotonicity assumptions in related settings \citep[cf.][]{Sweda2017}, we assume that  
$$t_{c}(\pi_1) < t_{c}(\pi_2) \iff b_{c}(\pi_1) \geq b_{c}(\pi_2)$$
and state the necessary \gls{mdp} modifications which hold as follows
\begin{align}
\begin{split}\label{eq:stateExt}
\mathcal{S} ={} &\{(C, b, a) : C = (c_0, ..., c_k) , c_k \in \mathcal{C}\forall k, b \in [0, q_{max}], a \in \{0,1\}\} \;,
\end{split}
\\[2ex]
\begin{split}\label{eq:actionExt}
\mathcal{U}(x) ={}& \{ (c, w) : w \in \mathcal{W}, c \in \bar{\mathcal{C}}, \\
&\qquad (\mathcal{W} = \mathcal{W}^0 \land \delta=0)  \lor (\mathcal{W} = \mathcal{W}^1 \land   ( \delta = 0 \lor (\delta = 1 \land w=1))) \} \;,
\end{split}
\\[2ex]
\begin{split}\label{eq:costExt}
V^{\pi} ((C, b, a)) ={}& (1-a) w \beta_{c_k}  + a \gamma_{c_k} +(1-w)(1-a)  [  t_{c_k,c_{k+1}} \\
&\qquad + (1 - \tilde{p}_{c_{k+1}})  V^{\pi}(C', b - k_{c_k,c_{k+1}}, 0)  + \tilde{p}_{c_{k+1}} V^{\pi} (C', b - k_{c_k,_{k+1}}, 1)] \;.
\end{split}
\end{align}

First, we include a vehicle's \gls{Soc} into the state space~(Equation~\ref{eq:stateExt}). Second, we modify the action space such that it depends on $\bar{\mathcal{C}}(C)$, which denotes a restricted set of charging station vertices that are reachable from state $(C, a)$ in less than $\gls{Tmax} - t(\pi)$ time, with $k_{c,c'} + \underline{b} \leq b $ and $t(\pi) = \sum_{k=0}^{n}t_{c_k,c_{k+1}}$~(Equation~\ref{eq:actionExt}). The policy-specific cost function results straightforwardly from the modified action and state spaces (Equation~\ref{eq:costExt}).
To account for this new setting in our algorithms, the following modifications are necessary.
\paragraph{Variants \gls{A} and \gls{B}:}
The rollout algorithm requires no modification. For the labeling algorithms, sets $S_c$ and $R_c$ now denote vertices that are reachable within the remaining time budget and within the remaining energy budget $b_c$. Accordingly, we add to the existing dominance conditions \eqref{eq:dom_1}--\eqref{eq:dom_5_new} or \eqref{eq:dom_1}--\eqref{eq:dom_5} that
\mathleft
\begin{equation}\label{eq:dom_7}
b_c(\pi_1) \geq b_c(\pi_2),
\end{equation}
\mathcenter
must hold.
The heuristic dominance relation remains unchanged, since the modified setting does not affect conditions~\eqref{eq:dom_1}\&\eqref{eq:dom_2}. 
\paragraph{Variants \gls{C} and \gls{D}:}
Here, the charging duration $L_c$ at station $c$ now depends on the amount of time needed to recharge from $b_{0}$ up to the maximum \gls{Soc}~$\underline{b}$. 
We introduce $\delta  L_c$ to denote the additional charging time due to the battery depletion $\delta  b_c$ during the search, where  $\delta  b_c = b_0 - b_c = \sum_{l=0}^{l=i}k_{c_l,c_{l+1}}$. We then account in each algorithm for the charging duration as
\begin{equation}
 L_c ' = L_c + \delta  b_c \frac{L_c\overline{b}}{(1 - b_{0})}.
 \end{equation}
\subsection{Computational complexity improvements}\label{subsec:speedup}
In this section, we proof additional characteristics that allow to improve the computational complexity of certain problem variants. We first introduce three action space reductions, before we focus on a sharpened dominance relation for both the exact and the heuristic labeling algorithm. 
\subsubsection{Action space reductions}
In the following, we discuss some action space reductions, which we summarize in Table~\ref{tab:graphs}. We refer to the initially defined action space as \textit{complete}. In addition, we create the following reduced action spaces.
\begin{description}
	\item[\gls{direct}:] To further reduce the search space, we restrict the visits from the last visited station $c_k$ to feasible neighbor stations $c$ such that there does not exist any station $c'$ on the shortest path from $c_k$ to $c$ and denote $\widetilde{\mathcal{A}}(C)$ the set of all feasible arcs $(c_k,c)$. We however allow $c$ to be visited multiple times and show with Proposition~\ref{prop:neighborOnly} that accordingly \gls{direct} doesn't lead to a loss of generality for problem variants \gls{A} and \gls{B}.
	\item[\gls{direct_res}:] For very large instances of the problem, we combine the \gls{direct} action space with the visits restrictions from \gls{all}. The setting significantly reduces the search space but at the expense of optimality for problem variants \gls{A} and \gls{B}.
	\item[\gls{res_5}:] Finally, we restrict visits from station $c_k$ to stations $c$ reachable in less than $T^r$ time units, i.e., $t_{c_k,c} \leq T^r$ while preserving the visit restriction of \gls{all}.
\end{description}
\begin{prop}\label{prop:neighborOnly}
	Action space $\mathcal{U}$ can be modified such that $\mathcal{U}(C, 0) = \{ (c) : (c_k, c) \in \widetilde{\mathcal{A}} \} $ without loss of generality for $\gamma_c = 0$ $\forall c \in \mathcal{C}$.
\end{prop}

\begin{table}[bp]
	\centering
	\caption{Modified action spaces\label{tab:graphs}}
	{\setlength{\tabcolsep}{0.08cm}\footnotesize
		\begin{tabular}{p{2.5cm} p{13cm} }\toprule
			\gls{all} & \glsdesc{all}\\
			\gls{direct} & \glsdesc{direct}\\
			\gls{direct_res} & \glsdesc{direct_res}\\
			\gls{res_5} & \glsdesc{res_5}\\
			\bottomrule
	\end{tabular}}
	{}
\end{table}

\subsubsection{Sharpened dominance relation for the dynamic programming algorithms}
In the following, we aim to sharpen the dominance relation, i.e., we aim to discard labels faster without dropping optimality. In our initial problem setting, we assume that both travel times and charging station availability probabilities are unbounded. We now account for a bound on each of these values that still reflects a real-world application. Specifically, we assume that \textit{i}) the \gls{ev} driver must travel at least a certain amount of time between two charging stations 
$$0<  \underline{t} \leq t_{c,c'}, \forall{c,c'}\in \mathcal{C} \times \mathcal{C},$$
and \textit{ii}) that one can never be entirely sure that a charging station is available
$$ p_c \leq \overline{p} < 1, \forall c \in \mathcal{C}.$$
These bounds can be computed during a preprocessing step and allow for a sharpened dominance relation for the dynamic programming based labeling algorithms without invalidating our generic model. 
We note that charging and waiting times at a station are bounded as well. Let $W_{min}$ and $L_{min}$ (respectively $W_{max}$ and $L_{max}$) be the minimal (respectively maximal) waiting and charging times.

We then consider two partial policies $\pi_1$ and $\pi_2$ that end with the same vertex visit and their associated labels $L_1$ and $L_2$ and we refine the initial dominance criterion \eqref{eq:dom_1}--\eqref{eq:dom_5_new} (resp. \eqref{eq:dom_1}--\eqref{eq:dom_5}) into stronger dominance checks. Let $\Delta A_c = A_{c}(\pi_1) - A_{c}(\pi_2)$, $\Delta \bar{\rho}_c = \bar{\rho}_c(\pi_1) -  \bar{\rho}_c(\pi_2)$ and $\Delta \alpha_c = \alpha_c(\pi_1) -  \alpha_c(\pi_2)$

In the initial setting, if $\Delta \bar{\rho}_c \leq 0$ and $\Delta A_c > 0$, i.e., Equation~\eqref{eq:dom_1} holds  but Equation~\eqref{eq:dom_2} not, we cannot conclude that $L_1$ dominates $L_2$. In the new setting, we can ensure that if quantity $\Delta \alpha_c$ is small enough while quantity $\bar{\rho}_c$ is large enough, then $L_1 \succ  L_2$, as the lower bounds on $t_{c,c'}$ and upper bounds on $p_c$ bound the propagated values of $\alpha_c$ and $\rho_c$.
We then say $L_1 \succ L_2$, if  \eqref{eq:dom_1}\&\eqref{eq:dom_3}--\eqref{eq:dom_5_new}(resp. \eqref{eq:dom_5}) still hold and 
\mathleft
\begin{equation}\label{eq:dom_8}
\overline{p} \Delta \alpha_c   \leq (-\Delta \bar{\rho}_c)( \underline{t} + \overline{p}(\gamma_{min} - \beta_{max}))  \;.
\end{equation}
\mathcenter
Table \ref{tab:min_param} summarizes all variant-specific parameters $\beta_{max}$, $\gamma_{min}$. 

\begin{table}[!hb]
	\centering
	\caption{Parameter affectation for each problem variant\label{tab:min_param}}
	{\setlength{\tabcolsep}{0.08cm}\footnotesize
		\begin{tabular}{ p{2cm}  p{2cm} p{2cm} p{2cm} p{3cm}}
			\toprule
			Parameter & \gls{A} & \gls{B}   & \gls{C} & \gls{D}   \\
			\midrule
			$\beta_{max}$  & $\beta$ & $W_{max} $& $\beta$  & $W_{max}+ L_{max}$   \\
			$\gamma_{min}$ & 0 & 0 & $L_{min}$ & $L_{min}$ \\
			\bottomrule
	\end{tabular}}
	{}
\end{table}
\section{Design of experiments}\label{sec:exper_design}

\begin{figure}[pt]
	\centering
	\begin{tabular}{c @{\qquad} c @{\qquad} c}
		\scalebox{0.895}{\includegraphics[width=.32\linewidth]{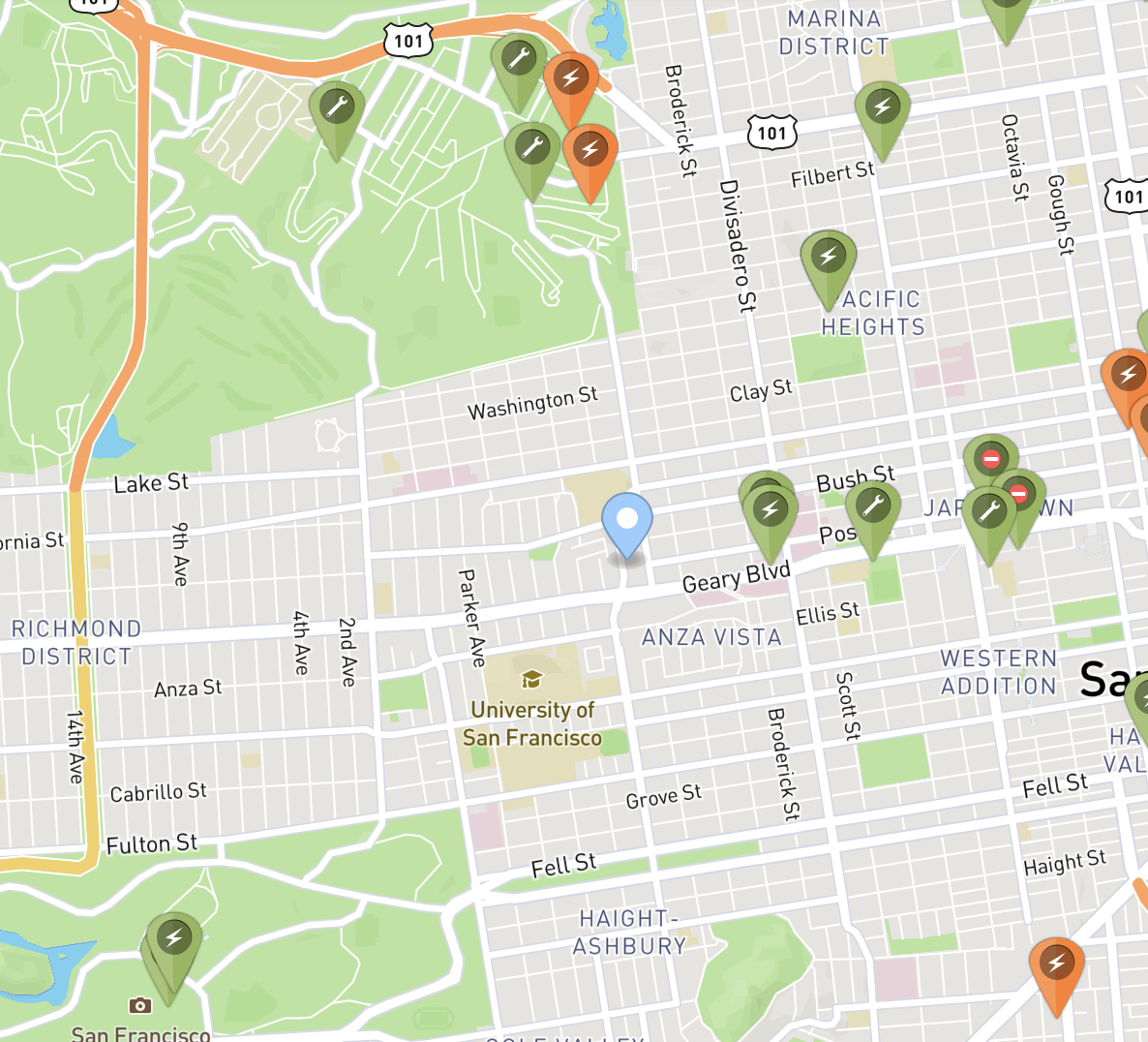}} &
		\scalebox{0.9}{\includegraphics[width=.32\linewidth]{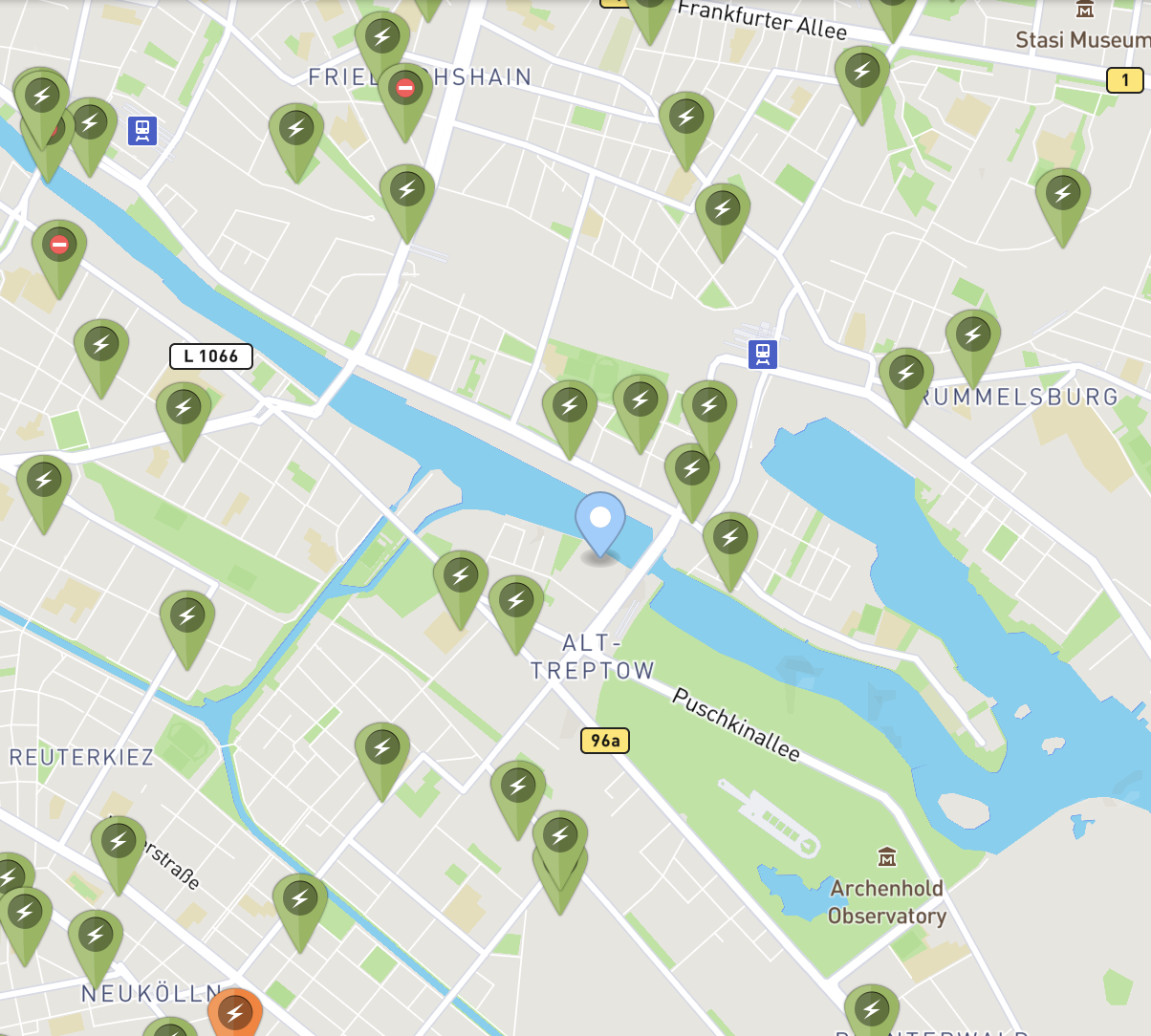}} &
		\scalebox{0.975}{\includegraphics[width=.32\linewidth]{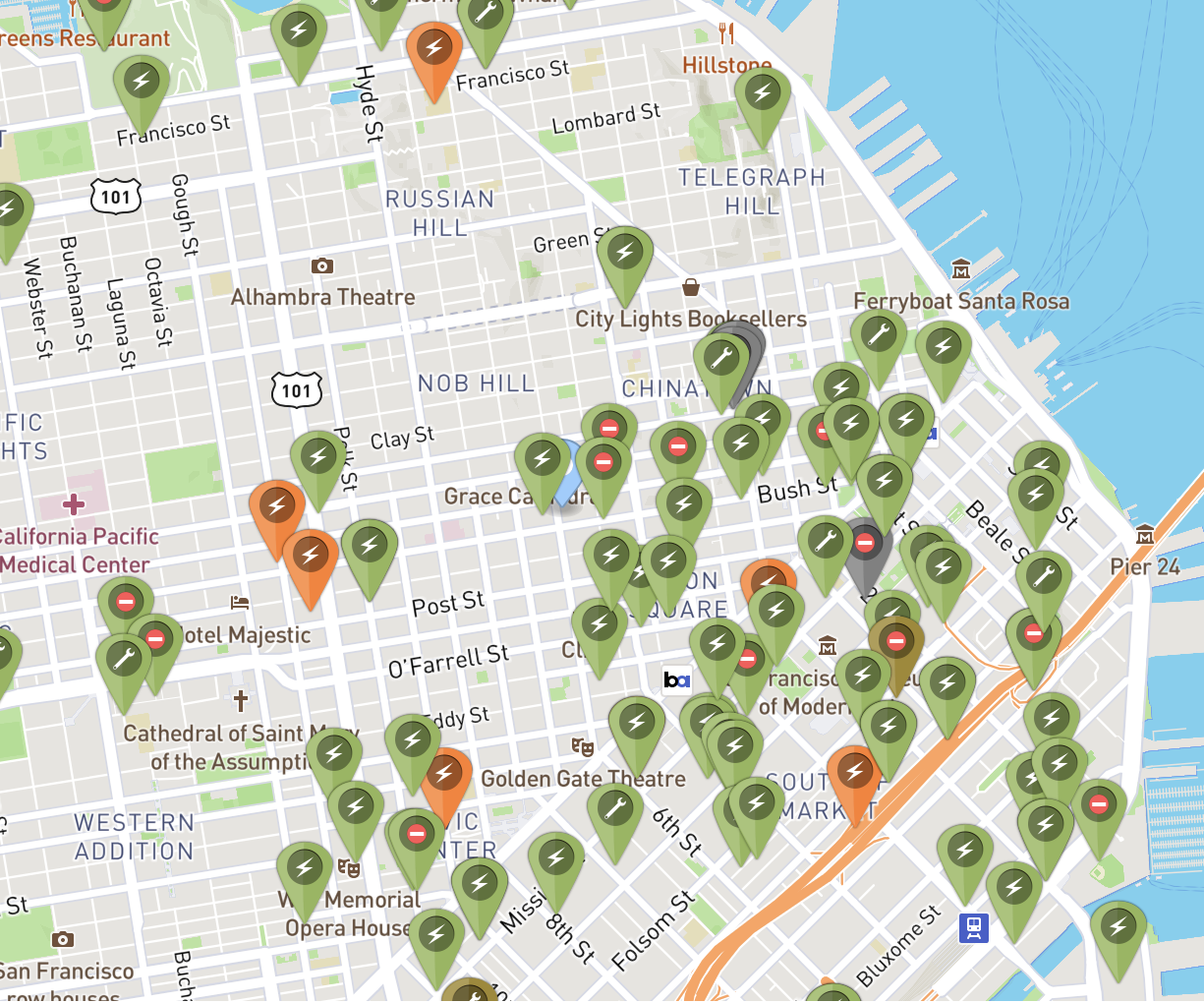}} \\
		\small (a) \gls{sf-1} & \small(b) \gls{ber-1} & \small(c) \gls{sf-2}
	\end{tabular}
	\fnote{\footnotesize Each subplot shows the geographic area used to build the respective instance graph. \gls{sf-1} represents the west side of San Francisco, USA, \gls{ber-1} the city center of Berlin, Germany, and \gls{sf-2} the financial district of San Francisco. This figure bases on \cite{OpenChargeMap} data, which is data licensed under CC BY-SA 4.0\protect\footnotemark.}
	\caption{\textbf{City maps}\label{fig:city_scenario}}
\end{figure}

To benchmark our algorithms, we develop real-world instances that allow for extensive simulation experiments. We consider three different spatial patterns (see Figure~\ref{fig:city_scenario}), based on the west side of San Francisco, USA (\gls{sf-1}), the city center of Berlin, Germany (\gls{ber-1}), and the financial district of San Francisco (\gls{sf-2}).  Here, we account for free-flow speeds to calculate travel times $t_{c,c'}$  that denote the time-shortest path between two stations  $c$ and $c'$. As our search algorithms appear to be rather insensitive to the search's starting point, we randomly choose one starting point for each pattern and use this starting point in every instance that builds on the respective pattern. 

\begin{figure}[pt]
	\centering
	\scalebox{0.8}{
\begin{tikzpicture}[font=\small]

\definecolor{color0}{rgb}{0.12156862745098,0.466666666666667,0.705882352941177}
\definecolor{color1}{rgb}{1,0.498039215686275,0.0549019607843137}
\definecolor{color2}{rgb}{0.172549019607843,0.627450980392157,0.172549019607843}

\begin{axis}[
height=8cm,
legend cell align={left},
legend style={fill opacity=0.8, draw opacity=1, text opacity=1, at={(0.03,0.97)}, anchor=north west, draw=white!80!black},
tick align=outside,
tick pos=both,
width=12cm,
x grid style={white!69.0196078431373!black},
xlabel={Search Radius \gls{surface} [meters]},
xmin=670, xmax=3530,
xtick style={color=black},
y grid style={white!69.0196078431373!black},
ylabel={Number of Stations},
ymin=-2.4, ymax=94.4,
ytick style={color=black},
scale only axis,
scale=0.7
]
\addplot [semithick, color0]
table {%
800 15
1000 20
1200 29
1400 36
1600 49
1800 62
2000 67
2200 72
2400 77
2600 80
2800 80
3000 85
3200 88
3400 90
};
\addlegendentry{\gls{sf-2}}
\addplot [semithick, color1, mark=+, mark size=1.5, mark options={solid}]
table {%
800 7
1000 9
1200 10
1400 10
1600 14
1800 16
2000 20
2200 26
2400 36
2600 43
2800 49
3000 56
3200 60
3400 64
};
\addlegendentry{\gls{ber-1}}
\addplot [semithick, color2, mark=*, mark size=1.5, mark options={solid}]
table {%
800 2
1000 3
1200 3
1400 4
1600 7
1800 9
2000 10
2200 15
2400 23
2600 28
2800 30
3000 33
3200 36
3400 43
};
\addlegendentry{\gls{sf-1}}
\end{axis}

\end{tikzpicture}}
	\caption{Amount of stations per spatial scenario\label{fig:density}}
\end{figure}
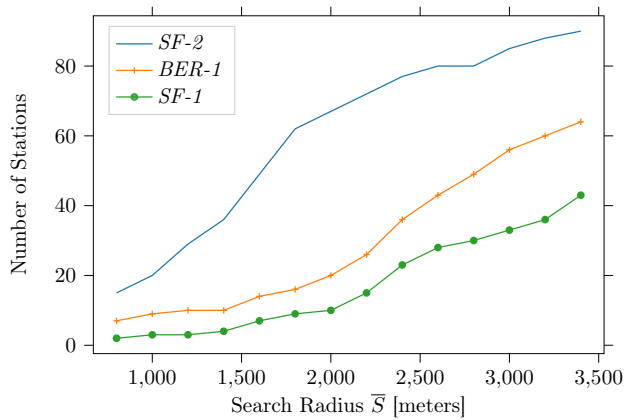

Besides the significant sensitivity to the instance size given by \gls{surface} and \gls{Tmax}, we found during preliminary studies that our search algorithms are sensitive to two general instance characteristics: the search area's charging station density and the charging station availability probability. Accordingly, we use the patterns described above to create a set of instances that covers a broad parameter range for these characteristics. Figure~\ref{fig:density} shows the amount of charging stations for each pattern, depending on the search radius $\overline{S}$ around each patterns starting point. 

Our TomTom internal availability study in the city of Berlin (cf. Table~\ref{tab:parking})  shows on average high charging stations availability in areas with a large number of on-street parking spots. In areas with few available parking spots, drivers often use available stations as free parking spots, thus blocking access to \gls{ev} drivers. In the study, the sole parking availability factor largely impacts the station availability. To reflect these amplitudes, we introduce three availability settings, drawing probabilities $p_c$ for each charging station $c$ from a $\beta-$distribution, which is centered on an expected availability of $[0.15,0.60,0.90]$ to consider a low- (\gls{low_avail}), medium- (\gls{avg_avail}) and high-availability (\gls{high_avail}) scenario. The \gls{avg_avail} and \gls{high_avail} settings represent areas with average to high parking availability. The \gls{low_avail} setting depicts a fictionous extreme case scenario prospectively corresponding to stricter parking policies and allows us to evaluate our algorithm behavior in such an environment.

\footnotetext{For more information see https://creativecommons.org/licenses/by-sa/4.0/}

For problem variants \gls{B} and \gls{D}, we consider a waiting duration $W_c$ for each station $c$, which is uniformly distributed in $[3, 15, 60, 120]$ minutes. For problem variants \gls{C} and \gls{D}, we consider a charging duration $L_c$, uniformly distributed in $[30, 60, 120]$ minutes, to account for heterogeneous stations. 

With this setup, we create a total of 9 scenarios by combining each area (\gls{ber-1}, \gls{sf-1}, \gls{sf-2}) with each availability probability distributions (\gls{low_avail}, \gls{avg_avail}, \gls{high_avail}). We then consider a small and a large-size instance for each combination, defined by the search's time budget and search radius $(\gls{Tmax} [\text{min}]/\gls{surface} [\text{meters}])$. 
For denser areas, \gls{ber-1} and \gls{sf-2}, this corresponds to $(5/800)$, resp.  $(10/2000)$. For the sparse area \gls{sf-1}, since an $800 \text{ meters}$-area only comprises $2$ stations we consider slightly wider search spaces $(10/2000)$, respectively ($15/2600$) for \gls{sf-1}, which derives a set of $18$ instances per variant, $72$ in total. We denote for each area, the smaller-size instance by $area/1$ (e.g., $\gls{sf-1}/1$) and the larger-size instance by $area/2$.

For our studies, we set the penalty cost to $\overline{\beta} = 120~\text{minutes}$ for \gls{A} and to $\overline{\beta} = 200~\text{minutes}$ for \gls{C} and refer to Section~\ref{subsec:sensitivities} for a discussion on its selection and sensitivity. To evaluate our algorithms, we conduct $N=1000$ simulation runs, each with a different station availability realization, drawn from the respective probability distribution.
For all average values reported in our results discussion, we applied two-tailed Wilcoxon signed-rank tests to verify its significance.  

\section{Results}\label{sec:experiments_v2}
In the following, we discuss our results. We first detail the performance of our algorithms with respect to several quality metrics as well as their computational effort (Section~\ref{sec:perform_analy}), before we discuss results of additional sensitivity analysis and modeling assumptions (Section~\ref{subsec:sensitivities}). 
We implemented the proposed algorithms single-threaded with Python 3.6.9, using PyPy 7.3.0 with GCC 7.3.1 and performed all experiments on a Virtual Machine on a Hypervisor, with 19 cores and 60 GB of RAM, running Ubuntu 18.04.3 LTS. For the following discussion we note that all algorithms can be implemented more efficiently, e.g., in a C\texttt{++} environment. However, as this work stems from an industry project, we are not allowed to disclose the C\texttt{++} related computational times for confidentiality reasons and use the Python implementation for a fair discussion of the algorithms'  complexity.

\subsection{Performance analysis}\label{sec:perform_analy}
We discuss the algorithms' performance from two perspectives. First, we analyze the practical benefit of our algorithms and compare their solution quality based on their performance in our simulation environment. We then relate these results to a technically focused discussion on their computational complexity.

During algorithm testing and development, we noticed that both the labeling heuristic and the rollout algorithm are sensitive to the used action space (cf. Section~\ref{subsec:speedup}) in between different problem variants. For the following discussion, we report results for the respective best-performing reduced action space and refer to Appendix~\ref{app:res_figures} for a detailed discussion of this impact.

\paragraph{Applicability}~\\ In practice, more than one key performance indicator exists to evaluate the performance of our algorithms. In the following we refer to $\widehat{\alpha}$ as the realized search cost, which corresponds to the realized driving time needed to find a station for variants \gls{A} and \gls{B} and to complete charging for \gls{C} and \gls{D}. For W variants, $\widehat{\alpha}$ includes the waiting time whenever the last visited station is occupied.
Specifically, we analyze for each algorithm its
\begin{description}
	\item[average search cost deviation] which we calculate as $$\Delta\widehat{\alpha} = \sum_{i = 1}^{1.000} \frac{\nicefrac{(\widehat{\alpha}_i  - \widehat{\alpha}_i^*)}{\widehat{\alpha}_i^*}}{1.000}$$
	with $\widehat{\alpha}_i$ being the realized search cost of the respective algorithm for simulation run $c$ and $\widehat{\alpha}_i^*$ being the best realized search cost for simulation run $c$ out of all algorithms;
	\item[success rate] that results straightforwardly from the number of simulations, for which  the respective search strategy finally led to a free charging station;
	\item[maximum search cost deviation] that results to $\Delta\widehat{\alpha}^{\text{max}} = \nicefrac{\widehat{\alpha}^{\text{max}}}{\widehat{\alpha}^{\text{max}^*}}$ with  $\widehat{\alpha}^{\text{max}} = \max \widehat{\alpha}_i$ being the maximum cost out of all simulation runs for a single algorithm and with  $\widehat{\alpha}^{\text{max}^*}$ being the maximum cost over all simulation runs and all algorithms.
\end{description}
As the exact labeling algorithm (LE) cannot solve large instances of our problem to optimality, we limit the comparison to our heuristic algorithms (LH and RO) and add a naive and a greedy solution as myopic benchmarks.
The greedy search (G) creates a sequence of charging station visits based on a greedy cost combining the travel time weighted by a station's availability probability and a penalty cost weighted by its occupancy probability. The naive search (N) mimics a driver without assistance by selecting the closest non-visited station with highest availability probability. 

Tables~\ref{tab:simul_search_avg}--\ref{tab:simul_search_max_BD} detail the average search cost deviations between all algorithms~(Table~\ref{tab:simul_search_avg}), the success rate of each algorithm (Table~\ref{tab:simul_search_success}), and the maximum search cost deviation between all algorithms~(Table~\ref{tab:simul_search_max_BD}), based on 1.000 simulation runs for each problem variant and scenario to avoid a statistical bias. 

\begin{table}[!bp]
	\centering
	\begin{threeparttable}	
		\caption{Average search cost deviations\label{tab:simul_search_avg}}
		{\setlength{\tabcolsep}{0.08cm}\footnotesize
		\begin{tabular}{rlrrrrrrrrrrrrrr}
			\toprule
			&       & \multicolumn{4}{c}{\gls{low_avail}} &       & \multicolumn{4}{c}{\gls{avg_avail}} &       & \multicolumn{4}{c}{\gls{high_avail}} \\
			\cmidrule{3-6}\cmidrule{8-11}\cmidrule{13-16}
			&       & \multicolumn{1}{l}{N} & \multicolumn{1}{l}{G} & \multicolumn{1}{l}{RO} & \multicolumn{1}{l}{LH} &       & \multicolumn{1}{l}{N} & \multicolumn{1}{l}{G} & \multicolumn{1}{l}{RO} & \multicolumn{1}{l}{LH} &       & \multicolumn{1}{l}{N} & \multicolumn{1}{l}{G} & \multicolumn{1}{l}{RO} & \multicolumn{1}{l}{LH} \\
			\midrule
			\gls{A} & \gls{sf-1}/1 & 1.38  & 0.78  & \textbf{0.00}     & 0.77  &       & 2.63  & 0.41  & \textbf{0.00}     & \textbf{0.00}     &       & 1.33  & 0.35  & 0.19  & \textbf{0.00} \\			
			& \gls{sf-1}/2 & 1.13  & 0.59  & \textbf{0.09}  & 0.14  &       & 2.66  & 2.55  & \textbf{0.00} & \textbf{0.00} &       & 2.52  & 1.08  & \textbf{0.00} & \textbf{0.00} \\
			& \gls{ber-1}/1 & 0.08  & \textbf{0.02}  & \textbf{0.02}  & 0.08  &       & 0.39  & 0.39  & \textbf{0.25}  & 0.27  &       & 0.28  & 0.98  & \textbf{0.00} & \textbf{0.00} \\
			& \gls{ber-1}/2 & 0.18  & 0.08  & \textbf{0.01}  & \textbf{0.01} &       & 3.38  & 0.16  & \textbf{0.03}  & \textbf{0.03}  &       & 4.14  & 0.24  & \textbf{0.00} & \textbf{0.00} \\
			& \gls{sf-2}/1 & 4.74  & 4.72  & 0.57  & \textbf{0.28}  &       & 5.98  & 5.98  & \textbf{0.01}  & \textbf{0.01}  &       & 6.51  & 3.18  & \textbf{0.01} & \textbf{0.01} \\
			& \gls{sf-2}/2 & 8.31  & 6.52  & 0.25  & \textbf{0.16}  &       & 10.7 & 6.16  & 0.17  & \textbf{0.10}  &       & 6.43  & 6.43  & \textbf{0.01}  & \textbf{0.01} \\
			\up\down
			\gls{B} & \gls{sf-1}/1 & 5.67  & 6.61  & 6.68  & \textbf{0.60}  &       & 2.71  & 1.72  & \textbf{0.01} & \textbf{0.01} &       & 1.08  & 0.12  & \textbf{0.00} & \textbf{0.00} \\
			& \gls{sf-1}/2 & 2.30   & 6.19  & 0.67  & \textbf{0.41} &       & 2.71  & 1.69  & \textbf{0.01}  & \textbf{0.01}  &       & 2.53  & 0.11  & \textbf{0.00} & \textbf{0.00} \\
			& \gls{ber-1}/1 & 0.41  & 1.32  & \textbf{0.20}   & \textbf{0.20}   &       & 0.15  & 0.36  & \textbf{0.04} & \textbf{0.04}  &       & 0.25  & \textbf{0.00} & \textbf{0.00} & \textbf{0.00} \\
			& \gls{ber-1}/2 & 4.39  & 1.35  & \textbf{0.20}   & \textbf{0.20}  &       & 4.36  & 0.62  & \textbf{0.24}  & \textbf{0.24}  &       & 4.12  & \textbf{0.00} & \textbf{0.00} & \textbf{0.00} \\
			& \gls{sf-2}/1 & 58.1  & 4.55  & \textbf{0.42}  & \textbf{0.42}  &       & 6.41  & 0.55  & \textbf{0.01}& \textbf{0.01}  &       & 6.52  & 0.04  & \textbf{0.01} & \textbf{0.01} \\
			& \gls{sf-2}/2 & 33.1 & 4.72  & \textbf{0.19} & \textbf{0.19}  &       & 10.8 & 0.73  & 0.17  & \textbf{0.09}  &       & 6.40   & 0.05  & \textbf{0.01}  & 0.05 \\
			\up\down
			\gls{C} & \gls{sf-1}/1 & 0.12  & \textbf{0.02} & 0.05  & 0.03  &       & 1.10   & 0.01  & \textbf{0.00} & \textbf{0.00} &       & 0.98  & 0.07  & \textbf{0.00} & \textbf{0.00} \\
			& \gls{sf-1}/2 & 1.41  & 0.18  & 0.11  & \textbf{0.20}  &       & 2.85  & 0.02  & 0.01  & \textbf{0.00} &       & 0.06  & 0.05  & \textbf{0.00} & \textbf{0.00} \\
			& \gls{ber-1}/1 & 0.20   & 0.19  & 0.07  & \textbf{0.04} &       & 0.13  & \textbf{0.01}  & \textbf{0.01}  & \textbf{0.01}  &       & 2.76  & \textbf{0.00} & \textbf{0.00} & \textbf{0.00} \\
			& \gls{ber-1}/2 & 0.42  & 0.35  & \textbf{0.05}  & 0.21  &       & 0.94  & 0.03  & \textbf{0.01}  & \textbf{0.01}  &       & 2.92  & 0.06  & \textbf{0.00} & \textbf{0.00} \\
			& \gls{sf-2}/1 & 1.24  & \textbf{0.01}  & 0.12  & 0.09  &       & 0.21  & 0.03  & \textbf{0.00} & \textbf{0.00} &       & 1.02  & 0.02  & \textbf{0.00} & \textbf{0.00} \\
			& \gls{sf-2}/2 & 0.19  & 0.06  & 0.03  & \textbf{0.02}  &       & 0.99  & 0.03  & \textbf{0.00} & \textbf{0.00} &       & 1.03  & 0.03  & \textbf{0.00} & 0.03 \\
			\up\down
			\gls{D} & \gls{sf-1}/1 & 2.37  & 1.17  & \textbf{0.00} & \textbf{0.00} &       & 1.14  & 0.04  & \textbf{0.00} & \textbf{0.00} &       & 0.98  & \textbf{0.00} & \textbf{0.00} & \textbf{0.00} \\
			& \gls{sf-1}/2 & 2.11  & 4.24  & \textbf{0.01}  & \textbf{0.01}  &       & 2.83  & 0.04  & \textbf{0.00} & \textbf{0.00} &       & 0.06  & \textbf{0.00} & \textbf{0.00} & \textbf{0.00} \\
			& \gls{ber-1}/1 & 0.74  & 0.01  & \textbf{0.00} & \textbf{0.00} &       & 0.12  & 0.02  & \textbf{0.00} & \textbf{0.00} &       & 2.79  & 0.02  & \textbf{0.00} & \textbf{0.00} \\
			& \gls{ber-1}/2 & 1.37  & \textbf{0.01}  & \textbf{0.01} & \textbf{0.01}  &       & 0.94  & 0.02  & \textbf{0.00} & \textbf{0.00} &       & 2.92  & \textbf{0.00} & \textbf{0.00} & \textbf{0.00} \\
			& \gls{sf-2}/1 & 3.48  & 0.03  & \textbf{0.01}  & \textbf{0.01}  &       & 0.17  & 0.01  & \textbf{0.00} & \textbf{0.00} &       & 1.02  & \textbf{0.00} & \textbf{0.00} & \textbf{0.00} \\
			& \gls{sf-2}/2 & 0.89  & 0.04  & \textbf{0.02}  & \textbf{0.02}  &       & 0.98  & 0.01  & \textbf{0.00} & \textbf{0.00} &       & 1.02  & \textbf{0.00} & \textbf{0.00} & \textbf{0.00} \\
			\bottomrule
	\end{tabular}}
	\begin{tablenotes}
		\small
		\item The table compares the average search cost deviation $\Delta\widehat{\alpha}$ for \gls{label}, \gls{rollout}, G and N for each of the 72 instances.
	\end{tablenotes}
	\end{threeparttable}
\end{table}%

Table~\ref{tab:simul_search_avg} shows that the RO and the LH algorithm significantly outperform the naive and the greedy benchmark in terms of average search costs. Here, an average deviation of zero implies that an algorithm always found the best search strategy, while increasing deviations indicate that the best search strategy was not found by the respective algorithm. For scenarios with high and average charging station availability, RO and LH show a similar performance. For scenarios with a low charging station availability, we observe higher differences that result in significant deviations for specific problem variants and classes.  As can be seen, higher search cost deviations occur for the W variants, especially for \gls{B}, due to the heterogeneous penalty costs that result from waiting at an occupied station. Particularly, on the \gls{sf-1} instances at low-availability, RO performs significantly worse compared to LH with respect to the average search time. In this case, a single varying station visit between two search strategies can cause such differences due to the limited amount of candidate stations and the large amplitude between penalty costs.

Table~\ref{tab:simul_search_success} shows the success rate of each algorithm for each problem variant and scenario. We disclose only $\neg W$ problem variants in this table, because  $W$ problem variants always end successfully; in the worst case with a long waiting time at the last visited charging station. As can be seen, the LH algorithm shows much higher success rates compared to the RO algorithm, which highlights the superiority of the LH algorithm compared to the RO algorithm as both algorithms show comparable performances with respect to average search cost. Remarkably, for some scenarios, the naive approach yields the highest success rates. This highlights that there exists a trade off between search costs and success rates for $\neg W$ problem variants. Analyzing Tables~\ref{tab:simul_search_avg}\&\ref{tab:simul_search_success} jointly, we conclude that for some scenarios the naive algorithm outperforms the LH algorithm in terms of success rates at the price of significantly higher search cost. Vice versa, the RO algorithm outperforms the LH algorithm for some scenarios  in terms of average search costs at the price of significantly reduced success rates (see, e.g., \gls{A}, scenario $SF-1/1$).

\begin{table}[!tp]
	\centering
	\begin{threeparttable}	
		\caption{Success rate\label{tab:simul_search_success}}
		{\setlength{\tabcolsep}{0.08cm}\footnotesize
		\begin{tabular}{rlrrrrrrrrrrrrrr}
			\toprule
			&       & \multicolumn{4}{c}{\gls{low_avail}} &       & \multicolumn{4}{c}{\gls{avg_avail}} &       & \multicolumn{4}{c}{\gls{high_avail}} \\
			\cmidrule{3-6}\cmidrule{8-11}\cmidrule{13-16}
			&       & \multicolumn{1}{l}{N} & \multicolumn{1}{l}{G} & \multicolumn{1}{l}{RO} & \multicolumn{1}{l}{LH} &       & \multicolumn{1}{l}{N} & \multicolumn{1}{l}{G} & \multicolumn{1}{l}{RO} & \multicolumn{1}{l}{LH} &       & \multicolumn{1}{l}{N} & \multicolumn{1}{l}{G} & \multicolumn{1}{l}{RO} & \multicolumn{1}{l}{LH} \\
			\midrule
			\gls{A} & \gls{sf-1}/1 & 0.81  & 0.85  & 0.36  & 0.86  &      & 1.00     & 1.00     & 1.00     & 1.00     &     & 1.00     & 1.00     & 1.00     & 1.00 \\
			& \gls{sf-1}/2 & 0.95  & 0.85  & 0.81  & 0.85  &      & 1.00     & 1.00     & 1.00     & 1.00     &     & 1.00     & 1.00     & 1.00     & 1.00 \\
			& \gls{ber-1}/1 & 0.78  & 0.77  & 0.77  & 0.80   &      & 1.00     & 1.00     & 1.00     & 1.00     &     & 1.00     & 1.00     & 1.00     & 1.00 \\
			& \gls{ber-1}/2 & 0.95  & 0.80   & 0.78  & 0.80   &      & 1.00     & 1.00     & 1.00     & 1.00     &     & 1.00     & 1.00     & 1.00     & 1.00 \\
			& \gls{sf-2}/1 & 0.70   & 0.80   & 0.78  & 0.86  &      & 1.00     & 1.00     & 1.00     & 1.00     &     & 1.00     & 1.00     & 1.00     & 1.00 \\
			& \gls{sf-2}/2 & 0.90   & 0.86  & 0.87  & 0.88  &      & 1.00     & 1.00     & 1.00     & 1.00     &     & 1.00     & 1.00     & 1.00     & 1.00 \\
			\up\down
			\gls{C} & \gls{sf-1}/1 & 0.80   & 0.55  & 0.79  & 0.83  &      & 1.00     & 1.00     & 1.00     & 1.00     &     & 1.00     & 1.00     & 1.00     & 1.00 \\
			& \gls{sf-1}/2 & 0.94  & 0.74  & 0.69  & 0.96  &      & 1.00     & 1.00     & 1.00     & 1.00     &     & 1.00     & 1.00     & 1.00     & 1.00 \\
			& \gls{ber-1}/1 & 0.78  & 0.73  & 0.78  & 0.78  &      & 1.00     & 1.00     & 0.99  & 1.00     &      & 1.00     & 1.00     & 1.00     & 1.00 \\
			& \gls{ber-1}/2 & 0.93  & 0.92  & 0.68  & 0.93 &      & 1.00     & 1.00     & 1.00     & 1.00     &     & 1.00     & 1.00     & 1.00     & 1.00 \\
			& \gls{sf-2}/1 & 0.74  & 0.51  & 0.42  & 0.77  &      & 1.00     & 1.00     & 1.00     & 1.00     &     & 1.00     & 1.00     & 1.00     & 1.00 \\
			& \gls{sf-2}/2 & 0.89  & 0.89  & 0.91  & 0.99 &      & 1.00     & 1.00     & 1.00     & 1.00     &     & 1.00     & 1.00     & 1.00     & 1.00 \\
			\bottomrule
		\end{tabular}}
	\begin{tablenotes}
		\small
		\item The table compares the success rate $\widehat{\rho}$ for \gls{label}, \gls{rollout}, G and N for each instance of $\neg W$ problem variants.
		\end{tablenotes}
	\end{threeparttable}
\end{table}

Table~\ref{tab:simul_search_max_BD} shows the deviation between the maximum search cost of all algorithms for $W$ problem variants, which are guaranteed to be feasible at the price of high waiting cost. Here, we observe a trade-off between the average search cost and the maximum search cost. As one can see, the LH and the RO algorithm show a similar performance for scenarios with medium and high charging station availability, but the LH algorithm performs better in the SF-1/1 scenario with low charging station availability. Across all scenarios the greedy algorithm sometimes yields the least maximum cost. However, this improvement stems from significantly increased search costs (cf. Table~\ref{tab:simul_search_avg}).

\begin{figure}[!hb]
	\begin{subfigure}{1.0\textwidth}
\begin{tikzpicture}
\small
\begin{groupplot}[group style={group size=3 by 1,  horizontal sep=1.6cm}]
\nextgroupplot[
height=5.5cm,
legend cell align={left},
legend style={fill opacity=0.8, draw opacity=1, text opacity=1, at={(0.03,0.97)}, anchor=north west, draw=white!80!black},
tick align=outside,
tick pos=both,
title={\gls{low_avail}},
width=5.5cm,
x grid style={white!69.0196078431373!black},
xlabel={$\widehat{\rho}$},
xmin=0.4, xmax=1.05,
xtick style={color=black},
y grid style={white!69.0196078431373!black},
ylabel={$\widehat{\alpha}$ [min]},
ymin=0, ymax=85,
scale only axis,
scale=0.7
]
\addplot [only marks, mark=x, draw=red, fill=red, colormap/viridis]
table{%
x                      y
0.841666666666667 1.95666666666667
0.728333333333333 1.645
0.821666666666667 2.69833333333333
0.848333333333333 3.50833333333333
};
\addlegendentry{\gls{A}}
\addplot [only marks, mark=*, draw=blue, fill=blue, colormap/viridis]
table{%
x                      y
0.876666666666667 58.9083333333333
0.711666666666667 50.6166666666667
0.723333333333333 50.0783333333333
0.846666666666667 78.775
};
\addlegendentry{\gls{C}}
\draw (axis cs:0.860855148607749,0.4) node[
  scale=0.4,
  anchor=south west,
  text=black,
  rotate=0.0
]{\large LH};
\draw (axis cs:0.7,1.29344558989847) node[
  scale=0.4,
  anchor= south east,
  text=black,
  rotate=0.0
]{\large RO};
\draw (axis cs:0.84,1.5) node[
  scale=0.4,
  anchor=south east,
  text=black,
  rotate=0.0
]{\large G};
\draw (axis cs:0.860855148607749,2.95605940634538) node[
  scale=0.4,
  anchor=south west,
  text=black,
  rotate=0.0
]{\large N};
\draw (axis cs:0.871790023634409,59.1644313003078) node[
  scale=0.4,
  anchor=south east,
  text=black,
  rotate=0.0
]{\large LH};
\draw (axis cs:0.70689961111828,50.8702178030303) node[
  scale=0.4,
  anchor=south east,
  text=black,
  rotate=0.0
]{\large RO};
\draw (axis cs:0.725470289268817,50.331884469697) node[
  scale=0.4,
  anchor=south west,
  text=black,
  rotate=0.0
]{\large G};
\draw (axis cs:0.848332396421505,79.0285511363636) node[
  scale=0.4,
  anchor=south west,
  text=black,
  rotate=0.0
]{\large N};

\nextgroupplot[
height=5.5cm,
tick align=outside,
tick pos=both,
title={\gls{avg_avail}},
width=5.5cm,
x grid style={white!69.0196078431373!black},
xlabel={$\widehat{\rho}$},
xmin=0.4, xmax=1.05,
xtick style={color=black},
y grid style={white!69.0196078431373!black},
ylabel={$\widehat{\alpha}$ [min]},
ymin=0, ymax=85,
scale only axis,
scale=0.7
]
\addplot [only marks, mark=x, draw=red, fill=red, colormap/viridis]
table{%
x                      y
1 0.846666666666667
1 0.846666666666667
1 1.66
1 2.72
};
\addplot [only marks, mark=*, draw=blue, fill=blue, colormap/viridis]
table{%
x                      y
1 31.1483333333333
0.998333333333333 31.1783333333333
1 31.6866666666667
1 63.2666666666667
};
\draw (axis cs:0.985,0.01) node[
  scale=0.4,
  anchor=south east,
  text=black,
  rotate=0.0
]{\large LH};
\draw (axis cs:0.985,2.5) node[
  scale=0.4,
  anchor=south east,
  text=black,
  rotate=0.0
]{\large RO};
\draw (axis cs:0.985,5.5) node[
  scale=0.4,
  anchor=south east,
  text=black,
  rotate=0.0
]{\large G};
\draw (axis cs:0.985,7.6) node[
  scale=0.4,
  anchor=south east,
  text=black,
  rotate=0.0
]{\large N};
\draw (axis cs:0.999,27.7973566660292) node[
  scale=0.4,
  anchor=north east,
  text=black,
  rotate=0.0
]{\large LH};
\draw (axis cs:0.995327899156997,30.8907971073366) node[
  scale=0.4,
  anchor=north east,
  text=black,
  rotate=0.0
]{\large RO};
\draw (axis cs:0.997103835083528,31.9442424242424) node[
  scale=0.4,
  anchor=south east,
  text=black,
  rotate=0.0
]{\large G};
\draw (axis cs:0.997742476577929,63.5242424242424) node[
  scale=0.4,
  anchor=south east,
  text=black,
  rotate=0.0
]{\large N};

\nextgroupplot[
height=5.5cm,
tick align=outside,
tick pos=both,
title={\gls{high_avail}},
width=5.5cm,
x grid style={white!69.0196078431373!black},
xlabel={$\widehat{\rho}$},
xmin=0.4, xmax=1.05,
xtick style={color=black},
y grid style={white!69.0196078431373!black},
ylabel={$\widehat{\alpha}$ [min]},
ymin=0, ymax=85,
ytick style={color=black}
]
\addplot [only marks, mark=x, draw=red, fill=red, colormap/viridis]
table{%
x                      y
1 0.57
1 0.57
1 1.07833333333333
1 1.81333333333333
};
\addplot [only marks, mark=*, draw=blue, fill=blue, colormap/viridis]
table{%
x                      y
1 31.3816666666667
1 31.23
1 32.415
1 76.7883333333333
};
\draw (axis cs:0.985,6.24722495956576e-13) node[
  scale=0.4,
  anchor=south east,
  text=black,
  rotate=0.0
]{\large LH};
\draw (axis cs:0.985,3.11936891232821) node[
  scale=0.4,
  anchor=south east,
  text=black,
  rotate=0.0
]{\large RO};
\draw (axis cs:1.00475798641742,3.11936891232821) node[
  scale=0.4,
  anchor=south west,
  text=black,
  rotate=0.0
]{\large G};
\draw (axis cs:1.00475798641742,6) node[
  scale=0.4,
  anchor=south,
  text=black,
  rotate=0.0
]{\large N};
\draw (axis cs:0.985,31.23) node[
  scale=0.4,
  anchor=north east,
  text=black,
  rotate=0.0
]{\large LH};
\draw (axis cs:1.0499917915069,30.23) node[
  scale=0.4,
  anchor=north east,
  text=black,
  rotate=0.0
]{\large RO};
\draw (axis cs:1.01123728413306,32.9047024740904) node[
  scale=0.4,
  anchor=south east,
  text=black,
  rotate=0.0
]{\large G};
\draw (axis cs:0.997743059205767,77.0459090909072) node[
  scale=0.4,
  anchor=south east,
  text=black,
  rotate=0.0
]{\large N};
\end{groupplot}

\end{tikzpicture}
		\vspace{-0.2cm}
		\subcaption{Averaged search cost vs. success rate for problem variants \gls{A} and \gls{C}\label{fig:no_wait_simul}}
		\vspace{0.4cm}
	\end{subfigure}
	\begin{subfigure}{1.0\textwidth}
\begin{tikzpicture}
\small
\definecolor{color0}{rgb}{0.75,0,0.75}

\begin{groupplot}[group style={group size=3 by 1,  horizontal sep=1.6cm}]
\nextgroupplot[
height=5.5cm,
legend cell align={left},
legend style={fill opacity=0.8, draw opacity=1, text opacity=1, at={(0.03,0.97)}, anchor=north west, draw=white!80!black},
tick align=outside,
tick pos=both,
title={\gls{low_avail}},
width=5.5cm,
x grid style={white!69.0196078431373!black},
xlabel={$\widehat{\alpha}^{\text{max}}$ [min]},
xmin=0, xmax=240,
xtick style={color=black},
y grid style={white!69.0196078431373!black},
ylabel={$\widehat{\alpha}$ [min]},
ymin=0, ymax=100,
ytick style={color=black}
]
\addplot [only marks, mark=asterisk, draw=color0, fill=color0, colormap/viridis]
table{%
x                      y
4.85 2.76
14.6583333333333 5.41666666666667
9 7.76833333333333
91.8633333333333 17.2283333333333
};
\addlegendentry{\gls{B}}
\addplot [only marks, mark=*, draw=green!50!black, fill=green!50!black, colormap/viridis]
table{%
x                      y
34.0733333333333 33.7966666666667
34.0733333333333 33.7966666666667
79.28 65.74
208.341666666667 95.205
};
\addlegendentry{\gls{D}}
\draw (axis cs:7.2873016076138,2.76) node[
  scale=0.4,
  anchor= west,
  text=black,
  rotate=0.0
]{\large LH};
\draw (axis cs:13,4.92583333393157) node[
  scale=0.4,
  anchor=south west,
  text=black,
  rotate=0.0
]{\large RO};
\draw (axis cs:0.0599525419354814,8.56219696909874) node[
  scale=0.4,
  anchor=south west,
  text=black,
  rotate=0.0
]{\large G};
\draw (axis cs:92.695849837507,17.5313636360814) node[
  scale=0.4,
  anchor=south west,
  text=black,
  rotate=0.0
]{\large N};
\draw (axis cs:25.0304301075269,34.2550169101732) node[
  scale=0.4,
  anchor=south west,
  text=black,
  rotate=0.0
]{\large LH};
\draw (axis cs:35.0275268817204,33.4936363636364) node[
  scale=0.4,
  anchor=north west,
  text=black,
  rotate=0.0
]{\large RO};
\draw (axis cs:79.9113470967742,66.0240909090909) node[
  scale=0.4,
  anchor=south west,
  text=black,
  rotate=0.0
]{\large G};
\draw (axis cs:207.849539698925,95.4890909090909) node[
  scale=0.4,
  anchor=south east,
  text=black,
  rotate=0.0
]{\large N};

\nextgroupplot[
height=5.5cm,
tick align=outside,
tick pos=both,
title={\gls{avg_avail}},
width=5.5cm,
x grid style={white!69.0196078431373!black},
xlabel={$\widehat{\alpha}^{\text{max}}$ [min]},
xmin=0, xmax=100,
xtick style={color=black},
y grid style={white!69.0196078431373!black},
ylabel={$\widehat{\alpha}$ [min]},
ymin=0, ymax=85,
ytick style={color=black}
]
\addplot [only marks, mark=asterisk, draw=color0, fill=color0, colormap/viridis]
table{%
x                      y
4.88333333333333 0.846666666666667
4.47666666666667 0.85
9.76333333333333 1.545
9.43333333333333 2.75
};
\addplot [only marks, mark=*, draw=green!50!black, fill=green!50!black, colormap/viridis]
table{%
x                      y
35.4483333333333 31.0333333333333
35.49 31.035
45.675 31.7066666666667
96.4483333333333 62.945
};
\draw (axis cs:11,0.01) node[
  scale=0.4,
  anchor=south west,
  text=black,
  rotate=0.0
]{\large LH};
\draw (axis cs:11,1.5) node[
  scale=0.4,
  anchor=south west,
  text=black,
  rotate=0.0
]{\large RO};
\draw (axis cs:14,6.5) node[
  scale=0.4,
  anchor= west,
  text=black,
  rotate=0.0
]{\large G};
\draw (axis cs:14,7.5) node[
  scale=0.4,
  anchor=south west,
  text=black,
  rotate=0.0
]{\large N};
\draw (axis cs:40,27.682261228355) node[
  scale=0.4,
  anchor=north,
  text=black,
  rotate=0.0
]{\large LH};
\draw (axis cs:40,30.7757575757576) node[
  scale=0.4,
  anchor=north,
  text=black,
  rotate=0.0
]{\large RO};
\draw (axis cs:46.1205520956031,31.9642424242424) node[
  scale=0.4,
  anchor=south west,
  text=black,
  rotate=0.0
]{\large G};
\draw (axis cs:96.1010336074272,63.2025757575757) node[
  scale=0.4,
  anchor=south east,
  text=black,
  rotate=0.0
]{\large N};

\nextgroupplot[
height=5.5cm,
tick align=outside,
tick pos=both,
title={\gls{high_avail}},
width=5.5cm,
x grid style={white!69.0196078431373!black},
xlabel={$\widehat{\alpha}^{\text{max}}$ [min]},
xmin=0, xmax=100,
xtick style={color=black},
y grid style={white!69.0196078431373!black},
ylabel={$\widehat{\alpha}$ [min]},
ymin=0, ymax=85,
ytick style={color=black}
]
\addplot [only marks, mark=asterisk, draw=color0, fill=color0, colormap/viridis]
table{%
x                      y
1.72833333333333 0.571666666666667
1.63333333333333 0.57
1.77 0.621666666666667
2.71166666666667 1.81166666666667
};
\addplot [only marks, mark=*, draw=green!50!black, fill=green!50!black, colormap/viridis]
table{%
x                      y
35.7466666666667 31.0816666666667
35.8183333333333 31.0766666666667
48.84 31.2633333333333
77.7116666666667 76.7966666666667
};
\draw (axis cs:5,0.00311028882861353) node[
  scale=0.4,
  anchor=south west,
  text=black,
  rotate=0.0
]{\large LH};
\draw (axis cs:5,2.5) node[
  scale=0.4,
  anchor=south west,
  text=black,
  rotate=0.0
]{\large RO};
\draw (axis cs:0.820408602150542,2) node[
  scale=0.4,
  anchor=south west,
  text=black,
  rotate=0.0
]{\large G};
\draw (axis cs:0.79310817204301,4) node[
  scale=0.4,
  anchor=south west,
  text=black,
  rotate=0.0
]{\large N};
\draw (axis cs:35.7466666666667 ,31.0766666666667) node[
  scale=0.4,
  anchor=south east,
  text=black,
  rotate=0.0
]{\large LH};
\draw (axis cs:35.8183333333333,31.0816666666667) node[
  scale=0.4,
  anchor=north east,
  text=black,
  rotate=0.0
]{\large RO};
\draw (axis cs:50.0355661290323,31.2633333333333) node[
  scale=0.4,
  anchor=south west,
  text=black,
  rotate=0.0
]{\large G};
\draw (axis cs:78.0589784946236,77.0542424242424) node[
  scale=0.4,
  anchor=south west,
  text=black,
  rotate=0.0
]{\large N};
\end{groupplot}

\end{tikzpicture}
		\vspace{-0.2cm}
	\subcaption{Averaged search cost vs. worst search cost for problem variants \gls{B} and \gls{D} \label{fig:wait_simul}}
	\end{subfigure}
	\caption{Trade off between the average search cost and the success rate, resp. the worst search cost}
	\label{fig:avail_impact}	
\end{figure}
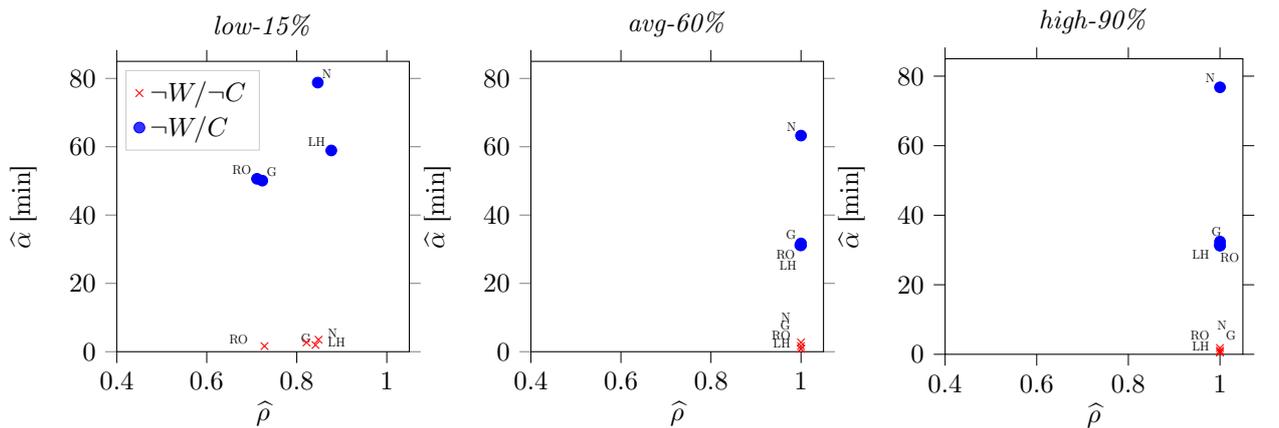
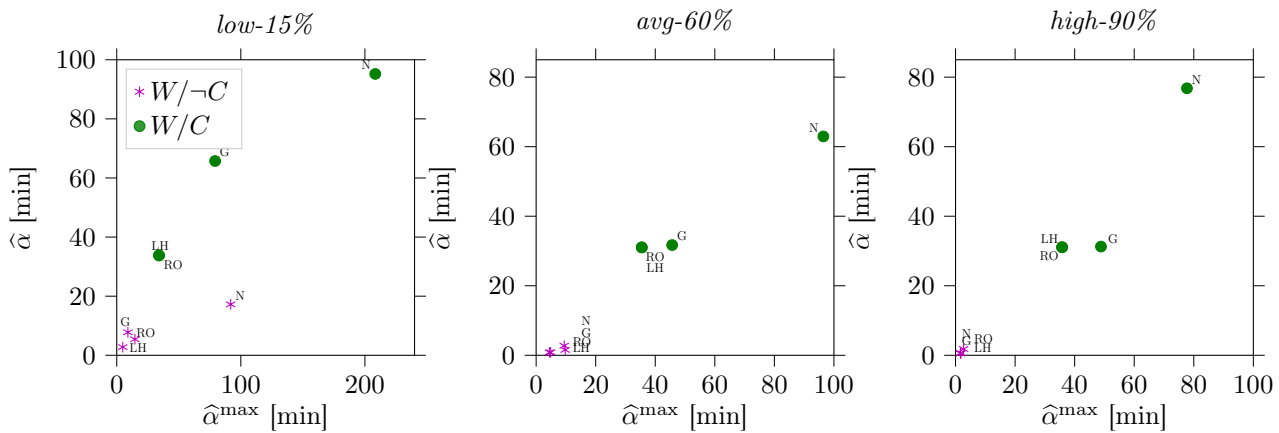

Figure~\ref{fig:avail_impact} illustrates the trade-offs discussed above by showing for all algorithms the trade off between  average search costs and i) the success rate, aggregated over all scenarios for $\neg W$ problem variants (Figure~\ref{fig:no_wait_simul}), or ii) the maximum search cost, aggregated over all scenarios for $W$ problem variants (Figure~\ref{fig:wait_simul}).
\begin{table}[!hb]
	\centering
	\begin{threeparttable}	
		\caption{Maximal Search cost \label{tab:simul_search_max_BD}}
		{\setlength{\tabcolsep}{0.08cm}\footnotesize
			\begin{tabular}{rlrrrrrrrrrrrrrr}
				\toprule
				&       & \multicolumn{4}{c}{\gls{low_avail}} &       & \multicolumn{4}{c}{\gls{avg_avail}} &       & \multicolumn{4}{c}{\gls{high_avail}} \\
				\cmidrule{3-6}\cmidrule{8-11}\cmidrule{13-16}
				&       & \multicolumn{1}{l}{N} & \multicolumn{1}{l}{G} & \multicolumn{1}{l}{RO} & \multicolumn{1}{l}{LH} &       & \multicolumn{1}{l}{N} & \multicolumn{1}{l}{G} & \multicolumn{1}{l}{RO} & \multicolumn{1}{l}{LH} &       & \multicolumn{1}{l}{N} & \multicolumn{1}{l}{G} & \multicolumn{1}{l}{RO} & \multicolumn{1}{l}{LH} \\
				\midrule
				\gls{B} & \gls{sf-1}/1 & 22.6 & 2.29  & 10.8 & \textbf{0.00}    &       & 1.42  & \textbf{0.00}     & 0.10   & 0.24  &       & 2.11  & \textbf{0.00}     & 0.13  & 0.13 \\
				& \gls{sf-1}/2 & 4.45  & 2.29  & \textbf{0.00}     & \textbf{0.00}     &       & 2.01  & 0.11  & \textbf{0.00}     & \textbf{0.00}     &       & 0.69  & \textbf{0.00}     & 0.13  & 0.13 \\
				& \gls{ber-1}/1 & 1.12  & \textbf{0.00}     & 0.32  & 0.32  &       & 0.28  & 2.08  & \textbf{0.00}     & \textbf{0.00}     &       & 0.27  & 0.34  & \textbf{0.00}     & \textbf{0.00} \\
				& \gls{ber-1}/2 & 33.7 & \textbf{0.00}     & 0.32  & 0.32  &       & 0.60   & 2.08  & \textbf{0.00}     & \textbf{0.00}     &       & 0.71  & 0.34  & \textbf{0.00}     & \textbf{0.00} \\
				& \gls{sf-2}/1 & 29.0 & 0.27  & \textbf{0.00}     & \textbf{0.00}     &       & 0.89  & 1.05  & \textbf{0.00}     & \textbf{0.00}     &       & 0.74  & \textbf{0.00}     & 0.70   & 0.70 \\
				& \gls{sf-2}/2 & 30.2 & 0.27  & \textbf{0.00}     & \textbf{0.00}     &       & 3.60   & 2.04  & \textbf{0.00}     & 1.07  &       & 0.02  & 0.55  & \textbf{0.00}     & 0.55 \\
				\up\down
				\gls{D} & \gls{sf-1}/1 & 6.07  & 1.22  & \textbf{0.00}     & \textbf{0.00}     &       & 2.60   & \textbf{0.00}     & 0.01  & 0.01  &       & 0.85  & \textbf{0.00}     & \textbf{0.00}     & \textbf{0.00} \\
				& \gls{sf-1}/2 & 2.78  & 6.24  & \textbf{0.00}     & \textbf{0.00}     &       & 2.61  & 0.01  & \textbf{0.00}     & \textbf{0.00}     &       & \textbf{0.00}     & 0.03  & 0.03  & 0.03 \\
				& \gls{ber-1}/1 & 2.69  & \textbf{0.00}     & \textbf{0.00}     & \textbf{0.00}     &       & 0.73  & 0.36  & \textbf{0.00}     & \textbf{0.00}     &       & 1.63  & 1.70   & \textbf{0.00}     & \textbf{0.00} \\
				& \gls{ber-1}/2 & 6.41  & \textbf{0.00}     & \textbf{0.00}     & \textbf{0.00}     &       & 0.96  & 1.38  & \textbf{0.00}     & \textbf{0.00}     &       & 2.39  & 0.01  & \textbf{0.00}     & \textbf{0.00} \\
				& \gls{sf-2}/1 & 6.34  & 0.06  & \textbf{0.00}     & \textbf{0.00}     &       & 2.57  & \textbf{0.00}     & \textbf{0.00}     & \textbf{0.00}     &       & 0.99  & \textbf{0.00}     & 0.01  & 0.01 \\
				& \gls{sf-2}/2 & 6.49  & 0.20   & \textbf{0.00}     & \textbf{0.00}     &       & 0.85  & \textbf{0.00}     & 0.02  & 0.01  &       & 0.95  & \textbf{0.00}     & 0.01  & \textbf{0.00} \\
				\bottomrule
		\end{tabular}}
		\begin{tablenotes}
			\small
			\item The table compares the maximal search costs $\Delta\widehat{\alpha}^{\text{max}}$ for \gls{label}, \gls{rollout}, G and N for each instance of $W$ problem variants.
		\end{tablenotes}
	\end{threeparttable}
\end{table}

For $\neg W$ problem variants and scenarios with medium or high charging station availability, all algorithms yield a 1.00\% success rate but the advanced algorithms yield significantly lower cost, with LH being the best-performing algorithm. In these cases the greedy algorithm performs close to the advanced algorithms for problem variant \gls{C}. For scenarios with low charging station availability the LH algorithm yields a better success rate but the RO algorithm yields lower search cost. For $W$ problem variants, we observe that the advanced algorithms outperform the myopic algorithms significantly, independent of the charging station availability. This performance difference is higher for problem variant \gls{B}. For problem variant \gls{D}, G performs close to the advanced algorithms for scenarios with medium or high charging station availability.

Concluding, our results show that the developed search algorithms can significantly improve the search quality across all problem variants and scenarios.
 Compared to G, the advanced algorithms decrease the search cost by 21 \% in average and up to 44\% for areas with a scarce number of charging stations and low charging station availability. For $\neg W$ variants, the failure rate decreases by 30 \% with low charging station availability. Moreover, advanced algorithms allow to reduce search times by 5 (\gls{B}) to 31 (\gls{D}) minutes compared to myopic approaches for W problem variants with low station availability. This reduction potential decreases for scenarios with average to high station availability. However, advanced algorithms appear to be more robust in these cases and lower the worst search costs by 20 \% (\gls{B}) and 12 \% (\gls{D}). Comparing the LH and RO algorithm among each other, LH prolongs the search compared to RO but obtains a significantly higher success rate for $\neg W$ variants. For scenarios \gls{avg_avail} and \gls{high_avail} RO and LH show a similar performance.

\paragraph{Computational tractability}~\\
To compare the performance of the LH and RO algorithm against the exact labeling algorithm, we derive a set of 8064 test instances by varying \gls{Tmax} $\in [5,10,15,20]~\text{minutes}$, $\gls{surface} \in [800,1.000,...,3400]~\text{meters}$, and the availability distribution in $\{\gls{low_avail}, \gls{avg_avail}, \gls{sf-2}\}$ for all problem variants and each search area. Here, we use a large time limit of 15,000 seconds to obtain a maximum set of solutions that are eligible for our comparison, i.e., solutions for instances that could be solved with all three algorithms.

Table~\ref{tab:qualityAggregExact}  shows the rate of instances solved within this time limit ($\hat{n}$), and the average computational time of the successful runs ($\hat{t}$) for both heuristic algorithms (RO, LH) and for the exact labeling algorithm (LE), as well as the average ratio ($g_\alpha$) between each heuristic result and the optimal solution.

\begin{table}[!t]
	\centering
	\begin{threeparttable}	
		\caption{Aggregated computational results over all tested instances per scenario\label{tab:qualityAggregExact}}
		{\setlength{\tabcolsep}{0.08cm}\footnotesize
		\begin{tabular}{rlrrrrrrrrrr p{0.02\textwidth} rrrrrrrrrr}
			\toprule
			& \multicolumn{1}{l}{ } & \multicolumn{3}{c}{ LH }  &       & \multicolumn{3}{c}{ RO } &       & \multicolumn{2}{c}{ L-E} && \multicolumn{3}{c}{ LH }  &       & \multicolumn{3}{c}{ RO } &       & \multicolumn{2}{c}{ L-E}  \\
			\cmidrule{3-5}\cmidrule{7-9}\cmidrule{11-12} \cmidrule{14-16}\cmidrule{18-20}\cmidrule{22-23}
			& \multicolumn{1}{r}{} & \multicolumn{1}{l}{ $g_\alpha$ } & \multicolumn{1}{l}{ $\hat{t}$ } & \multicolumn{1}{l}{ $ \hat{n}$  } &       & \multicolumn{1}{l}{ $g_\alpha$ } & \multicolumn{1}{l}{ $\hat{t}$ } & \multicolumn{1}{l}{ $ \hat{n}$ } &       & \multicolumn{1}{l}{ $\hat{t}$ } & \multicolumn{1}{l}{ $ \hat{n}$} && \multicolumn{1}{l}{ $g_\alpha$ } & \multicolumn{1}{l}{ $\hat{t}$ } & \multicolumn{1}{l}{ $ \hat{n}$  } &       & \multicolumn{1}{l}{ $g_\alpha$ } & \multicolumn{1}{l}{ $\hat{t}$ } & \multicolumn{1}{l}{ $ \hat{n}$} &       & \multicolumn{1}{l}{ $\hat{t}$ } & \multicolumn{1}{l}{ $ \hat{n}$} \\
			\midrule
			\up\down
			&       & \multicolumn{10}{c}{\gls{A}}                                  & \multicolumn{10}{c}{\gls{C}} \\
			\cmidrule{3-12}\cmidrule{14-23}
			\gls{low_avail}  & \gls{sf-1} & 1.01  & 1.78  & 1.00   &       & 1.32   & 0.09  & 1.00   &       & 165 & 0.64    && 1.00& 104 & 1.00   &       & 1.10& 0.04  & 1.00   &       & 87.7 & 0.64 \\
			& \gls{ber-1}  & 1.04  & 0.49  & 1.00   &       & 1.43  & 0.28  & 1.00   &       & 1386 & 0.46    &&  1.00 & 412 & 1.00   &       & 1.43  & 0.18  & 1.00   &       & 1338  & 0.46 \\
			& \gls{sf-2}  & 1.03 & 445 & 0.66    &       & 1.22   & 1.38  & 1.00   &       & 8646  & 0.04    && 1.06  & 1088  & 0.70    &       & 1.48  & 0.86  & 1.00   &       & 5313  & 0.02 \\
			&       &       &       &       &       &       &       &       &       &       &     &       &       &       &       &       &       &       &       &       & \\
			\gls{avg_avail}   & \gls{sf-1} & \textbf{1.00} & 266 & 0.96    &       & 1.04  & 0.11  & 1.00   &       & 7.02 & 0.64    && \textbf{1.00} & 9.34  & 1.00   &       & 1.05  & 0.05  & 1.00   &       & 7.12  & 0.64 \\
			& \gls{ber-1}  & 1.00 & 84.5 & 1.00   &       & 1.01  & 0.23  & 1.00   &       & 473 & 0.59  && 1.00 & 0.54  &1.00  &       & 1.03  & 0.15  & 1.00   &       & 1057 & 0.64 \\	
			& \gls{sf-2}  & \textbf{1.00} & 803 & 0.34    &       & 1.11  & 1.76  & 1.00   &       & 4816  & 0.05    & & \textbf{1.00} & 355 & 0.57    &       & 1.00 & 0.83  & 1.00   &       & 5560  & 0.25\\
			&       &       &       &       &       &       &       &       &       &       &     &       &       &       &       &       &       &       &       &       & \\
			\gls{high_avail}   & \gls{sf-1} & \textbf{1.00} & 106 & 0.66    &       & 1.03  & 0.10   & 1.00   &       & 16.4  & 0.64    && 1.05  & 75.1 & 0.93    &       & 1.01  & 0.05  & 1.00   &       &8.48   & 0.64 \\
			& \gls{ber-1}  & \textbf{1.00} & 66.7 & 0.66    &       & \textbf{1.00} & 0.35  & 1.00   &       & 517 & 0.57    && \textbf{1.00} & 2.82  & 1.00   &       & 1.00 & 0.15  & 1.00   &       & 266  & 0.59\\
			& \gls{sf-2}  & \textbf{1.00} & 1.70   & 0.25    &       & \textbf{1.00} & 0.95  & 1.00   &       & 4156  & 0.04     && \textbf{1.00} & 750 & 0.34    &       & \textbf{1.00} & 1.06  & 1.00   &       & 2303  & 0.04 \\
			\midrule
			\up\down
			&       & \multicolumn{10}{c}{\gls{B}}                                  & \multicolumn{10}{c}{\gls{D}} \\
			\cmidrule{3-12}\cmidrule{14-23}
			\gls{low_avail}  & \gls{sf-1} & 1.04  & 0.05  & 1.00   &       & 1.09  & 0.01  & 1.00   &       & 482 & 0.68    && \textbf{1.00} & 1.90  & 1.00   &       & \textbf{1.00} & 0.12  & 1.00   &       & 493 & 0.68 \\
			& \gls{ber-1}  & \textbf{1.00} & 1.60   & 1.00   &       & \textbf{1.00} & 0.04  & 1.00   &       & 283 & 0.64    && \textbf{1.00} & 5.98  & 1.00   &       & \textbf{1.00} & 0.77  & 1.00   &       & 803 & 0.66 \\
			& \gls{sf-2}  & \textbf{1.00} & 160 & 0.64    &       & 1.03  & 0.10   & 1.00   &       & 2111  & 0.25    && \textbf{1.00} & 358 & 0.79    &       & \textbf{1.00} & 2.95  & 1.00   &       & 2565  & 0.25 \\
			&       &       &       &       &       &       &       &       &       &       &     &       &       &       &       &       &       &       &       &       &  \\
			\gls{avg_avail}   & \gls{sf-1} & 1.00 & 0.05  & 1.00   &       & 1.03  & 0.01  & 1.00   &       & 2.33  & 0.64    && \textbf{1.00} & 0.25  & 1.00   &       & 1.00 & 0.12  & 1.00   &       & 473 & 0.68 \\
			& \gls{ber-1}  & \textbf{1.00} & 0.72  & 1.00   &       & \textbf{1.00} & 0.01  & 1.00   &       & 1046  & 0.64    && \textbf{1.00} & 0.83  & 1.00   &       & \textbf{1.00} & 0.57  & 1.00   &       & 495 & 0.64 \\
			& \gls{sf-2}  & \textbf{1.00} & 490 & 0.54    &       & 1.09  & 0.03  & 1.00   &       & 4219  & 0.25    && \textbf{1.00} & 527 & 0.64    &       & 1.00 & 2.59  & 1.00   &       & 2031  & 0.25 \\
			&       &       &       &       &       &       &       &       &       &       &      &       &       &       &       &       &       &       &       &       &  \\
			\gls{high_avail}   & \gls{sf-1} & \textbf{1.00} & 0.06  & 1.00   &       & 1.00 & 0.01  & 1.00   &       & 1.68  & 0.64    && \textbf{1.00} & 1.99  & 0.89    &       & \textbf{1.00} & 0.08  & 1.00   &       & 352 & 0.68 \\
			& \gls{ber-1}  & \textbf{1.00} & 0.38  & 1.00   &       & \textbf{1.00} & 0.01  & 1.00   &       & 476 & 0.64    && \textbf{1.00} & 0.05  & 0.89    &       & \textbf{1.00} & 0.34  & 1.00   &       & 630 & 0.66 \\
			& \gls{sf-2}  & \textbf{1.00} & 0.43  & 0.25    &       & \textbf{1.00} & 0.05  & 1.00   &       & 2024  & 0.25    && \textbf{1.00} & 4.27  & 0.29    &       & \textbf{1.00} & 1.88  & 1.00   &       & 2661  & 0.25 \\
			\bottomrule
		\end{tabular}}
	\begin{tablenotes}
		\small
		\item Abbreviations hold as follows: $g_\alpha$ - averaged optimality gap over all tested instances with $g_\alpha = \frac{\alpha^{\text{heur}}}{\alpha^{\text{opt}}}$, $\hat{t}$ - averaged computational time[s], $\hat{n}$ - rate of instances that can be computed in less than 15000 seconds. We note that an average $g_\alpha$ of 1.00 indicates that an algorithm (almost) always finds the optimal solution. If it always finds the optimal solution we highlight the respective $g_\alpha$ in bold font, whereas we leave it in normal font if some solutions remain heuristic but are not reflected in the value of $g_\alpha$ due to rounding.
	\end{tablenotes}
\vspace{-0.4cm}
	\end{threeparttable}
\end{table}

 As can be seen, our LH algorithm provides optimal or close to optimal solutions. The RO algorithm shows a similar solution quality for problem variants of type $W$ but shows a significantly worse solution quality for problem variants of type $\neg W$ with a low charging station availability. While the RO algorithm  succeeds in solving all instances with computational times of a few seconds, the LH algorithm improves upon its exact counterpart with respect to the number of instances solved and computational times but can neither solve all instances nor preserve computational times at the order of seconds.
 
Figure~\ref{fig:opt_gap} illustrates the solution quality deviations between RO and LH for a representative subset of scenarios (\gls{ber-1}, $\gls{Tmax} = 10$) with low and medium charging station availability and varying search radii $\gls{surface}\in [800,1.000,...,3400]$. As can be seen, the observed differences are sensitive to the problem variant. Significant differences occur for $\neg W$ problem variants with low charging station availability and varying search radii, while the algorithms perform similarly on all other problem variants. These high deviations result from penalty costs for unsuccessful searches, which can only result for $\neg W$ problem variants and are more likely to occur at low charging station availability.
 
 Figure~\ref{fig:explosion} shows the computational time of LH, RO, and LE for a representative subset of instances (\gls{ber-1}, $\gls{Tmax} = 20$, \gls{high_avail}) for different search radii. As can be seen, RO and LH remain equally efficient for search radii up to 1400m but the computational time of LH increases exponentially for bigger search radii. Synthesizing Table~\ref{tab:qualityAggregExact} and Figures~\ref{fig:opt_gap}\&\ref{fig:explosion}, we observe a trade-off between LH and RO: while RO yields robust computational times at the price of varying solution quality, LH yields a robust solution quality at the price of exponentially increasing computational time.

From a practitioner's perspective, computational times of a few milliseconds are imperative to deploy a search algorithm in practice, e.g., embedded into a navigation application. Here, one could resolve the trade-off between RO and LH in two different ways. On the one hand, one could apply both algorithms selectively, using LH for tractable problem sizes and RO for larger problem sizes. On the other hand, one could always apply LH and terminate its search after a given time limit. To analyze which strategy appears to be more promising in our case, we compare the performance of RO and LH against each other, limiting the computation time of LH to 1 second, which equals a sufficiently small computational time when using an efficient implementation. 
 \begin{figure}[!ht]
	\centering
	\scalebox{0.57}{\input{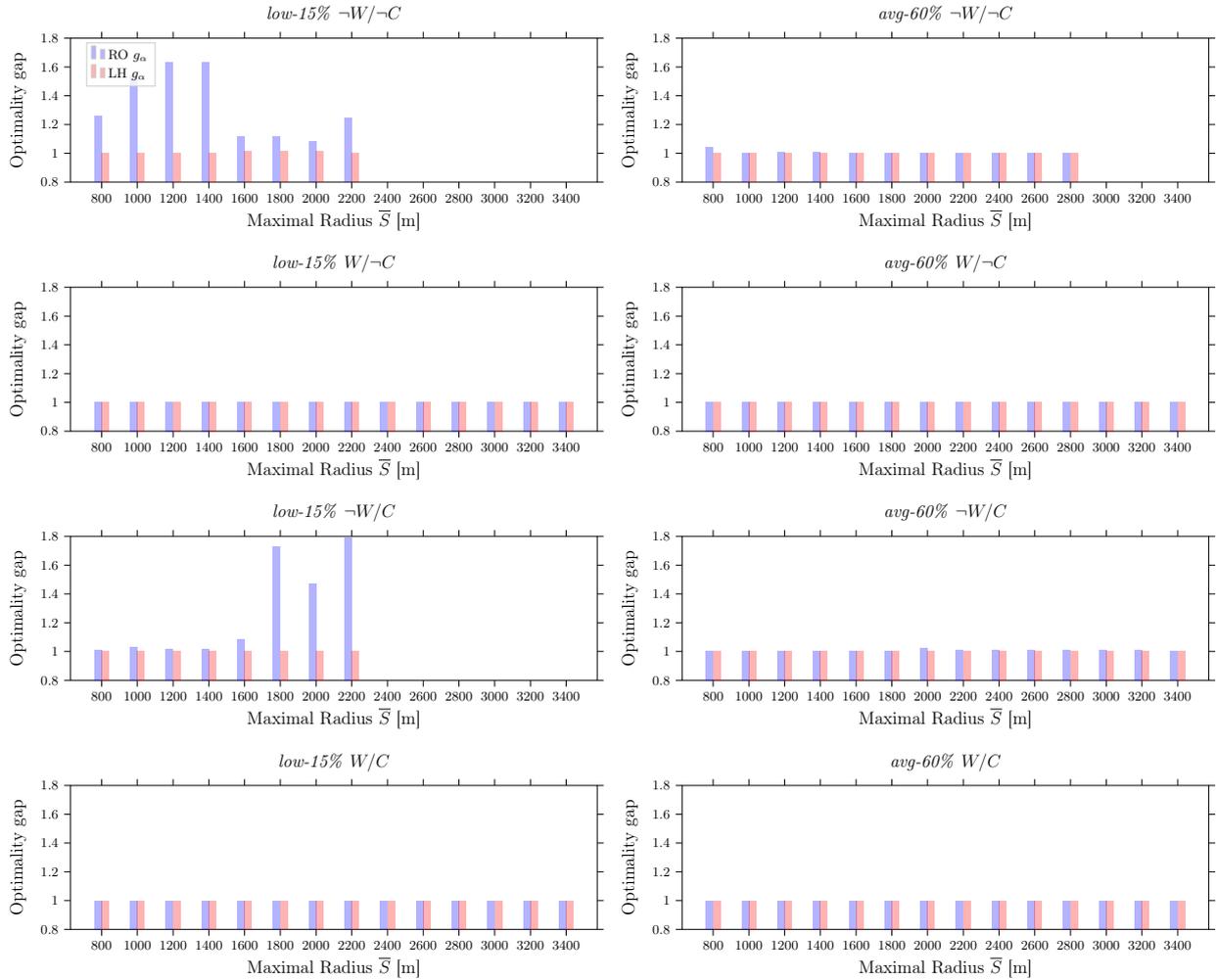}}
	\caption{Optimality gap for \gls{label} and \gls{rollout} for the \gls{ber-1} scenario with fixed \gls{Tmax}=10 min for 15 \% and 60 \% availability scenarios\label{fig:opt_gap}}
\end{figure}
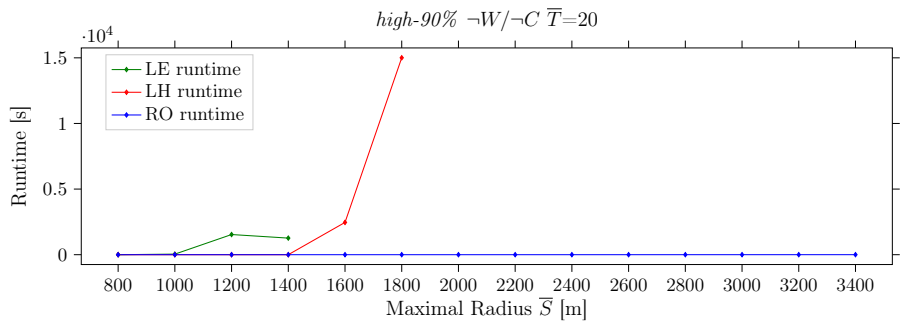
\begin{figure}[!ht]
	\centering
	\scalebox{0.65}{
\begin{tikzpicture}[font=\normalsize]

\begin{axis}[
height=6cm,
legend cell align={left},
legend style={fill opacity=0.8, draw opacity=1, text opacity=1, at={(0.03,0.97)}, anchor=north west, draw=white!80!black},
tick align=outside,
tick pos=both,
title={\gls{high_avail} \gls{A} \gls{Tmax}=20},
width=18cm,
x grid style={white!69.0196078431373!black},
xlabel={Maximal Radius \gls{surface} [m]},
xmin=-0.65, xmax=13.65,
xtick style={color=black},
xtick={0,1,2,3,4,5,6,7,8,9,10,11,12,13},
xticklabels={800,1000,1200,1400,1600,1800,2000,2200,2400,2600,2800,3000,3200,3400},
y grid style={white!69.0196078431373!black},
ylabel={Runtime [s]},
ymin=-750.2155, ymax=15754.5255,
ytick style={color=black},
title style={font=\large},
label style={font=\large}
]
\addplot [semithick, green!50!black, mark=diamond*, mark size=1.25, mark options={solid}]
table {%
0 0.25
1 43.82
2 1536.25
3 1264.32
4 nan
5 nan
6 nan
7 nan
8 nan
9 nan
10 nan
11 nan
12 nan
13 nan
};
\addlegendentry{\gls{exact} runtime}
\addplot [semithick, red, mark=diamond*, mark size=1.25, mark options={solid}]
table {%
0 0.01
1 0.02
2 0.02
3 0.02
4 2459.81
5 15004.31
6 nan
7 nan
8 nan
9 nan
10 nan
11 nan
12 nan
13 nan
};
\addlegendentry{\gls{label} runtime}
\addplot [semithick, blue, mark=diamond*, mark size=1.25, mark options={solid}]
table {%
0 0
1 0.02
2 0.03
3 0.04
4 0.01
5 0.04
6 0.05
7 0.13
8 0.34
9 0.51
10 0.81
11 1.1
12 1.38
13 1.74
};
\addlegendentry{\gls{rollout} runtime}
\end{axis}

\end{tikzpicture}}
	\caption{Computational times for \gls{label}, \gls{exact} and \gls{rollout} for the \gls{ber-1} scenario with fixed \gls{Tmax}=20 min with 90 \% availability \label{fig:explosion}}
\end{figure}

Figure~\ref{fig:A_comparison} shows this comparison for problem variant \gls{A}, while we provide figures for all other problem variants in Appendix~\ref{app:res_heur} to keep this paper concise. In general, the time-limited LH outperforms RO (blue areas) for small search radii, in particular for instances with low charging station availability, whereas RO outperforms the time-limited LH (red areas) in some cases for large search time budgets and bigger search radii. Accordingly, using both algorithms selectively appears to be a reasonable deployment strategy in practice.
\begin{figure}[!ht]
	\centering
	\vspace{-0.4cm}
	\scalebox{0.35}{\input{./chapters/asset_results/NO_CHARGE_NO_WAIT_Run1.0_new.pgf}}
	\vspace{-0.4cm}
	\caption{Extensive heuristic comparison for problem variant \gls{A}\label{fig:A_comparison}}
	\fnote{Each subplot shows $\Delta \alpha = \alpha^{LH} - \alpha^{RO}$ (with $\alpha^{LH}$, resp. $\alpha^{RO}$ being the solution cost for LH, resp. RO) as a function of \gls{Tmax} and \gls{surface}, where we limit LH computational times to 1 second. Each subplot corresponds to one of the 9 scenarios resulting from the combination of each area (\gls{sf-1}, \gls{ber-1}, \gls{sf-2}) with each availability distribution (\gls{low_avail}, \gls{avg_avail}, \gls{high_avail}). Over all subplots, availability increases from left to right and station density increases from top to bottom.}
\end{figure}

\subsection{Extended analysis}\label{subsec:sensitivities}
In the following, we analyze the sensitivity of our algorithms towards parameter and design decisions. We first analyze the algorithms' sensitivity towards the penalty cost $\bar{\beta}$ for $\neg W$ problem variants, before we study the impact of a time dependent recovery function, substantiate the heuristic dominance decisions and study the impact of additional charging times due to battery depletion during the search.

\begin{figure}[htbp]
	\centering
	\begin{subfigure}{1.0\textwidth}
		\centering
		\scalebox{0.46}{
\begin{tikzpicture}[font=\large]

\definecolor{color0}{rgb}{0.12156862745098,0.466666666666667,0.705882352941177}
\definecolor{color1}{rgb}{1,0.498039215686275,0.0549019607843137}
\definecolor{color2}{rgb}{0.172549019607843,0.627450980392157,0.172549019607843}

\begin{groupplot}[group style={group size=3 by 1, horizontal sep=2.65cm}]
\nextgroupplot[
height=8cm,
tick align=outside,
tick pos=both,
title={ },
width=10cm,
x grid style={white!69.0196078431373!black},
xlabel={$\bar{\beta}$ [minute]},
xmin=-12.5, xmax=262.5,
xtick style={color=black},
y grid style={white!69.0196078431373!black},
ylabel={Runtime [s]},
ymin=-6.74575, ymax=141.66075,
ytick style={color=black}
]
\addplot [semithick, color0]
table {%
0 0
1 0
2 0
3 0
4 0
5 0
10 0
15 0
20 0
25 0
30 0
35 0
40 0
45 0
60 0
75 0
90 0
105 0.005
120 0
150 0
175 0
200 0
220 0
250 0.005
};
\addplot [semithick, color1]
table {%
0 0
1 0
2 0
3 0
4 0
5 0
10 0
15 0
20 0
25 0
30 0
35 0
40 0
45 0
60 0
75 0
90 0
105 0
120 0
150 0
175 0
200 0
220 0
250 0
};
\addplot [semithick, color2]
table {%
0 134.915
1 122.1925
2 105.5375
3 89.465
4 76.2275
5 61.75
10 36.9975
15 26.8225
20 20.605
25 16.8875
30 14.4075
35 13.0825
40 11.5825
45 10.6
60 8.7525
75 7.11
90 5.955
105 4.63
120 4.3375
150 3.165
175 2.2375
200 1.82
220 1.465
250 1.185
};

\nextgroupplot[
height=8cm,
log basis y={10},
tick align=outside,
tick pos=both,
legend cell align={left},
legend style={fill opacity=0.8, draw opacity=1, text opacity=1, draw=white!80!black, at={(-0.8,1.1)}, anchor=south, legend columns=3, /tikz/every even column/.append style={column sep=0.5cm}, /tikz/every odd column/.append style={column sep=0.15cm}},
width=10cm,
x grid style={white!69.0196078431373!black},
xlabel={$\bar{\beta}$ [minute]},
xmin=-12.5, xmax=262.5,
xtick style={color=black},
y grid style={white!69.0196078431373!black},
ylabel={Failure Probability},
ymin=0.299402935537643, ymax=1.05910785220117,
ymode=log,
ytick style={color=black}
]
\legend{LH, RO, LE};
\addplot [semithick, color0]
table {%
0 0.91
1 0.91
2 0.91
3 0.91
4 0.5977
5 0.55085
10 0.4883
15 0.4883
20 0.48545
25 0.48545
30 0.48545
35 0.48545
40 0.48545
45 0.48545
60 0.48495
75 0.48495
90 0.48495
105 0.48495
120 0.48455
150 0.48455
175 0.48455
200 0.48455
220 0.48455
250 0.48455
};
\addplot [semithick, color1]
table {%
0 0.5047
1 0.5047
2 0.4874
3 0.4874
4 0.4874
5 0.4874
10 0.4874
15 0.4874
20 0.4874
25 0.4874
30 0.4874
35 0.4874
40 0.4874
45 0.4874
60 0.4874
75 0.4874
90 0.4874
105 0.48455
120 0.48455
150 0.48455
175 0.48455
200 0.48455
220 0.48455
250 0.48455
};
\addplot [semithick, color2]
table {%
0 1
1 1
2 0.65715
3 0.6229
4 0.430075
5 0.430075
10 0.3447
15 0.331275
20 0.325275
25 0.325275
30 0.322025
35 0.322025
40 0.322025
45 0.322025
60 0.319175
75 0.318975
90 0.318975
105 0.3171
120 0.3171
150 0.3171
175 0.3171
200 0.3171
220 0.3171
250 0.3171
};

\nextgroupplot[
height=8cm,
tick align=outside,
tick pos=left,
width=10cm,
x grid style={white!69.0196078431373!black},
xlabel={$\bar{\beta}$ [minute]},
xmin=-12.5, xmax=262.5,
xtick style={color=black},
y grid style={white!69.0196078431373!black},
ylabel={Search cost [minutes]},
ymin=0.740050940390142, ymax=2.34810802958462,
ytick style={color=black}
]
\addplot [semithick, color0]
table {%
0 0.813144444444444
1 0.813144444444444
2 0.813144444444442
3 0.813144444444437
4 1.64882814161769
5 1.75907837119441
10 2.07207915683784
15 2.0671032370998
20 2.12131977279714
25 2.11546272117355
30 2.11552002961623
35 2.10966297799265
40 2.10380592636909
45 2.10386323481176
60 2.11879259749962
75 2.11008879012897
90 2.1226959010489
105 2.10808472127751
120 2.1311719351408
150 2.1315330465517
175 2.11396493759836
200 2.12076951027119
220 2.13401754420594
250 2.13437865561685
};
\addplot [semithick, color1]
table {%
0 2.04932014837595
1 2.06366653868549
2 2.07360137034309
3 2.08456340349604
4 2.06830714595764
5 2.07926917911058
10 2.07370099284497
15 2.06813280657935
20 2.09572468165273
25 2.09015649538711
30 2.0845883091215
35 2.07902012285589
40 2.07939370723794
45 2.07382552097232
60 2.08433925286682
75 2.07357646471762
90 2.09003196725977
105 2.11738223408965
120 2.1311719351408
150 2.1315330465517
175 2.11111932853321
200 2.11792390120603
220 2.13117193514079
250 2.13743484069711
};
\addplot [semithick, color2]
table {%
0 nan
1 nan
2 nan
3 nan
4 nan
5 nan
10 1.72644782473491
15 1.79103421645392
20 1.84056791606386
25 1.83716584247352
30 1.86901049721061
35 1.86732761165792
40 1.8625306254575
45 1.86380491993792
60 2.07387603201458
75 2.07344744159843
90 2.07972743722976
105 2.26737220430398
120 2.27441936383406
150 2.27501452553033
175 2.26404612149854
200 2.26973304290258
220 2.27452935181586
250 2.27501452553033
};
\end{groupplot}

\end{tikzpicture}}
		\subcaption{\gls{A}\label{fig:beta_A}}
		\vspace{0.2cm}
	\end{subfigure}
	\begin{subfigure}{1.0\textwidth}
		\centering
		\scalebox{0.46}{
\begin{tikzpicture}[font=\large]

\definecolor{color0}{rgb}{0.12156862745098,0.466666666666667,0.705882352941177}
\definecolor{color1}{rgb}{1,0.498039215686275,0.0549019607843137}
\definecolor{color2}{rgb}{0.172549019607843,0.627450980392157,0.172549019607843}

\begin{groupplot}[group style={group size=3 by 1, horizontal sep=2.65cm}]
\nextgroupplot[
height=8cm,
tick align=outside,
tick pos=both,
title={ },
width=10cm,
x grid style={white!69.0196078431373!black},
xlabel={$\bar{\beta}$ [minute]},
xmin=-12.5, xmax=262.5,
xtick style={color=black},
y grid style={white!69.0196078431373!black},
ylabel={Runtime [s]},
ymin=-9.235125, ymax=193.937625,
ytick style={color=black}
]
\addplot [semithick, color0]
table {%
0 0.005
1 0.005
2 0.025
3 0.015
4 0.01
5 0.01
10 0.01
15 0.01
20 0.01
25 0.01
30 0.005
35 0.005
40 0.005
45 0.005
60 0.005
75 0.005
90 0.01
105 0.01
120 0.005
150 0.01
175 0.01
200 0.01
220 0.01
250 0.01
};
\addplot [semithick, color1]
table {%
0 0
1 0
2 0
3 0
4 0
5 0
10 0
15 0
20 0
25 0
30 0
35 0
40 0
45 0
60 0
75 0
90 0
105 0
120 0
150 0
175 0
200 0
220 0
250 0
};
\addplot [semithick, color2]
table {%
0 140.46
1 141.6825
2 176.9175
3 173.2175
4 175.3625
5 174.845
10 184.7025
15 175.4
20 155.9525
25 156.4525
30 157.5475
35 133.7175
40 104.1025
45 89.1175
60 58.3025
75 29.545
90 25.4225
105 23.3975
120 22.6225
150 6.5125
175 4.9125
200 4.01
220 3.52
250 3.1425
};

\nextgroupplot[
height=8cm,
log basis y={10},
tick align=outside,
tick pos=left,
width=10cm,
x grid style={white!69.0196078431373!black},
xlabel={\(\displaystyle \beta\) [minute]},
xmin=-12.5, xmax=262.5,
xtick style={color=black},
y grid style={white!69.0196078431373!black},
ylabel={Failure Probability},
ymin=0.308381454157568, ymax=1.05761872383335,
ymode=log,
ytick style={color=black}
]
\addplot [semithick, color0]
table {%
0 0.9024
1 0.9024
2 0.9024
3 0.9024
4 0.9024
5 0.9024
10 0.9024
15 0.9024
20 0.9024
25 0.9024
30 0.9024
35 0.9024
40 0.9024
45 0.9024
60 0.8515
75 0.6196
90 0.59785
105 0.59785
120 0.59785
150 0.48495
175 0.48495
200 0.48495
220 0.48495
250 0.48495
};
\addplot [semithick, color1]
table {%
0 0.81365
1 0.81365
2 0.81365
3 0.81365
4 0.81365
5 0.81365
10 0.81365
15 0.81365
20 0.74265
25 0.77575
30 0.77575
35 0.77575
40 0.77575
45 0.77575
60 0.59785
75 0.59785
90 0.59785
105 0.6048
120 0.51425
150 0.52645
175 0.5047
200 0.5136
220 0.5136
250 0.5136
};
\addplot [semithick, color2]
table {%
0 1
1 1
2 1
3 1
4 1
5 1
10 1
15 1
20 1
25 1
30 1
35 1
40 0.95935
45 0.9235
60 0.877825
75 0.421225
90 0.41775
105 0.41775
120 0.41775
150 0.33205
175 0.32615
200 0.32615
220 0.32615
250 0.32615
};

\nextgroupplot[
height=8cm,
tick align=outside,
tick pos=left,
width=10cm,
x grid style={white!69.0196078431373!black},
xlabel={\(\displaystyle \beta\) [minute]},
xmin=-12.5, xmax=262.5,
xtick style={color=black},
y grid style={white!69.0196078431373!black},
ylabel={Search cost [minutes]},
ymin=57.2451542779073, ymax=93.3244598642872,
ytick style={color=black}
]
\addplot [semithick, color0]
table {%
0 85.1397379517645
1 85.2150261106246
2 85.2630959787934
3 85.3383841376536
4 85.4136722965137
5 85.4617421646825
10 85.3489637689048
15 85.2634036638184
20 85.1506252680409
25 85.4726294809589
30 85.3598510851813
35 85.2470726894038
40 85.1342942936259
45 85.4835167972356
60 75.2971060889694
75 75.8690183344406
90 75.9011130173602
105 75.8880591146188
120 75.9022235025686
150 84.2668018230081
175 84.2736063956809
200 84.2531926776624
220 84.2628985194587
250 84.266801823008
};
\addplot [semithick, color1]
table {%
0 91.6753466589188
1 91.6827719238291
2 91.6629788980481
3 91.6704041629583
4 91.6778294278687
5 91.6844914285426
10 91.6671811717114
15 89.5360294172118
20 81.9123328274788
25 75.2001296911259
30 75.1847176423862
35 75.1881877688731
40 75.1727757201335
45 75.2034641373117
60 77.4733761685225
75 77.823944453986
90 77.8461655612398
105 77.0063092899476
120 82.2387768254877
150 84.107839541666
175 83.9396103813306
200 83.2806066665358
220 83.2927612493185
250 83.2910476456161
};
\addplot [semithick, color2]
table {%
0 nan
1 nan
2 nan
3 nan
4 nan
5 nan
10 nan
15 nan
20 nan
25 nan
30 nan
35 nan
40 nan
45 nan
60 nan
75 58.8954674143711
90 58.8917065427566
105 58.9005623444336
120 58.8851227136518
150 70.6548572851366
175 71.3684827693418
200 71.3558882877679
220 71.3626513067178
250 71.3614436997262
};
\end{groupplot}

\end{tikzpicture}}
		\subcaption{\gls{C}\label{fig:beta_C}}
	\end{subfigure}
	\caption{Impact of $\bar{\beta}$ on averaged computational time~$t^s$ and failure rate for the \gls{low_avail} instances}
	\label{fig:beta_sensitivity}	
\end{figure}
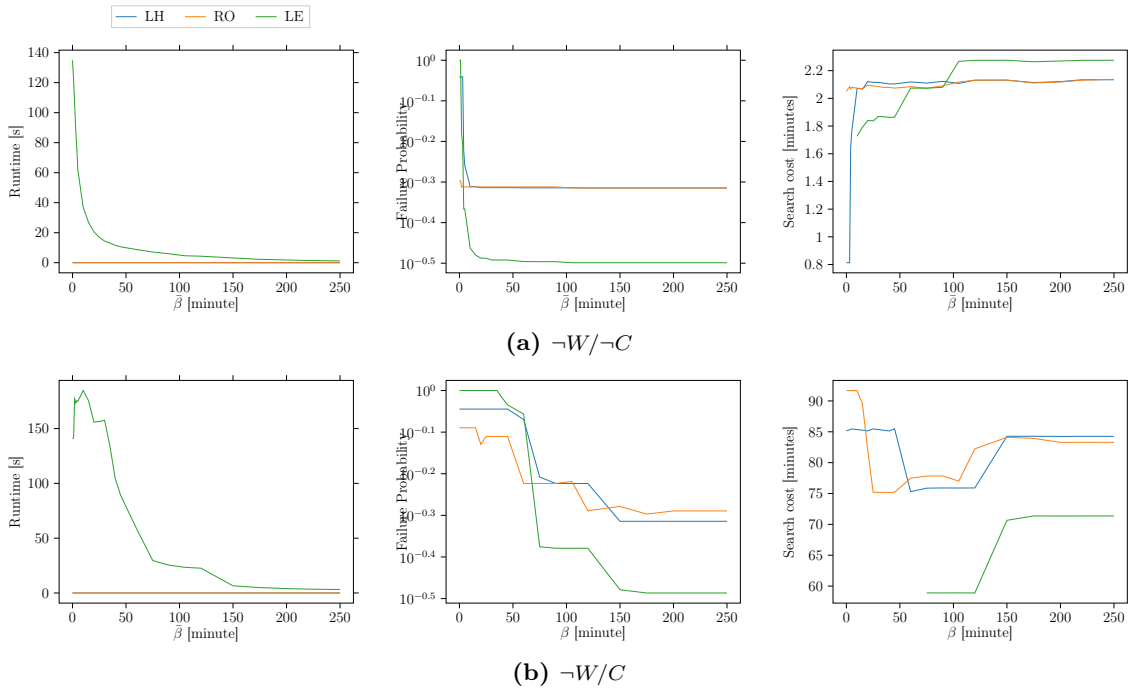

\paragraph{Termination penalty}~\\
We limit the discussion of termination penalty sensitivities to low charging station availability instances as these appear to be the most sensitive. Apparently, analyzing the search cost $\alpha$ does not allow for a meaningful interpretation of $\bar{\beta}$ sensitivities as $\alpha$ naturally increases with increasing $\bar{\beta}$. To circumvent this issue we decompose $\alpha$ into user-relevant metrics (see Appendix~\ref{sec:alpha_decomp}), and analyze the computational time, the expected search time $t_s$ and the probability $\bar{\rho}$ that the search unsuccessfully terminates, averaged over six \gls{low_avail}-instances for problem variant \gls{A} (\ref{fig:beta_sensitivity}a) and \gls{C} (\ref{fig:beta_sensitivity}b) in Figure~\ref{fig:beta_sensitivity}. 

As can be seen, we observe a goal conflict between the expected search time and the search's success rate: with increasing $\beta$ we obtain better success rates at the price of higher expected search times. We note that one must choose $\beta \geq 30$ minutes for problem variant \gls{C}, as otherwise the cost for not visiting any station is lower than the cost for visiting at least one station (i.e., minimal charging time). We observe that the $\beta$ values which are necessary to obtain the best possible success rate are significantly higher (120 minutes respectively 200 minutes for \gls{A} and \gls{C}) than this lower bound. 

While the computational times of LE significantly decrease with increasing $\beta$, the computational times of LH and RO remain insensitive to changes in $\beta$ for low charging station availability scenarios. However, additional analysis show that LH computational times increase with increasing $\beta$ in high charging station availability scenarios, such that varying $\beta$ for risk-adverse searches in a real-world implementation (i.e., choosing a higher $\beta$) should be chosen with respect to the computational overhead for LH.

\paragraph{Time-dependent recovery function}~\\
To analyze the impact of considering time-dependent recovering probabilities, we compare the objective values obtained in our main model (without recovering) with the model introduced in Section~\ref{subsec:timeDependent} (with recovering). 
In the former we compute the search path with a persistent charging station occupancy but calculate the objective value $\alpha$ considering time-dependent recovery for a fair comparison of both models. 
Based on preliminary studies, we set the average occupancy time to two hours, $\frac{1}{\mu} = 120 $ minutes.

Table~\ref{table:timeexp} compares the value of $\alpha$ for both models. As can bee seen no significant difference exists between the initial model and the updated model at the exception of smaller areas with large time budget, particularly in case of low station availability. 
\begin{table}[!ht]
	\centering
	\begin{threeparttable}	
		\caption{Potential solution improvement for the time-dependent probability recovery function for problem variant \gls{A} \label{table:timeexp}}
		{\setlength{\tabcolsep}{0.15cm}\footnotesize
			\begin{tabular}{rrrrrrrrrrrrrr}
				\toprule
				&       &       & \multicolumn{5}{c}{\gls{low_avail}} &       & \multicolumn{5}{c}{\gls{avg_avail}} \\
				\cmidrule{4-8}\cmidrule{10-14}
				&       &       & \multicolumn{2}{c}{\gls{label}} &       & \multicolumn{2}{c}{\gls{rollout}} &       & \multicolumn{2}{c}{\gls{label}} &       & \multicolumn{2}{c}{\gls{rollout}} \\
				\cmidrule{4-5}\cmidrule{7-8}\cmidrule{10-11}\cmidrule{13-14}
				\multicolumn{1}{l}{ $\gls{Tmax}$} & \multicolumn{1}{l}{\gls{surface}} &       & \multicolumn{1}{l}{$\alpha^{ref}$} & \multicolumn{1}{l}{$\alpha^{new}$} &       & \multicolumn{1}{l}{$\alpha^{ref}$} & \multicolumn{1}{l}{$\alpha^{new}$} &       & \multicolumn{1}{l}{$\alpha^{ref}$} & \multicolumn{1}{l}{$\alpha^{new}$} &       & \multicolumn{1}{l}{$\alpha^{ref}$} & \multicolumn{1}{l}{$\alpha^{new}$} \\
				\midrule
				5     & 800   &       & 25.3 & 25.1 &       & 28.5 & 28.7 &       & 1.41  & 1.41  &       & 1.55  & 1.54 \\
				5     & 2000  &       & 13.9 & 13.9 &       & 19.9 & 19.8 &       & 1.34  & 1.34  &       & 1.34  & 1.34 \\
				5     & 3400  &       & 9.63  & 9.52  &       & 19.9 & 19.8 &       & 1.34  & 1.34  &       & 1.34  & 1.34 \\
				10    & 800   &       & 23.9 & 24.8 &       & 28.5 & 29.3 &       & 1.35  & 1.35  &       & 1.57  & 1.55 \\
				10    & 2000  &       & 6.84  & 6.84  &       & 15.2 & 15.2 &       & 1.22  & 1.22  &       & 1.23  & 1.22 \\
				10    & 3400  &       & 3.22  & 3.22  &       & 8.60   & 4.65  &       & 1.22  & 1.22  &       & 1.22  & 1.22 \\
				15    & 800   &       & 23.9 & 23.4 &       & 28.5 & 28.6 &       & 1.35  & 1.32  &       & 1.55  & 1.53 \\
				15    & 2000  &       & 4.23  & 4.23  &       & 8.95  & 6.84  &       & 1.22  & 1.22  &       & 1.22  & 1.22 \\
				15    & 3400  &       & 2.44  & 2.44  &       & 3.11  & 2.86  &       & 1.22  & 1.22  &       & 1.22  & 1.22 \\
				20    & 800   &       &\textbf{23.9}& \textbf{17.2} &       & 28.5 & 27.8 &       & 1.35  & 1.33  &       & 1.55  & 1.52 \\
				20    & 2000  &       & 3.56  & 3.52  &       & 8.95  & 9.21  &       & 1.22  & 1.22  &       & 1.22  & 1.22 \\
				20    & 3400  &       & 2.41  & 2.41  &       & 2.95  & 2.83  &       & 1.22  & 1.33  &       & 1.22  & 1.22 \\
				\bottomrule
		\end{tabular}}
		\begin{tablenotes}
			\small
			\item The table compares for \gls{ber-1} combined with \gls{low_avail} and \gls{avg_avail} the objective value obtained in the updated setting ($\alpha^{new}$) and the initial setting ($\alpha^{ref}$). The table excludes \gls{high_avail} results as these do not show any deviations. Significant differences are shown in bold characters.
		\end{tablenotes}
	\end{threeparttable}
\end{table}
This is the only particular case where visiting stations multiple times within the time budget might be worth an extra detour, as one could expect at least one of the observed stations to be freed.

\paragraph{Relaxed dominance criteria}~

To design our labeling heuristic, we relaxed the dominance check of LE as described in Section~\ref{sec:method}. In the following, we substantiate the design decision from Section~\ref{sec:method}. Here, we identify each possible variant of the dominance check with a boolean quintuple that signifies whether an equation (\ref{eq:dom_1}, \ref{eq:dom_2}, \ref{eq:dom_3}, \ref{eq:dom_4}, \ref{eq:dom_5_new} resp. \ref{eq:dom_5}) is active (= 1) or not (=0) in the respective dominance check, e.g., quintuple (1,0,1,0,0) identifies the dominance check variant in which only equations~\ref{eq:dom_1} and~\ref{eq:dom_3} are active.

\begin{figure}[htb!]
	\centering
	\scalebox{0.78}{\input {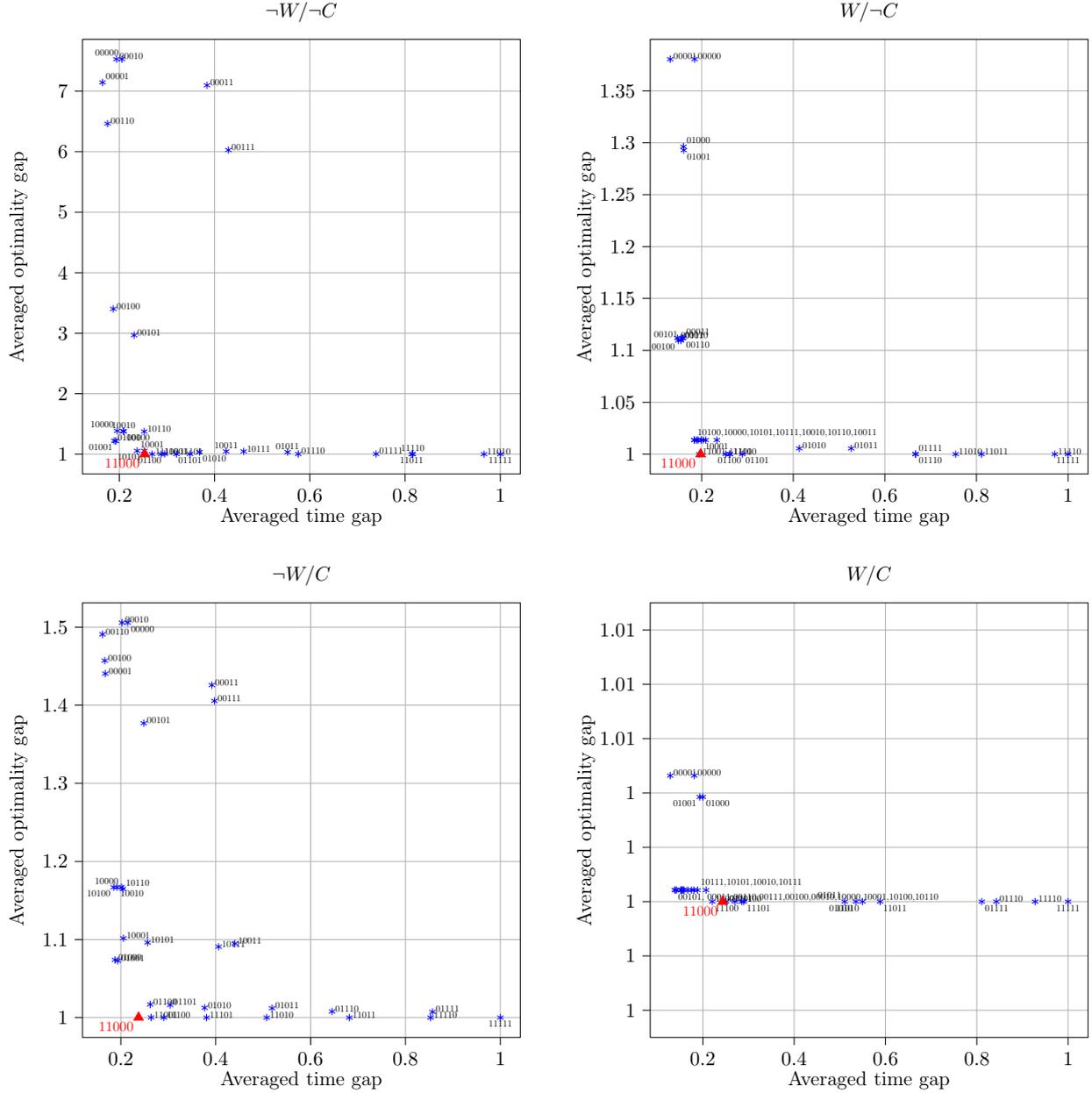}}
	\caption{Comparison of heuristic dominance criteria for all problem variants\label{fig:ref_dom}}
	\fnote{The Figure shows the averaged optimality gap $g_\alpha = \sum_{i} \nicefrac{\alpha_i}{\alpha^{\text{opt}}_i}$ as a function of the averaged computational time gap $g_t = \sum_{i} \nicefrac{t_i}{t^{\text{opt}}_i}$ for each possible heuristic criterion for all variants.
	For each variant, both values are averaged over 16 instances corresponding to \gls{ber-1} and \gls{sf-1} combined resp. with \gls{low_avail}, \gls{avg_avail} and \gls{high_avail} for $\gls{surface} \in [1200,1400,1600, 1800]$ and fixed $\gls{Tmax} = 10$. Red triangles show results for our selected dominance criteria.}
\end{figure}

Figure~\ref{fig:ref_dom} shows the trade-off between the optimality gap and the computational times for all dominance criterion and problem variants.
As can be seen, the (heuristic) dominance criterion as chosen in Section~\ref{sec:method} -- (1,1,0,0,0) -- yields the lowest computational times possible to achieve the best possible solution quality among all heuristic dominance criteria for problem variants \gls{A}, \gls{B}, and \gls{C}. For \gls{D}, (1,1,0,0,1) yields the best trade-off by slightly decreasing computational times obtained with (1,1,0,0,0) but selecting (1,1,0,0,0) allows the best possible generic implementation for LH.

\paragraph{Battery depletion}~\\
For the following experiments, we consider a flat topography and base additional charging time calculations due to battery depletion on the technical characteristics of a Renault Zoe (battery: Z.E.50 [52 kWh], engine: R110/R135, \citep{Renault2020}).
Furthermore, we assume that the battery level remains at minimum at $20\%$ of its maximum capacity to account for safety considerations or driver anxiety. 
We approximate the extra charging time based on a linear charge curve since the additional depletion adds only a limited amount to a (partial) recharge and remains in the linear part of the overall charge curve. To calculate energy consumption, we assume a constant speed of $v=50km/h$, which remains a worst case estimate in an urban context. We define $L_c$ as the expected time to charge the battery from its initial state to a full state at station $c$. Then, the depleted amount of energy after $t$ (additional) minutes of driving results to $\delta b_c(v) = \frac{t}{T_{a}(v)} \times 52 $.

In this setting, our results show that the impact of additional charging times due to an extended search and the resulting battery depletion is very limited in urban areas. For LH, we observe an average objective value change of 0.02\%, which amounts to a maximum deviation of 1.5\% for single instances. The RO algorithm appears to be even less sensitive to the changed objective and shows no significant differences. 
\section{Conclusion and outlook}\label{sec:conclusion}
In this paper, we studied charging station search algorithms for stochastic environments, motivated by real-world applications in today's navigation system applications. We introduced the underlying problem as a finite horizon \gls{mdp} that covers several real-world problem variants. In this setting, we aim to find cost minimal search paths for an individual driver. We developed three solution algorithms: an exact labeling algorithm as well as a heuristic labeling algorithm and a rollout heuristic. We benchmarked these algorithms using an extensive real-world case study with instances for the cities of San Francisco and Berlin. Our results show that the heuristic algorithms allow for a significant speed-up compared to the exact algorithm at a price of a reasonable performance loss. Moreover, we show that our algorithms significantly improve a driver's success to find a free charging station compared to myopic and greedy search approaches. This is in particular the case if the number of free charging stations in the search area is scarce. In this case, our algorithms reduce a driver's search time by up to 44\%. 

In future work, we aim to leverage this work to study the impact of coordination and information sharing between multiple drivers. By so doing, we can study the impact of additional coordination that reduces the amount of uncertainty in the system. Moreover, studying the charging station search problem from a system perspective with a perfect information setting may yield an interesting upper bound that allows for an improved assessment of the solution quality of our algorithms in a stochastic setting.


\section*{Acknowledgements}
This work was partially funded by the German Federal Ministry for Economic Affairs and Energy within project iMove (01ME16003B). The authors would like to thank Marko Rosenm{\"u}ller and Andreas Linscheid for their support in the Berlin availability study.

%
\singlespacing{
\bibliographystyle{model5-names}
\bibliography{main}} 
%
\onehalfspacing
\begin{appendices}
	\normalsize
	\section{Proofs}\label{sec:appendixProofs}

\begin{proof}[Proof of Proposition \ref{prop:explicitCost}]
	Let $\pi$ be the policy associated to sequence $C = (c_0, c_1, ..., c_n)$. 
	We consider state $x_k= (c_k,0)$ with charging station $c_k$ not being available and $c_k \neq c_n$.  Thus, for $i < n$, $\prod_{j=i}^{i} (1-p_{c_j}) = 1 $. For $i = n$, $\prod_{j=i}^{i} (1-p_{c_j}) = 1-p_{c_n}$.   
	
	We then introduce $F$ as follows
	\begin{equation}
		\begin{aligned}
		F^{ \pi}(x_k) &= \prod_{i=k}^{n} (1-\tilde{p}_{c_i}(C_{[k:i-1]})) \beta_{c_n}   + \sum_{i = k}^{n-1} \left[ t_{c_i,c_{i+1}}  \prod_{j=k}^{i} (1-\tilde{p}_{c_j}(C_{[k:j-1]}))\right] \\
		&+ \sum_{i = k}^{n} \left[ \gamma_{c_i} \tilde{p}_{c_i}(C_{[k:i-1]}) \prod_{j=k}^{i-1} (1-\tilde{p}_{c_j}(C_{[k:i-1]}))\right] \;.\\
		\end{aligned}
	\end{equation}
	
	We notice that $F^{\pi}(x_n) = (1- p_{c_n}) \beta_{c_n} + p_{c_n} \gamma_{c_n}  = V^{ \pi}(x_n) $. \\
	We now show that $F$ fulfills the recursive definition of the policy specific cost function and by recursion that $F=V$. From state $x_k = (C_{[0,k]},0)$, we let the cost for being in next state $x_{k+1} = (C_{[0,k+1]}, 0)$ and seek to express $F^{\pi}(x_k)$ as a function of $F^{ \pi}(x_{k+1}) $ to fulfill the recursive definition \ref{eq:cost_null}. 
	\begin{equation}
	\begin{split}
	F^{\pi}(x_{k+1}) =~&\prod_{j=k+1}^{n} (1-p_{c_j}) (\beta_{c_n}) + \sum_{i = k+1}^{n-1} [ t_{c_i,c_{i+1}}  \prod_{j=k+1}^{i} (1-p_{c_j})] \\
	&+ \sum_{i = k+1}^{n} [ \gamma_{c_i} p_{c_i} \prod_{j=k+1}^{i-1} (1-p_{c_j})]
	\end{split}
	\end{equation}
	
	\begin{equation}
	\begin{split}
	F^{ \pi}(x_k) =~& (1-p_{c_k})\prod_{j=k+1}^{n} (1-p_{c_j}) (\beta_{c_n})   + t_{c_k,c_{k+1}}  + (1-p_{k}) \sum_{i = k+1}^{n-1} [ t_{c_i,c_{i+1}}  \prod_{j=k+1}^{i} (1-p_{c_j})] \\
	&+ p_{c_k}\gamma_{c_k} + (1-p_{c_k}) \sum_{i = k}^{n} [ \gamma_{c_i} p_{c_i} \prod_{j=k}^{i-1} (1-p_{c_j})] \\
	F^{\pi}(x_k)  =~& t_{c_k,c_{k+1}}  + (1-p_{c_k}) F^{ \pi}(x_{k+1}) + p_{c_k}\gamma_{c_k}
	\end{split}
	\end{equation}
	Accordingly, $F$ fulfills the recursive definition \ref{eq:cost_null} for $w = 0 $, which concludes the proof.
\end{proof}

\begin{proof}[Proof of Proposition \ref{prop:neighborOnly}]
	We consider two simple search sequences $(c)$ and $(c,c')$ extended with the same visit sequence $C=(c_0, ..., c_n) $. Let $C' = (c) \circ C $, resp. $C'' = (c,c')  \circ C $, associated to policies $\pi'$, resp. $\pi''$. Let $t_{c,c_0} = t_{c,c'} +  t_{c',c_0}$ and let $c'$ be a direct neighbor of $c$ (i.e., there is no station $d$ such that  $t_{c,c'} = t_{c,d} +  t_{d,c'}$) and let $c_0$ not be a direct neighbor. 
	
	We now show that from the considered state~$x_0 = ((c),0)$, visiting~$c'$ before $c_0$ is always better than straightforwardly visiting~$c_0$. We get
	\[V^{\pi'}((c),0)  = t_{c,c_0} + (1 - \tilde{p}_{c_0}) V^{\pi'}((c,c_0) ,0), \]
	\[V^{\pi''}((c),0) =  t_{c,c'} + (1 - \tilde{p}_{c'}) t_{c',c_0}  +  (1 - \tilde{p}_{c'}) (1 - \tilde{p}_{c_0}) V^{\pi''}((c,c',c_0) ,0) \]
	and distinguish two cases: 
	\begin{description}
		\item[Case 1 ($c' \notin C$):] In this case, the valuation of any unexplored station after $c_0$ does not depend on preceding visits in the respective sequence, i.e., $ V^{\pi''}((c,c_0) ,0)  = V^{\pi''}((c,c',c_0) ,0) $.
		Given that $(1 - \tilde{p}_{c}) \leq 1$, we straightforwardly obtain $V^{\pi''}((c),0) \leq V^{\pi'}((c),0)$. 
		\item[Case 2 ($c' \in C$):] In this case, the valuation of any unexplored station after $c_0$ depends on preceding visits in the respective sequence and we obtain the following cost expansions, visiting $c'$ at position $k$:
		
		 For path  $C'$, $c'$ is visited for the first time such that $\tilde{p}_{c_k} = p_{c'}$ and we get
			\begin{multline}
			V^{\pi'}(C' ,0) =  \prod_{j=0}^{n} (1-\tilde{p}_{c_j}) (\overline{\beta})   + \sum_{i = 0}^{k-1} t_{c_i,c_{i+1}}  \prod_{j=0}^{i} (1-\tilde{p}_{c_j}) +  \sum_{i = k}^{n-1} [ t_{c_i,c_{i+1}}  (1 - \tilde{p}_{c'}) \prod_{j=0, j \neq k}^{i}  (1 - \tilde{p}_{c_j})
			\end{multline}
			
		For path  $C''$, $c'$ is visited for the second time such that $\tilde{p}_{c_k} = 0$ and we get
			\begin{multline}
			(1 - \tilde{p}_{c'}) V^{\mathcal{\pi''}}(C'' ,0) = (1 - \tilde{p}_{c'}) \prod_{j=0}^{n} (1-p_{c_j}) (\overline{\beta})   + (1 - \tilde{p}_{c'}) \sum_{i = 0}^{k-1} [ t_{c_i,c_{i+1}}  \prod_{j=0}^{i} (1-p_{c_j}) \\
			+   \sum_{i = k}^{n-1} [ t_{c_i,c_{i+1}} (1 - \tilde{p}_{c'})  \prod_{j=0, j \neq k}^{i} (1-p_{c_j})
			\end{multline}
			
		Since  $ ( 1 - \tilde{p}_{c'}) \leq 1$, we have $$( 1 - \tilde{p}_{c'}) V^{\mathcal{\pi''}}((c,c',c_0) ,0)\leq V^{\pi'}((c,c_0) ,0) $$ and consequently $$V^{\pi''}(x_0)  \le V^{\pi'}(x_0). $$ In both cases, $\pi''$ is preferred over $\pi'$ (thus $C''$ over $C'$), such that candidate stations can be restricted to neighbor stations only, which concludes the proof.		
	\end{description}
\end{proof}

\section{Problem Complexity}\label{sec:hardness}
\begin{prop}\label{prop:hardness}
  The \gls{scps} problem is NP-hard, even with metric travel times.
\end{prop}
\begin{proof}
  We show hardness through reduction from the \gls{abk:tsp} with metric and integer travel times. The decision problem variant of the \gls{abk:tsp} can be defined as follows: we consider a set $\mathcal{C}$ of $n$ sites and travel times $t_{c,c'}\in\mathbb{N}$ between these. Travel times are bounded $1\leq t_{c,c'}\leq\Delta$ for all $c,c'\in\mathcal{C}$. 
We are asked for a tour (i.e., a Hamiltonian path) $c_0,\ldots,c_n=c_0$ whose length satisfies $\sum_{i=0}^{n-1}t_{c_i,c_{i+1}}\leq\theta$ for a given $\theta$. Given that travel times are integer, we can assume w.l.o.g.\ $\theta\in\mathbb{N}$. Further, we assume that triangle inequality holds, i.e., $t_{c,c'}+t_{c',c''}\geq t_{c,c''}$ for all $c,c',c''\in\mathcal{C}$. We note that hardness for this restricted metric case implies hardness for the generic case as well.

\textit{Step 1:} We construct an instance for the station search from the~\gls{abk:tsp} instance as follows: We select an arbitrary vertex $s\in\mathcal{C}$ and designate it as start vertex $c_0:=s$ for the search. We then create a duplicate $s'$ in the same location ($t_{s,s'}=0$), which serves as the termination vertex. Let $q$ be an arbitrary value satisfying
  \begin{equation}
    \label{eq:hardness:qdef}
    \left(1-\frac{1}{\Delta(n+1)}\right)^{\tfrac{1}{n-1}}\leq q<1
    \;.
  \end{equation}
  We parameterize the search as follows: All vertices $c$ have an availability of $p_i:=1-q$ without recovery. There is no penalty for successful charging ($\gamma_i:=0$ $\forall i$). For unsuccessfully terminating at $s'$, the penalty is
  \begin{equation}\begin{aligned}
      \beta_{s'}:=\frac{2\Delta}{q(1-q)}+1 \;.\label{eq:hardness:betasdef}
    \end{aligned}\end{equation}
  For all other vertices $c\neq s'$ the penalty is
  \begin{equation}\begin{aligned}
      \beta_{c}:=\frac{1}{q^{n+1}}(\beta_{s'}+n\Delta)+1 \;.\label{eq:hardness:betaidef}
    \end{aligned}\end{equation}
  Now for any search path $C=(c_0,\ldots,c_k)$ that does not visit any vertex multiple times, it holds that
  \begin{equation}\begin{aligned}
    \label{eq:hardness:pathcost}
    \alpha(C) &= \left(\prod_{i=0}^k(1-p_{c_i})\right)\beta_{c_k} + \sum_{i=0}^{k-1}t_{c_i,c_{i+1}}\prod_{j=0}^i(1-p_{c_j}) + \sum_{i=0}^k\gamma_{c_i}p_{c_i}\prod_{j=0}^{i-1}(1-p_{c_j}) \\
    &= q^{k+1}\beta_{c_k} + \sum_{i=0}^{k-1}q^{i+1} t_{c_i,c_{i+1}}\;.
  \end{aligned}\end{equation}

\textit{Step 2:} We now claim that the TSP instance possesses a solution with cost at most $\theta\in\mathbb{N}$ if, and only if, the station search admits a search path $C$ with $\alpha(C)\leq q\theta+q^{n+1}\beta_{s'}$. This is done by transforming solutions between the two problems and carefully mapping their objective values.

For the first direction, we assume that a TSP tour is given. We convert it to a search path $C=(c_0,\ldots,c_n)$ by cutting at $s$ such that $c_0=s$ and $c_n=s'$. Then
\[
  \alpha(C)= q^{n+1}\beta_{s'} + \sum_{i=0}^{n-1}q^{i+1} t_{c_i,c_{i+1}}
  \leq q^{n+1}\beta_{s'}+q \sum_{i=0}^{n-1} t_{c_i,c_{i+1}}
  \leq q^{n+1}\beta_{s'}+q\theta
  \;.
\]
Vice versa, we assume that we are given a search path $P$ with $\alpha(P)\leq q\theta+q^{n+1}\beta_{s'}$. Then for an optimal search path $C=(c_0,\ldots,c_k)$ it holds that $\alpha(P)\leq\alpha(C)\leq q\theta+q^{n+1}\beta_{s'}$. Given metric travel times and no recovery, we can assume that $C$ does not visit any vertex more than once. We construct a tour through the following observations:
\begin{enumerate}
\item $C$ visits $s'$: Assume it does not. Let $C'$ be $C$ extended by ending at $s'$. Then
  \begin{equation}
  \begin{split}
	\alpha(C')&=\alpha(C)-q^{k+1}\beta_{c_k}+q^{k+2}\beta_{s'}+q^{k+1} t_{c_k,s'} \\
  					&= \alpha(C)-q^{k+1}(\beta_{c_k}-q\beta_{s'}-t_{c_k,s'}) \stackrel{(*)}{<} \alpha(C)
  \end{split}
  \end{equation}
  using in (*) that $\beta_{c_k}>q\beta_{s'} + \Delta$ by~\eqref{eq:hardness:betaidef}. This contradicts the optimality of $C$.
\item $C$ visits $s'$ last: Assume it does not. We obtain $C'$ from $C$ by moving $s'$ to the end. Then it holds that
  \begin{equation}\begin{split}
      \alpha(C) &\geq q^{k+1}\beta_{c_k} \geq q^{n+1}\beta_{c_k}\;\text{and}\\
      \alpha(C') & \leq q^{k+1}\beta_{s'}+\sum_{i=0}^{k-1}q^{i+1}\Delta \leq \beta_{s'}+n\Delta \;.
    \end{split}\end{equation}
  Given that $q^{n+1}\beta_{c_k}>\beta_{s'}+n\Delta$~\eqref{eq:hardness:betaidef}, it holds that $\alpha(C)>\alpha(C')$, contradicting optimality.
\item $C$ visits every vertex: Assume $C=(c_0,\ldots,c_{k-1},s')$ omits some vertex $c'$. Let $C'=(c_0,\ldots,c_{k-1},c',s')$. Then
  \begin{equation}\begin{split}
      \alpha(C)-\alpha(C') &= \left( q^kt_{c_{k-1},s}+q^{k+1}\beta_{s'} \right) - \left( q^k t_{c_{k-1},c'}+q^{k+1}t_{c',s'}+q^{k+2}\beta_{s'}\right) \\
      &= q^k( t_{c_{k-1},s}-t_{c_{k-1},c'}-q t_{c',s'} + (q-q^2)\beta_{s'} ) \\
      &\geq q^k( -2\Delta + q(1-q)\beta_{s'}) \stackrel{\eqref{eq:hardness:betasdef}}{>} 0\;,
    \end{split}\end{equation}
  again contradicting optimality. 
\item It is clear now that $C$ corresponds to a TSP tour, by identifying $s$ with $s'$. It holds that
  \begin{equation}\label{eq:hardness:objclaim}
    \alpha(C) = \sum_{i=0}^{n-1}q^{i+1}t_{c_i,c_{i+1}} + q^{n+1}\beta_{s'} \leq q\theta+q^{n+1}\beta_{s'} \;.
  \end{equation}
  Assume the tour would violate the threshold, i.e., $\sum_{i=0}^{n-1}t_{c_i,c_{i+1}}>\theta$. Given integrality, the length then is at least $\theta+1$. It follows that
  \begin{equation}\begin{split}
      \sum_{i=0}^{n-1} q^{i+1}t_{c_i,c_{i+1}}
      &\geq \sum_{i=0}^{n-1} q^nt_{c_i,c_{i+1}} = q\sum_{i=0}^{n-1} q^{n-1}t_{c_i,c_{i+1}} \\
      &\stackrel{\eqref{eq:hardness:qdef}}{\geq} q\sum_{i=0}^{n-1}(1-\frac{1}{\Delta(n+1)}) t_{c_i,c_{i+1}} 
      =q\left[\sum_{i=0}^{n-1} t_{c_i,c_{i+1}} - \sum_{i=0}^{n-1}\frac{t_{c_i,c_{i+1}}}{\Delta(n+1)} \right] \\
      &\geq q\left[\theta+1 - \sum_{i=0}^{n-1}\frac{1}{n+1} \right] > q\theta
      \;.
    \end{split}\end{equation}
  This contradicts~\eqref{eq:hardness:objclaim}, thereby proving that $\sum_{i=0}^{n-1}t_{c_i,c_{i+1}}\leq\theta$.
\end{enumerate}
\end{proof}

	\section{Search cost decomposition}\label{sec:alpha_decomp}
In this section, we show by proving Proposition~\ref{prop:alpha_decomposition} that the total search cost $\alpha(\pi)$ as derived in Section~\ref{sec:model} can be expressed as a function $t(\pi)$, $\rho(\pi)$ and an additional quantity $t^s(\pi)$ representing the expected time to find a station assuming at least one station is available among $C$, the visits sequence associated with $\pi$. We derive the quantity $t_s(\pi)$ based on the work of \cite{Arndt2016} that describes $t^s(\pi)$ as a sum of the partial accumulated driving time to a station (parking spot in their work) weighted by the probability the driver charges exactly at this station.

\begin{prop}\label{prop:alpha_decomposition}
	Cost $\alpha$ can be decomposed to exhibit $t^s(\pi)$ as follows,
	\begin{equation}\label{eq:new_alpha}
		\alpha(\pi) = t^s(\pi)  \cdot \rho(\pi)  +  \bar{\rho}(\pi)  \cdot (t(\pi)  + \beta_{c_n}).
	\end{equation}
\end{prop}

\begin{proof}
	We recall that $\alpha(\pi) = A(\pi) + \bar{\rho}(\pi)\cdot \beta_{c_n}$ (cf. Equation~\ref{eq:def_A}).
	For the sake of conciseness, we simplify the notation for the remainder of this proof as follows: $C = (0,1,..,n)$ such that $t_{k,k+1} = t_{c_{k}, c_{k+1}}$,
	$\bar{\rho}_k = \prod_{i=0}^{k} \bar{p}_k$   We let $\rho_n = \rho(\pi) $, $A_n = A(\pi) $, $t_n = t(\pi)$, $t^s_n = t^s(\pi)$. \\
	We then define $t^s$ based on \cite{Arndt2016} as
	$$t^s_n = \frac{\sum_{k=0}^{n-1} \bar{\rho}_{k-1} p_{k} (t_{k} + \gamma_k)}{\rho_n}.$$ 
	We now note that $\bar{\rho}_{k} p_{k}$ represents the probability of station $k$ being available when visited, given that no station in $i \in (0,..,k-1)$ was available.
	
	We then introduce the quantity $B_n = \sum_{k=0}^{n-1} \bar{\rho}_{k-1} p_{k} (t_{k} + \gamma_k) $ such that $t^s_n \cdot \rho_n = B_n$ and note that to prove \ref{eq:new_alpha}, it is sufficient to show
	\begin{equation}
		 A_n = B_n  +  \bar{\rho}_n  \cdot t_n,
	\end{equation}
	which follows by recursion:\\
	Step 1:	For $n=0$, \ref{eq:new_alpha} holds : $A_0 = t_{0,1}$ and $B_0 = p_1 t_{0,1} + \bar{p}_1 t_{0,1} = A_0$.\\
	Step 2: We assume that \ref{eq:new_alpha} holds and show that $A_{n+1} = B_{n+1} + \bar{\rho}_{n+1}  \cdot t_{n+1}$ holds, too:
	\begin{equation}\label{eq:B_iterated}
		\begin{aligned}
		A_{n+1} &= A_n + \bar{\rho}_n \cdot (t_{n,n+1} + p_{n+1} \gamma_{n+1})\\
		B_{n+1} &=  B_n + \bar{\rho}_n p_{n+1} \cdot (t_{n+1} +  \gamma_{n+1})
		\end{aligned}
	\end{equation}
	Given \ref{eq:new_alpha}, 
	\begin{equation}\label{eq:A_iterated}
		\begin{aligned}
			A_{n+1} &=  B_n  +  \bar{\rho}_n  \cdot t_n  + \bar{\rho}_n (t_{n, n+1} +  p_{n+1} \gamma_{n+1})\\
			\Leftrightarrow A_{n+1} &=  B_n  +  \bar{\rho}_n  \cdot t_n  + \bar{\rho}_n (t_{n, n+1} +  p_{n+1} \gamma_{n+1})+ \bar{\rho}_n p_{n+1} \cdot (t_{n+1} +  \gamma_{n+1}) - \bar{\rho}_n p_{n+1} \cdot(t_{n+1} +  \gamma_{n+1})\\
		\end{aligned}
	\end{equation} 
	From \ref{eq:B_iterated} and \ref{eq:A_iterated}, we get :
	\begin{equation}\label{eq:A_B_iterated}
	\begin{aligned}
		A_{n+1} &=  B_{n+1}  +  \bar{\rho}_n  \cdot (t_n)  + \bar{\rho}_n (t_{n, n+1} +  p_{n+1} \gamma_{n+1}) - \bar{\rho}_n p_{n+1} \cdot (t_{n+1} + \gamma_{n+1})\\
		\Leftrightarrow A_{n+1} &=  B_{n+1}  +  \bar{\rho}_n  \cdot t_{n+1}   -   \bar{\rho}_n ( p_{n+1}) t_{n+1}   \\
		\Leftrightarrow A_{n+1} &=  B_{n+1}  +  \bar{\rho}_{n+1} \cdot t_{n+1} \\
		\end{aligned}
	\end{equation} 
	This concludes the proof. 
\end{proof}

\section{Reduced action spaces results}\label{app:res_figures}
Table~\ref{tab:qualityAggreg} compares the cost $\alpha$, the computational times, and the percentage share of instances that can be computed in less than $15000$ seconds by each heuristic. Further, for the instances that can be solved to optimality within $15000$ seconds, the table compares for both heuristics the averaged optimality gap and computational time gap.

\begin{table}[!hb]
	\centering
	\begin{threeparttable}	
		\caption{Aggregated computational results over all tested instances for each problem variant \label{tab:qualityAggreg}}
		{\setlength{\tabcolsep}{0.15cm}\footnotesize
		\begin{tabular}{ccrrrrrrrrrrrrrr}\toprule
			\multicolumn{2}{c}{\textbf{Graph setup}} & \multicolumn{5}{c}{\gls{label} }  & \multicolumn{5}{c}{\gls{rollout}}   \\	
			\cmidrule(lr{1em}){3-7}\cmidrule(lr{1em}){8-12}
			Problem variant & Action space & $g_t$ & $g_\alpha$ & $\hat{t}$ & $\hat{\alpha}$ & $ \hat{n}$  & $g_t$ & $g_\alpha$ & $\hat{t}$ & $\hat{\alpha}$ & $ \hat{n}$ \\
			\midrule
			\multirow{2}{*}{\gls{A}} & \gls{direct}  & 0.21&1.01&197&7.90&0.73 &0.35&1.27&0.08&11.2&1.00  \\
			& \gls{direct_res}  & 0.16&1.04&235&7.91&0.82 &-- & -- & -- & -- & --  \\
			& \gls{res_5}  &0.17&1.00&221&7.84&0.74 &-- & -- & -- & -- & -- \\
			& \gls{all} & -- & -- & -- & -- & --  &0.23&1.13&0.58&9.54&1.00\\
			\\
			\multirow{2}{*}{\gls{B}} & \gls{direct}  &0.20&1.00&72.7&2.48&0.83 &0.20&1.03&0.03&2.52&1.00  \\
			& \gls{direct_res}  &0.16&1.02&16.7&2.49&1.00 &-- & -- & -- & -- & --  \\
			& \gls{res_5}  &0.16&1.01&250&2.46&0.78 &-- & -- & -- & -- & --  \\
			& \gls{all} &-- & -- & -- & -- & --  &0.36&1.04&0.76&2.53&1.00\\
			\\
			\multirow{2}{*}{\gls{C}} & \gls{res_5}  &0.17&1.01&311&48.6&0.84 &0.18&1.12&0.31&53.1&1.00 \\
			& \gls{all}  &0.16&1.01&486&48.4&0.84 &0.19&1.12&0.38&53.6&1.00  \\
			\\
			\multirow{2}{*}{\gls{D}} & \gls{res_5}  &0.26&1.00&120&34.5&0.83 &0.31&1.00&0.88&34.5&1.00  \\
			& \gls{all}  &0.20&1.00&100&34.5&0.83 &0.33&1.00&1.05&34.5&1.00  \\				
			\bottomrule
	\end{tabular}}
	\begin{tablenotes}
	\small
	\item Abbreviations hold as follows: $g_\alpha$ - averaged optimality gap over all tested instances with $g_\alpha = \frac{\alpha^{\text{heur}}}{\alpha^{\text{opt}}}$, $g_t$ - averaged computational time gap with $g_t = \frac{t^{\text{heur}}}{t^{\text{opt}}}$, $\hat{\alpha}$ - averaged search cost $\alpha$ [min], $\hat{t}$ - averaged computational time[s], $\hat{n}$ - rate of instances that can be computed in less than 15000 seconds.
\end{tablenotes}
\end{threeparttable}
\end{table}

Preliminary results show that using the \gls{all} action space for problem variants \gls{A} and \gls{B} with LH is computationally too heavy to be of any practical use, such that we restrict results to the other restricted action spaces. As can be seen, while \gls{direct} only slightly helps saving computational times on average for \gls{A} and LH, it allows to solve~6\% more instances within the allocated time for \gls{B}. For \gls{rollout_des}, in \gls{B}, results show a~96\% decrease of the computational time. For \gls{C}, restricting the next station visits from the current location to the ones accessible in less than~5 minutes allows to save ~36\% of the computational times compared to \gls{all}. Accordingly, Table~\ref{table:graph_heur} shows the most appropriate action space for each heuristic and problem variant.

\begin{table}[htbp]
	\centering
	\begin{threeparttable}	
		\caption{Best action space /heuristic combination per problem variant \label{table:graph_heur}}
		{\setlength{\tabcolsep}{0.15cm}\footnotesize
			\begin{tabular} {p{4cm}  p{3cm}   p{3cm}}
			\toprule
			Problem variant & \gls{label} & \gls{rollout}  \\
			\midrule
			\gls{A} & \gls{direct} &  \gls{all}\\
			\gls{B} &\gls{direct} & \gls{direct} \\
			\gls{C} & \gls{res_5} &  \gls{all} \\
			\gls{D} & \gls{all} & \gls{all} \\		
			\bottomrule
	\end{tabular}}
	\begin{tablenotes}
		\small
		\item For all problem variants, the table shows for each heuristic (\gls{label}, \gls{rollout}) the graph setting to that provides the best trade-off between computational times and solution quality.
	\end{tablenotes}
	\end{threeparttable}
\end{table}

\section{Additional tractability analysis results}\label{app:res_heur}
Figure~\ref{fig:heurComparison} shows the extensive heuristic comparisons for the remaining problem variants \gls{B}, \gls{C}, \gls{D}. 
\begin{figure}[!bp]
	\centering
	\begin{subfigure}{1\textwidth}
		\centering
		\scalebox{0.27}{\input{./chapters/asset_results/WAIT_Run1.0_new.pgf}}
		\caption{Heuristic comparison for problem variant \gls{B} \label{fig:B_comparison}}
	\end{subfigure}
%
\end{figure}
\begin{figure}[!t]\ContinuedFloat
	\centering
	\begin{subfigure}{1\textwidth}
	\centering
	\scalebox{0.27}{\input {./chapters/asset_results/CHARGE_Run1.0_new.pgf}} 
	\caption{Heuristic comparison for problem variant \gls{C} \label{fig:C_comparison}}
	\end{subfigure}
	\begin{subfigure}{1\textwidth}
		\centering
		\scalebox{0.27}{\input{./chapters/asset_results/CHARGE_WAIT_Run1.0_new.pgf}}
		\caption{Heuristic comparison for problem variant \gls{D} \label{fig:D_comparison}}
	\end{subfigure}
	\centering
	\caption{Extensive heuristic comparison for problem variants \gls{B}, \gls{C}, \gls{D} \label{fig:heurComparison}}
	\fnote{Each subplot shows $\Delta \alpha = \alpha^{LH} - \alpha^{RO}$ (with $\alpha^{LH}$, resp. $\alpha^{RO}$, the solution cost for LH, resp. RO) as a function of \gls{Tmax} and \gls{surface}, where we limit LH computational times to 1 second. Each subplot corresponds to one of the 9 scenarios resulting from the combination of each area (\gls{sf-1}, \gls{ber-1}, \gls{sf-2}) with each availability distribution (\gls{low_avail}, \gls{avg_avail}, \gls{high_avail}). Over all subplots, availability increases from left to right and station density increases from top to bottom.}
\end{figure}
\end{appendices}

\end{document}